\documentclass[fleqn,usenatbib]{mnras}

\usepackage{newtxtext,newtxmath}

\usepackage[T1]{fontenc}

\DeclareRobustCommand{\VAN}[3]{#2}
\let\VANthebibliography\thebibliography
\def\thebibliography{\DeclareRobustCommand{\VAN}[3]{##3}\VANthebibliography}

\usepackage{graphicx}	
\usepackage[dvipsnames]{xcolor}
\usepackage{amsmath}	
\usepackage{pdflscape}
\usepackage{soul}

\newcommand{\be}{\begin{equation}}
\newcommand{\ee}{\end{equation}}

\newcommand{\Msun}{M$_\odot$}
\newcommand{\Ledd}{$L_{\rm Edd}$}
\newcommand\mbh{M_{\bullet}}

\newcommand{\juan}[1]{{\color{ForestGreen} Juan: #1}}

\newcommand{\chris}[1]{{\color{blue} Chris: #1}}

\newcommand{\raj}[1]{\textcolor{magenta}{\textbf{#1}}}

\usepackage{tikz,xcolor,hyperref}
\definecolor{lime}{HTML}{A6CE39}
\DeclareRobustCommand{\orcidicon}{%
	\begin{tikzpicture}
	\draw[lime, fill=lime] (0,0) 
	circle [radius=0.16] 
	node[white] {{\fontfamily{qag}\selectfont \tiny ID}};
	\draw[white, fill=white] (-0.0625,0.095) 
	circle [radius=0.007];
	\end{tikzpicture}
	\hspace{-2mm}
}
\foreach \x in {A, ..., Z}{%
	\expandafter\xdef\csname orcid\x\endcsname{\noexpand\href{https://orcid.org/\csname orcidauthor\x\endcsname}{\noexpand\orcidicon}}
}

\title[Echo Mapping of NGC~7469]{Echo mapping of the black hole accretion flow in NGC~7469}
\author[Prince, et~al.]{
Raj Prince\orcidA$^{1,2,3},$\thanks{E-mail: pr93@st-andrews.ac.uk, rajprince59.bhu@gmail.com}
Juan~V.~Hern\'andez Santisteban\orcidB$^{3}$,
Keith Horne\orcidC$^{3}$,
J. Gelbord\orcidF$^{4}$,
Ian McHardy\orcidG$^{5}$,
\newauthor{ R. Edelson\orcidE$^{6}$,
C.~A. Onken\orcidD$^{7}$,
F.~R. Donnan\orcidH$^{8}$,
M. Vestergaard\orcidV$^{9, 10}$,
S. Kaspi\orcidQ$^{11}$,
H. Winkler\orcidX$^{12}$, 
}
\newauthor{ E. M. Cackett\orcidK$^{13}$,
H. Landt$^{14}$,
A.~J. Barth\orcidJ$^{15}$,
T. Treu\orcidU$^{16}$,
S. Valenti\orcidI$^{17}$,
P. Lira$^{18}$, 
D. Chelouche\orcidL$^{19, 20}$,
}
\newauthor{ E. Romero Colmenero\orcidM$^{21}$,
M.~R. Goad\orcidN$^{22}$,
D. H. Gonzalez-Buitrago\orcidO$^{23}$,
E. Kara\orcidP$^{24}$, and C. Villforth\orcidW$^{24}$
}
\\\\
$^{1}$ Department of Physics, Institute of Science, Banaras Hindu University, Varanasi-221005, India \\
$^{2}$Center for Theoretical Physics, Polish Academy of Sciences, Al.Lonikow, 32/46, 02-668, Warsaw, Poland\\
$^{3}$SUPA School of Physics \& Astronomy, University of St~Andrews, North Haugh, St~Andrews KY16~9NS, Scotland, UK\\
$^{4}$Spectral Sciences Inc., 4 Fourth Ave., Burlington, MA 01803, USA\\
$^{5}$Department of Physics and Astronomy, University of Southampton, SO17~1BJ, UK\\
$^{6}$ Eureka Scientific Inc., 2452 Delmer St. Suite 100, Oakland, CA 94602, USA\\
$^{7}$Research School of Astronomy \& Astrophysics, The Australian National University, Canberra, ACT 2611, Australia\\
$^{8}$Department of Physics, University of Oxford, Keble Road, Oxford, OX1~3RH, UK\\
$^{9}$ Steward Observatory, University of Arizona, 933 North Cherry Avenue, Tucson, AZ 85721, USA\\
$^{10}$ DARK, The Niels Bohr Institute, University of Copenhagen, Jagtvej 155, DK-2200  Copenhagen N, Denmark\\
$^{11}$ School of Physics and Astronomy and Wise Observatory, Tel Aviv University, Tel Aviv 6997801, Israel\\
$^{12}$ Department of Physics, University of Johannesburg, P.O.Box 524, 2006 Auckland Park, South Africa\\
$^{13}$ Department of Physics and Astronomy, Wayne State University, 666 W. Hancock Street, Detroit, MI 48201, USA\\
$^{14}$ Centre for Extragalactic Astronomy, Department of Physics, Durham University, South Road, Durham DH1~3LE, UK \\
$^{15}$ Department of Physics and Astronomy, 4129 Frederick Reines Hall, University of California, Irvine, CA 92697-4575, USA \\
$^{16}$ Department of Physics and Astronomy, University of California, Los Angeles, CA 90095, USA \\
$^{17}$Department of Physics, University of California, 1 Shields Avenue, Davis, CA 95616-5270, USA\\
$^{18}$ Department of Astronomy, Faculty of Physical and Mathematical Science, University of Chile, Santiago, Chile\\
$^{19}$ Department of Physics, Faculty of Natural Sciences, University of Haifa, Haifa 3498838, Israel \\
$^{20}$ Haifa Research Center for Theoretical Physics and Astrophysics, University of Haifa, Haifa 3498838, Israel\\
$^{21}$ South African Astronomical Observatory, P.O Box 9, Observatory 7935, Cape Town, South Africa \\
$^{22}$ School of Physics and Astronomy, University of Leicester, University Road, Leicester, LE1~7RH, UK\\
$^{23}$ Instituto de Astronomía, Universidad Nacional Autónoma de México, Km 103 Carretera Tijuana-Ensenada, 22860 Ensenada B.C., México\\
$^{24}$ MIT Kavli Institute for Astrophysics and Space Research, Massachusetts Institute of Technology, Cambridge, MA 02139, USA\\
$^{25}$ Department of Physics, University of Bath, Claverton Down, Bath BA2~7AY, UK
}
\date{Accepted XXX. Received YYY; in original form ZZZ}

\pubyear{2024}

\begin{document}
\label{firstpage}
\pagerange{\pageref{firstpage}--\pageref{lastpage}}
\maketitle

\begin{abstract}
Reverberation mapping (RM) can measure black hole accretion disc sizes and radial structure through observable light travel time lags that should increase with wavelength as $\tau\propto\lambda^{4/3}$ due to the disc's $T\propto r^{-3/4}$ temperature profile.
Our 250-day RM campaign on NGC~7469 combines sub-day cadence 7-band photometry from the Las Cumbres Observatory robotic telescopes and weekly X-ray and UVOT data from {\it Swift}.
By fitting these light curves, we measure the spectral energy distribution (SED) of the variable accretion disc, and inter-band lags of just 1.5 days across the UV to optical range.
The disc SED is close to the expected $f_\nu\propto\nu^{1/3}$, and the lags are consistent with $\tau\propto \lambda^{4/3}$, but three times larger than expected.
{ We consider several possible modifications to standard disc assumptions. 
First, for a $9\times10^6$\,M$_\odot$ black hole and 2 possible spins $a^\star=(0,1)$, we fit the X-ray-UV-optical SED with a compact relativistic corona at height $H_x\sim(46,27)\,R_g$ 
irradiating a flat disc with accretion rate $\dot{m}_{\rm Edd}\sim(0.23,0.24)$ inclined to the line of sight by $i<20^\circ$.
To fit the lags as well as the SED, this model requires a low spin $a^\star\approx0$ and boosts disc color temperatures by a factor $f_{\rm col}\approx1.8$, which
shifts reprocessed light to shorter wavelengths.}
Our Bowl model with $f_{\rm col}=1$ neglects relativity near the black hole, but 
fits the UV-optical lags and SEDs using a flat disc with
{$\dot{m}_{\rm Edd}<0.06$} and a steep outer rim at $R_{\rm out}/c\sim5-10$ days with $H/R<1\%$.
 This rim occurs near the $10^3$K dust sublimation temperature in the disc atmosphere, supporting models that invoke dust opacity to thicken the disc and launch failed radiatively-driven dusty outflows at the inner edge of the broad line region (BLR).
Finally, the disc lags and SEDs exhibit a significant excess in the $u$ and $r$ bands, suggesting Balmer continuum and H$\alpha$ emission, respectively, from the BLR.

\end{abstract}




\section{Introduction}
The supermassive black holes (SMBH) that lie at the centers of massive galaxies are crucial building blocks in the evolution and assembly of galaxies \citep{Kormendy2013, Combes2023}. The Event Horizon Telescope has produced spectacular images of the accretion flow very close to the SMBHs M87$^*$ and Sgr~A$^*$ \citep{2019ApJ...875L...1E,EHT2022}. Despite these notable examples, the prospect of direct measurements of accretion disk for a large sample of SMBH across cosmic time remains out of reach. We currently rely on indirect measurements of rapidly accreting SMBH associated with Active Galactic Nuclei (AGN) to infer the mass of the black hole at its core and the size of the accretion disc that surrounds it. The most reliable technique to estimate these physical properties is known as reverberation mapping (\citealt{1982ApJ...255..419B, 2014SSRv..183..253P}), which uses light travel time between different continuum bands and/or emission line variations as measurements of distinct regions within the AGN. This technique has successfully allowed measuring the sizes of the accretion disc, and broad-line regions (BLR) and thus to infer SMBH masses for hundreds of AGN \citep[e.g.,][]{2004ApJ...613..682P, 2015PASP..127...67B,Shen2023}.


Intensive broadband reverberation mapping (IBRM) campaigns with the Neil Gehrels {\it Swift} Observatory \citep{Gehrels:2004} and the global robotic telescope network of the Las Cumbres Observatory \citep{Brown:2013} play an essential role in monitoring dozens of AGN at a high cadence 
 (1-7 days) with broadband photometry across a large energy range (e.g., \citealt{2014ApJ...788...48S, 2014MNRAS.444.1469M, 2015ApJ...806..129E, 2016ApJ...821...56F, 2016MNRAS.456.4040T, 2017ApJ...840...41E, 2018MNRAS.480.2881M, 2018MNRAS.474.5351P, 2019ApJ...870..123E, 2020MNRAS.494.4057P, Hernandez20, Cackett_2020}). These IBRM campaigns are probing the smaller region of the accretion disc (and thus faster variability). In particular, these studies have established for the first time that AGN inter-band continuum lags increase smoothly with wavelength throughout the optical/UV, consistent with the prediction of standard thin accretion disc theory (\citealt{1998ApJ...500..162C,2007MNRAS.380..669C}), and the discovery of excess lags especially around the Balmer jump, that are attributed to the diffuse continuum emission (DCE) from the broad-line region (BLR) \citep{Korista2001, Korista2019, Netzer2022}. However, the disc sizes appear too large compared to the expected size from the standard accretion theory \citep{1973SS} and there is no simple relation between optical and X-ray variations. This strongly challenges the standard reprocessing model \citep{2019ApJ...870..123E} and stimulates the development of new models for the structure of the accretion flow and the origin of variability in AGN central engines \citep[e.g.,][]{Gardner2017,2019ApJ...879L..24K, Sun2020, Mahmoud2020}. 

In this IBRM study, we revisit NGC~7469, a very bright \citep{Seyfert1943} and well-studied nearby \citep[$z=0.016268$,][]{Springob2005} AGN with a 260-day campaign of intensive sub-day cadence monitoring with LCO and weekly monitoring with {\it Swift}. Already in the pioneering intensive monitoring experiment using {\it IUE} and {\it RXTE}, clear inter-band delays were measured between X-rays and the UV \citep{Wanders1997, Nandra1998ApJ...505..594N,1998ApJ...500..162C}. In a recent re-analysis of those data, in combination with more recent {\it Swift} and ground-based monitoring, \citealt{2020MNRAS.494.4057P} derive cross-correlation time lags relative to X-ray variations that constrain the size and radial temperature profile of the accretion disc.  The lags increase with wavelength from $\sim1$~day in the UV to $\sim 3$~days in the optical, with uncertainties of 0.5 to 1~day,
a trend that is broadly consistent with the $\tau\propto\lambda^{4/3}$ 
relation expected for the $T\propto R^{-3/4}$ profile of an accretion disc. 
These encouraging results from previous studies help to motivate our more intensive and longer-duration monitoring of NGC~7469 with LCO, which extends to longer wavelengths and significantly improves the lag constraints across the optical range.



The paper is structured as follows:  Section~\ref{sec:observations} describes the observations and data reduction steps used to prepare the LCO and {\it Swift} data, including inter-telescope calibration of the LCO light curves. 
Section~\ref{sec:timeseries} details the time-series analysis, including inter-band lag measurements from the broadband light curves, determination of the constant (host galaxy) and variable (AGN disc) SEDs.
Interpretation of the results follows in Section~\ref{sec:interpretations}, 
with modeling the SED to assess the energy budget, and to fit the lag spectrum to test models with a lamp-post geometry. 
In Section~\ref{sec:discussion} 
we discuss possible scenarios to explain the larger-than-expected lag spectra, including DCE from the BLR, and reprocessing on a bowl-shaped accretion disc with a steep outer rim.
We summarise our conclusions in Sec.~\ref{sec:conclusions}. 
Throughout the work, we adopt the standard Planck cosmology with $H_0 = 69.6$~km~s$^{-1}$~Mpc$^{-1}$, $\Omega_{\rm M} = 0.286$, and $\Omega_{\Lambda} = 0.714$, for which NGC~7469's redshift $z = 0.016268$ gives a luminosity distance $D_{\rm L} = 71$~Mpc.

\section{Observations}\label{sec:observations}
\subsection{Las Cumbres Observatory}


We monitored the multi-wavelength flux variations of NGC~7469 nucleus in 7 photometric bands (\textit{u, B, g, V, r, i, z$_s$}) using Las Cumbres Observatory \citep[LCO,][]{Brown:2013} 1-m robotic telescope network through a Key Project (PI:~Hern{\'a}ndez~Santisteban). These telescopes are equipped with nearly-identical Sinistro cameras covering a $26.5\times26.5$ arcmin field of view with 4k~$\times$~4k CCD detectors ($0.389^{\prime\prime}$~pixel$^{-1}$).
Our photometric monitoring spans 257 days, from 2021~May~11 through 2022~Jan~23,
with 359 successful visits achieving a mean cadence of $dt=0.72$~days.
On each visit a pair of exposures was taken in each filter, providing an internal consistency check on statistical errors and a means of detecting outliers arising from cosmic ray hits or other issues.
Exposure times were 120~s in $u$,
30~s in $B,g,V,r$ and $i$, and 60~s in $z_{s}$, typically achieving a photometric uncertainty of $\sim1$\%, based on the light curves of comparison stars.

\begin{figure}\label{fig:hst}
    \centering
    \includegraphics[trim=0cm 15mm 0cm 1.5cm, clip, width=0.45\textwidth]{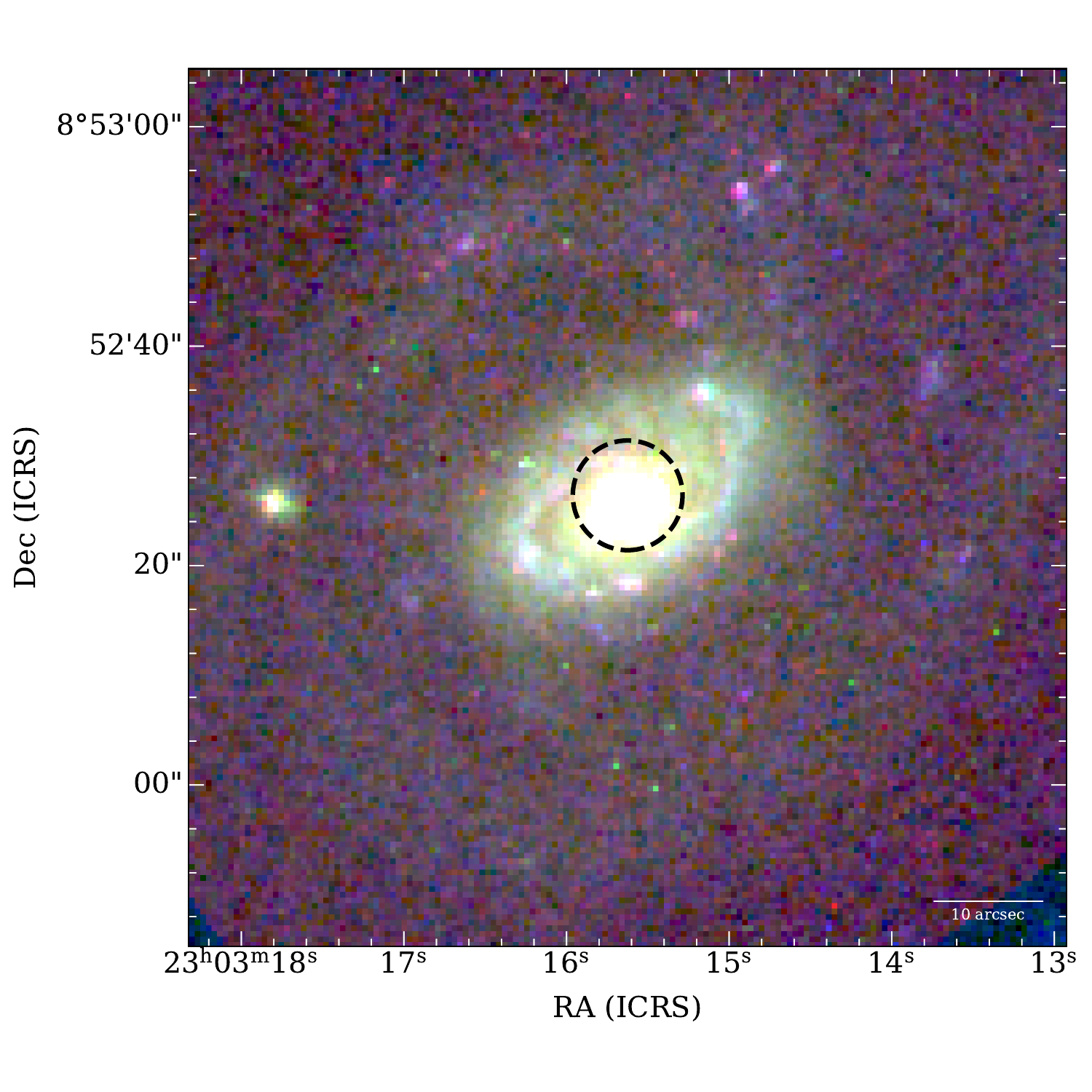}  
    \includegraphics[trim=-12mm 60mm 20mm 60mm, clip, width=0.45\textwidth]{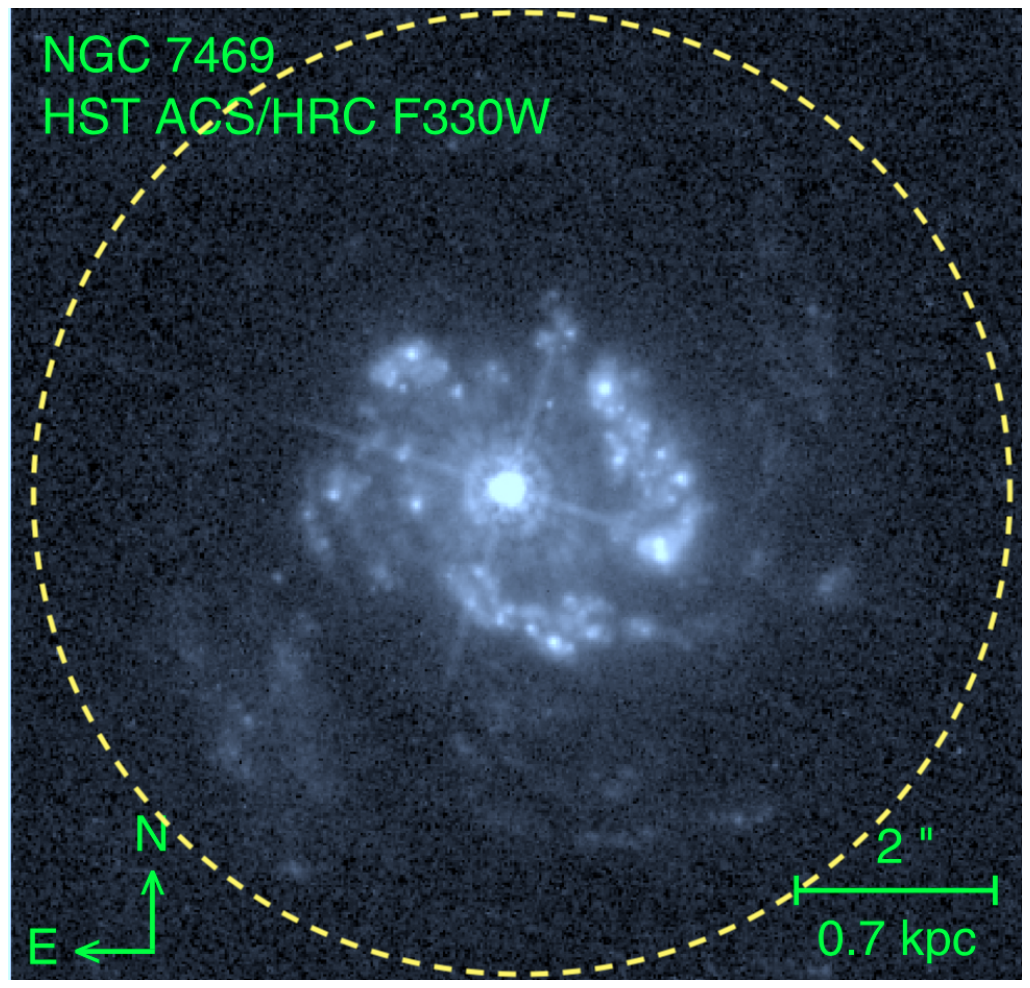}
    \caption{{Top:} Combined {\it HST} image of NGC~7469 using archival data for the following filters: F660N (red), F547M (green), and F336W (blue). The LCO and {\it Swift} 5\arcsec\ radius aperture is shown as the black dashed line.
    {Bottom: HST image from \citet{Mehdipour2018} showing the starburst ring well inside the 5\arcsec\ radius aperture.}}
\end{figure}

The processed CCD data downloaded from the LCO Archive\footnote{\url{http://archive.lco.global}} are bias and flat-field corrected images. The photometric extraction was performed through the AGN Variability Archive - AVA\footnote{\url{http://alymantara.com/ava}}. The details of the photometric extraction and absolute calibration can be found in \citet{Hernandez20}. We used {\sc SExtractor} \citep{bertin:1996} to perform aperture photometry on the stars in the field. We use a 5\arcsec\ radius aperture which provides a good compromise between signal-to-noise and seeing effects throughout the campaign. This aperture is shown in Fig.~\ref{fig:hst} in reference to a combined {\it HST} image of {the host galaxy and its nuclear starburst ring.
The aperture encloses significant host galaxy starlight including the starburst ring.}

We then obtain individual zero-points for each image by cross-matching the field with APASS \citep{henden:2018} and PAN-Starrs \citep[for $z_s$][]{Flewelling2020} and performing a bootstrapping simulation to obtain the zero-point and its $1\sigma$ uncertainty. This correction was applied to all light curves before the intercalibration described below.
\subsubsection{Inter-telescope calibration}
\begin{figure*}
    \centering
    \includegraphics[trim=25mm 8mm 18mm 25mm, clip,scale=0.45]{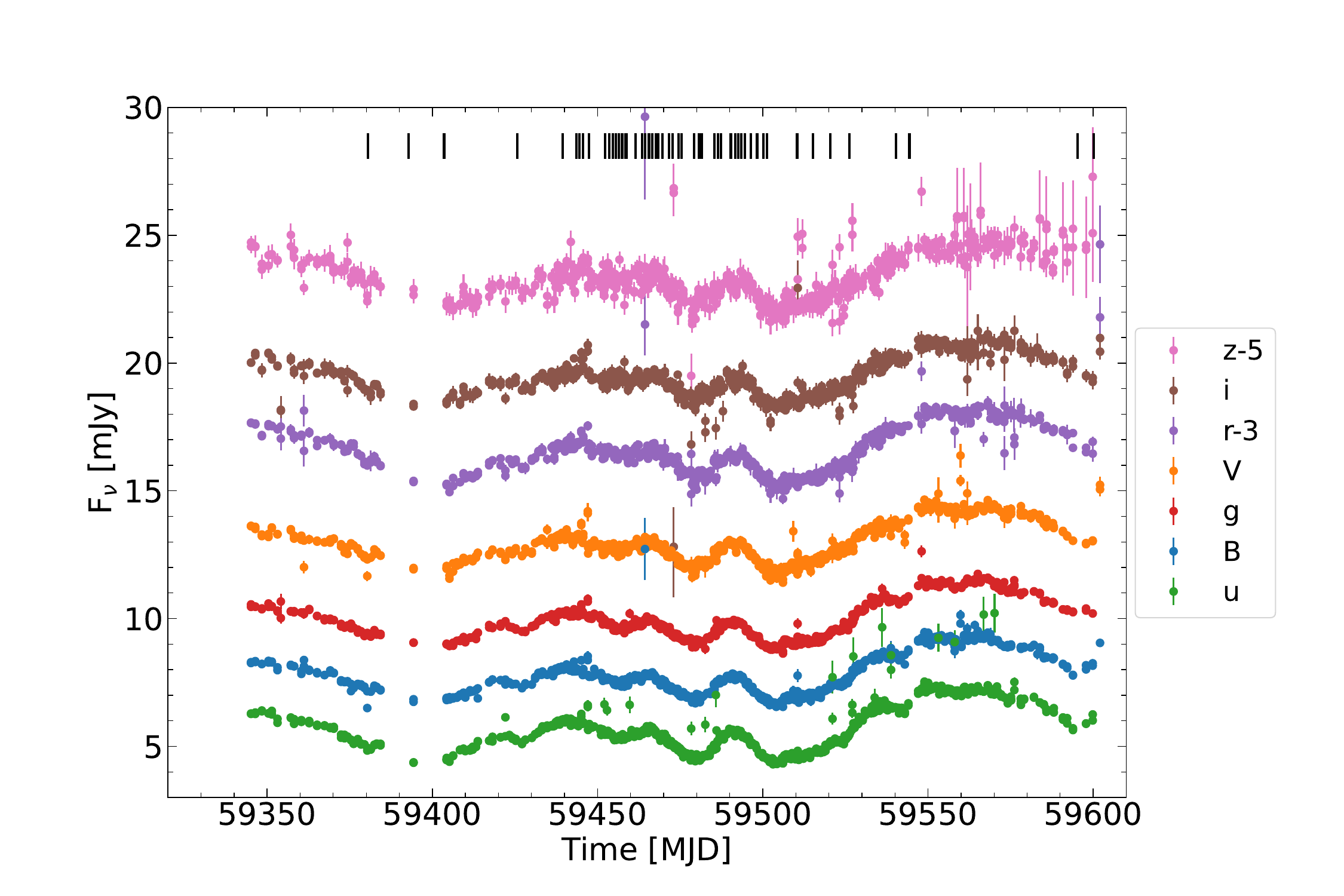}
    \caption{The inter-calibrated LCO light curves for NGC~7469 in
    7 broadband filters ($u,B,g,V,r,i, z_s$). The $r$ and $z_S$ light curves are shifted down by 5~mJy for clarity. The object is also monitored spectroscopically using a low-resolution spectrograph and the epochs at which FLOYDS spectra were taken are marked as black vertical lines along the top edge, but the FLOYDS data are not used in this work. 
    \label{fig:lco}}
\end{figure*}
The LCO light curves are measured from CCD images obtained with nine 1-m robotic telescopes. 
While these telescopes and CCD cameras are nearly identical, slight differences in their CCD response, filter bandpasses, and observing conditions produce small additive and multiplicative offsets that we calibrate to align the light curve data from different telescopes \citep[e.g.,][]{Hernandez20,Donnan2023}.

We used {\sc PyROA}\footnote{\url{https://github.com/FergusDonnan/PyROA}} to inter-calibrate these light curves,
as described in \cite{Donnan2021,Donnan2023}.
Briefly, {\sc PyROA} models the light curve shape, $X(t)$, as a running optimal average (ROA) of the inter-calibrated data from all the telescopes. The ROA model employs a Gaussian window function whose width $\Delta$ is one of the model parameters that is marginalized in the final calibration.  Small $\Delta$ improves the likelihood but incurs a penalty for increasing the flexibility of $X(t)$.  
MCMC methods based on the Bayesian information criterion (BIC) sample the joint posterior probability distribution over $3\times9+1=28$ inter-calibration parameters (additive offset $B_i$, multiplicative scaling $A_i$, and systematic uncertainty $\sigma_i$, for each of the 9 telescopes, plus $\Delta$). Default uniform priors are assumed and, because the LCO telescopes and detectors are nearly identical, constraints are applied to keep $\langle A_i \rangle = 1$ and 
$\langle B_i \rangle = 0$ averaged over all 9 telescopes.

The inter-calibration model includes extra variance parameters $\sigma_i$, unique to each telescope, which are added in quadrature with the nominal error bars. Thus the telescopes that produce light curve data noisier than expected are penalized. Their weight is reduced when we define the light curve shape.
Individual outliers on various epochs are also identified by expanding their error bars so that they deviate by no more than 4~$\sigma$ from the fitted model. This is a robust form of $\sigma$-clipping that reduces the need for human intervention to decide whether or not to omit specific outliers for subsequent analysis.

Fig.~\ref{fig:lco} presents the light curves and error bars that result from this inter-calibration procedure, showing that NGC~7469 exhibits a strongly correlated variability across the full optical range. The mean flux levels increase with wavelength and the variability amplitudes are approximately constant in $F_\nu$.
While inter-band time delays are not easily evident by eye, our subsequent analysis of these data defines inter-band time lags with
a precision of order 0.1 days.

\subsection{{\it Swift} Observatory}

\begin{figure}
    \centering
       \includegraphics[trim=20mm 12mm 20mm 20mm, clip, width=0.45\textwidth]{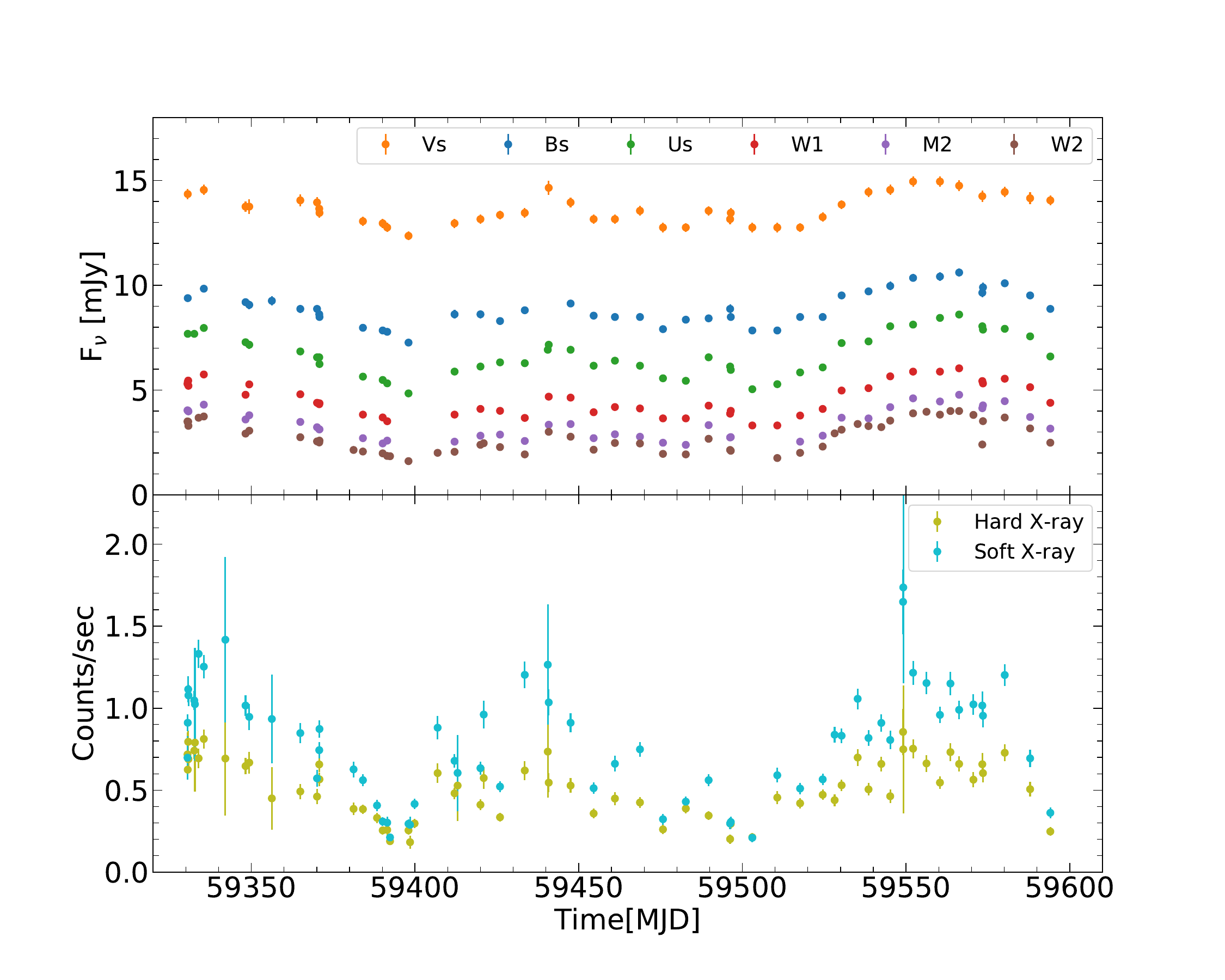}   
    \caption{The {\it Swift} UVOT and XRT light curves, shown in the upper and lower panels, respectively, for periods overlapping with the LCO light curves in Figure \ref{fig:lco}. 
    \label{fig:swift} }
\end{figure}

In parallel with our sub-day cadence LCO monitoring campaign, the Neil Gehrels {\it Swift} Observatory \citep[{\it Swift} hereafter,][]{Gehrels:2004} was monitoring NGC~7469 with a weekly cadence as part of a long-term variability program (PI: McHardy). We use the period (MJD 59320-59600) that overlaps with the LCO monitoring period.
Although the weekly cadence is insufficient to measure accurate time lags among the {\it Swift} light curves, we are able, by using the LCO lightcurves as a pattern, to estimate {\it Swift} lags relative to LCO with an accuracy of about 0.25~d. This helpfully extends useful lag measurements into the ultraviolet regime.  The {\it Swift} data also afford a wider view of the spectral energy distribution {(SED)} including the optical, UV, and X-ray regimes.
Accordingly, our analysis employs the ultraviolet and optical light curves from the UVOT \citep{2005SSRv..120...95R} and X-ray data from the XRT \citep{2005SSRv..120..165B} that were taken at times concurrent with our LCO monitoring.
These {\it Swift} light curves are shown in Fig.~\ref{fig:swift}.
 

The XRT observations were taken in photon counting (PC) mode, with an average exposure of $\sim1$~ks. The light curve and spectral extraction were performed following the standard procedure described in \citet{Evans:2009}, via the web {\it Swift} XRT tools\footnote{\url{https://www.swift.ac.uk/user_objects/}}. The light curves were split into two broad bands for their analysis: a soft X-ray band (0.3-2 keV) and a hard X-ray band (2-10 keV), as shown in the bottom panel of Fig.~\ref{fig:swift}.


The UVOT light curves of NGC~7469, shown in the top panel of Fig.~\ref{fig:swift}, were taken simultaneously with the X-ray observations, and make use of all six filters onboard {\it Swift} (3 UV bands-- UVW1, UVM2, UVW2, and 3 optical bands -- U, B, V). The light curves were extracted using a 5-arcsec radius aperture,
and the exposure sensitivity maps were used to clip dropout measurements from the light curves \citep[see details in][]{Hernandez20}. Despite the lower cadence of these observations, they provide an anchor at shorter wavelengths to probe AGN variability extending into the UV and to retrieve the SED as shown in Sec.~\ref{sec:interpretations}. 


\begin{figure*}
    \centering
    \includegraphics[width=15.8cm,trim=3.5cm 7cm 5cm 7cm,clip]{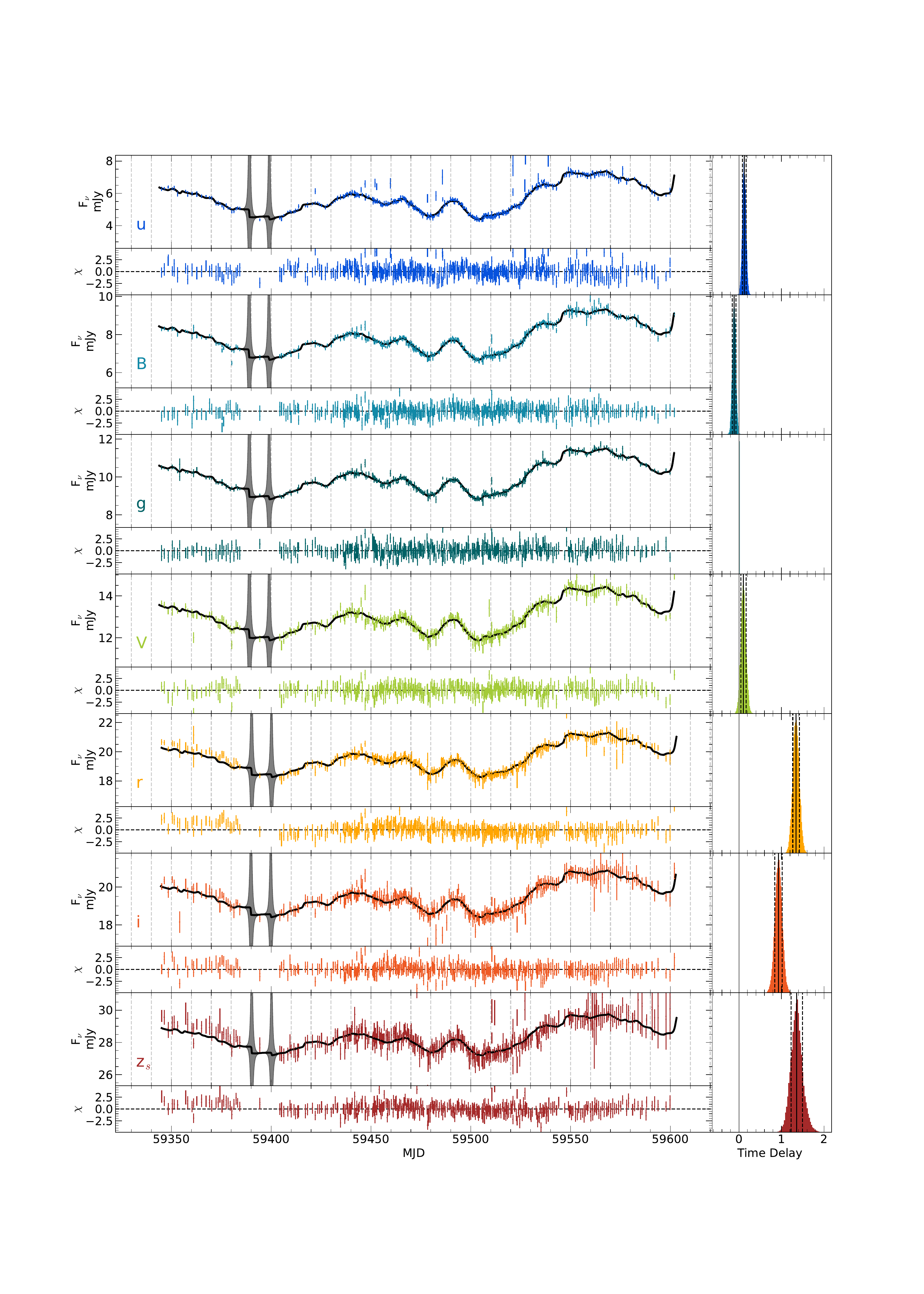}
    \caption{Simultaneous {\sc PyROA} fit to LCO light curves in all 7 photometric bands. Each light curve panel shows data (color-coded error bars) and the best-fit model light curve (solid black curve) with its $1\sigma$ uncertainty envelope (grey band). All fluxes are in mJy. The subpanel below each light curve shows the normalized residuals ($-5<\chi<+5)$ to the jointly fitted {\sc PyROA} model. The right panel shows the marginalized posterior distribution for the time lag $\tau$, relative to the $g$ band light curve, with the median value and 68\% confidence interval indicated by vertical solid and dashed black lines, respectively. 
    Time delays are in observed frame days. 
    The parameters of this fit, including the mean and RMS fluxes and the lag measurements in each band, are summarised in Table~\ref{tab:lags}.
    \label{fig:LCOlags}}
\end{figure*}

\section{Time Series Analysis}\label{sec:timeseries}

Our time-series analysis aims to measure for each band
a mean flux, a variability amplitude, and a time delay.
We do this using the {\sc PyROA} code \citep[see][for details]{Donnan2021,Donnan2023}. 
{\sc PyROA} models multi-band light curve data assuming that all light curves
have the same shape $X(t)$ 
but each light curve $\lambda$ has a different background level $\bar{F}(\lambda)$,
variation amplitude $\Delta F(\lambda)$,
and time shift $\tau(\lambda)$:
\be \label{eqn:pyroa}
    F(\lambda,t) = \bar{F}(\lambda) + \Delta F(\lambda)\, X\left( t - \tau(\lambda) \right)
    \ .
\ee
The light curve shape $X(t)$ is normalised to zero mean, $\left<X\right>=0$,
and unit variance, $\left<X^2\right>=1$, so that $\bar{F}(\lambda)$ is the mean and $\Delta F(\lambda)$ is the RMS of the model light curve at each wavelength.
This is the model introduced in \citet{Donnan2021} to estimate time delays from the light curves of multiply-lensed quasars.

The model in Eqn.~(\ref{eqn:pyroa}) approximates the delay distribution as a delta function, and thus neglects the temporal blurring caused by the range and distribution of delays that arise from
the finite size and specific geometry of the reprocessing region.
While this is clearly incorrect in detail, neglecting the width and shape of the delay distribution, it nevertheless provides useful lag measurements, especially when temporal blurring is not clearly detected in the light curves.
The {\sc PyROA} code can be used with finite width and different shapes for the delay distribution \citep{Donnan2023}, but we have not found it necessary to engage these additional parameters in our analysis.

\subsection{{\sc PyROA} fit to the 7-band LCO light curves}

Our {\sc PyROA} fit to the 7-band LCO light curve data is shown
in Fig.~\ref{fig:LCOlags}. 
The parameters, summarised in
Table~\ref{tab:lags}, include for each band the mean flux $\bar{F}$, the RMS flux $\Delta F$, the RMS of residual flux variations $\sigma$, and the time lag $\tau$ relative to the $g$ band. The parameter uncertainties are derived from the MCMC samples.
The ROA light curve shape $X(t)$, for a Gaussian window function with an RMS of $\Delta=0.98$
~d, is flexible enough to provide
a good match to the data in all bands.
Note that the uncertainty envelope (grey band) balloons out in two short data gaps,
but otherwise, $X(t)$ is tightly constrained.
The ROA fit has also coped gracefully with the small fraction of outliers, for example, particularly in the $u$ band in the top panel. 

Normalized residuals plotted below each light curve indicate a generally successful fit.
Slow trends remain in the residuals for some bands. For example,
the model is below the data near the start for the $r$, $i$, and $z_s$ bands,
suggesting that slow flux variations with 0.1--0.3~mJy amplitudes may be present that are relatively red compared with the faster variations modeled by $X(t)$ in Eqn.~(\ref{eqn:pyroa}).
The {\sc PyROA} model can be extended to fit these slow variations, but we have not pursued this here.
The model light curve turns up rapidly at the end to fit a high flux in the final epoch. This feature is probably not real, but we have not removed this final epoch.
It has little effect on the inferred lags.

Turning to the inter-band lags, the right column of panels shows posterior distributions from the MCMC samples of the lag parameters relative to the $g$ band.
Our fit includes two copies of the $g$-band data, one with and one without a lag parameter, to anchor the lags while providing a lag measurement and uncertainty for all bands, including $g$.
The time lags are tightly constrained in all bands. 
They generally increase with wavelength but lags larger than this trend are found for the $u$ and $r$ bands.
We attribute this to Balmer continuum emission in the $u$ band
and H$\alpha$ emission in the $r$ band.

\begingroup
\setlength{\tabcolsep}{3pt} 
\renewcommand{\arraystretch}{1.5} 
\begin{table*}
    \centering
    \caption{{\sc PyROA} fit parameters for 6 {\it Swift} bands (W2, M2, W1, Us, Bs, Vs) and 7 LCO bands ($u$, $B$, $g$, $V$, $r$, $i$, $z_s$). 
    For each band, $\lambda$ is the pivot wavelength, and $\Delta\lambda$ is the FWHM bandwidth (taken from \url{ http://svo2.cab.inta-csic.es/theory/fps/} for {\it Swift} and LCO).
    The light curve mean
    and RMS fluxes are $\bar{F}$ and $\Delta F$, respectively, and
    $\tau$ is the time lag (in the observed frame) relative to the $g$ band light curve. 
    The RMS residual parameter $\sigma$ is added in quadrature with the nominal error bars. We also use {\sc PyROA} to derive separately an X-ray time lag, $\tau_{\rm X-ray}$ = -1.87$^{+1.16}_{-1.03}$ days relative to the LCO $g$ band.
    }
    \begin{tabular}{lcccccc|cccc}
    \hline
    & & & \multicolumn{4}{|c|}{{\sc PyROA} fit to the LCO$+${\it Swift} light curves in Fig.~\ref{fig:lags}}
    &\multicolumn{4}{|c|}{ {\sc PyROA} fit to LCO light curves in Fig.~\ref{fig:LCOlags}}\\
    \hline
    Filter & $\lambda$& $\Delta\lambda$& $\Delta F$ & $\bar{F}$ & $\tau$ & $\sigma$&$\Delta F$ & $\bar{F}$ & $\tau$ & $\sigma$ \\
      & (\AA) & (\AA) &(mJy) & (mJy) & (day) & (mJy)&(mJy) & (mJy) & (day) & (mJy) \\
   \hline
{\it Swift} W2&1928 & 744&$  0.673^{+0.009}_{-0.009}$ & $  2.697^{+0.010}_{-0.009}$ & $ -0.630^{+0.128}_{-0.136}$ & $  0.080^{+0.008}_{-0.007}$ \\
{\it Swift} M2&2246 & 530 &$  0.716^{+0.017}_{-0.016}$ & $  3.255^{+0.014}_{-0.015}$ & $ -0.420^{+0.196}_{-0.190}$ & $  0.081^{+0.015}_{-0.016}$ \\
{\it Swift} W1&2600 & 801&$  0.773^{+0.013}_{-0.012}$ & $  4.495^{+0.012}_{-0.014}$ & $ -0.668^{+0.148}_{-0.154}$ & $  0.047^{+0.018}_{-0.022}$ \\
{\it Swift} Us&3465 & 662&$  0.972^{+0.021}_{-0.019}$ & $  6.647^{+0.017}_{-0.016}$ & $ -0.469^{+0.202}_{-0.186}$ & $  0.098^{+0.021}_{-0.019}$ \\
LCO $u$&3508 & 638 &$  0.854^{+0.003}_{-0.003}$ & $  5.733^{+0.003}_{-0.003}$ & $  0.125^{+0.045}_{-0.045}$ & $  0.032^{+0.004}_{-0.004}$ & $0.835^{+0.003}_{-0.003}$ & $   5.680^{+0.002}_{-0.002}$ & $   0.124^{+0.044}_{-0.044}$ & $   0.033^{+0.004}_{-0.005}$\\
LCO $B$ &4361 & 951&$  0.758^{+0.003}_{-0.003}$ & $  7.870^{+0.003}_{-0.003}$ & $ -0.124^{+0.044}_{-0.044}$ & $  0.006^{+0.006}_{-0.004}$ &$   0.741^{+0.003}_{-0.003}$ & $   7.823^{+0.002}_{-0.002}$ & $  -0.113^{+0.041}_{-0.041}$ & $   0.006^{+0.006}_{-0.004}$\\
{\it Swift} Bs &4392 & 866&$  0.762^{+0.025}_{-0.025}$ & $  8.883^{+0.023}_{-0.023}$ & $ -0.958^{+0.325}_{-0.298}$ & $  0.125^{+0.032}_{-0.032}$ \\
LCO $g$&4770 & 1476 &$  0.766^{+0.002}_{-0.003}$ & $ 10.031^{+0.002}_{-0.003}$ & $  0.003^{+0.038}_{-0.038}$ & $  0.005^{+0.005}_{-0.003}$ &$   0.749^{+0.002}_{-0.002}$ & $   9.983^{0.002}_{0.002}$ & $   -0.001^{+0.035}_{-0.037}$ & $   0.004^{+0.004}_{-0.003}$\\
{\it Swift} Vs &5468 & 655&$  0.689^{+0.029}_{-0.029}$ & $ 13.649^{+0.026}_{-0.026}$ & $  0.558^{+0.349}_{-0.412}$ & $  0.026^{+0.032}_{-0.018}$ \\
LCO $V$ &5468 & 836&$  0.730^{+0.004}_{-0.004}$ & $ 13.028^{+0.004}_{-0.004}$ & $  0.102^{+0.061}_{-0.061}$ & $  0.054^{+0.006}_{-0.006}$ &$   0.713^{+0.004}_{-0.004}$ & $  12.982^{+0.004}_{-0.004}$ & $   0.105^{+0.063}_{-0.063}$ & $   0.053^{+0.007}_{-0.007}$\\
LCO $r$&6215 & 1400&$  0.864^{+0.006}_{-0.006}$ & $ 19.630^{+0.006}_{-0.006}$ & $  1.364^{+0.075}_{-0.083}$ & $     0.116^{+0.006}_{-0.006}$ &$   0.845^{+0.005}_{-0.005}$ & $  19.576^{+0.005}_{-0.005}$ & $   1.342^{+0.075}_{-0.078}$ & $   0.115^{+0.006}_{-0.006}$\\
LCO $i$&7545 & 1311&$  0.704^{+0.006}_{-0.006}$ & $ 19.516^{+0.006}_{-0.005}$ & $  0.942^{+0.086}_{-0.082}$ & $  0.084^{+0.009}_{-0.009}$ &$   0.689^{0.005}_{0.005}$ & $  19.472^{+0.006}_{-0.006}$ & $   0.930^{+0.087}_{-0.086}$ & $   0.082^{+0.009}_{-0.010}$\\
LCO $z_s$&8700 & 1024&$  0.727^{+0.010}_{-0.010}$ & $ 28.351^{+0.010}_{-0.009}$ & $  1.331^{+0.142}_{-0.136}$ & $  0.122^{+0.016}_{-0.020}$&$   0.711^{+0.009}_{-0.009}$ & $  28.306^{+0.009}_{-0.009}$ & $   1.357^{+0.133}_{-0.134}$ & $   0.121^{+0.018}_{-0.023}$
\\ \hline \\
\end{tabular}
\label{tab:lags}
\end{table*}
\endgroup

\begin{figure*}
    \centering
    \includegraphics[width=16.8cm,trim=1.5cm 7.0cm 0.4cm 8cm,clip]{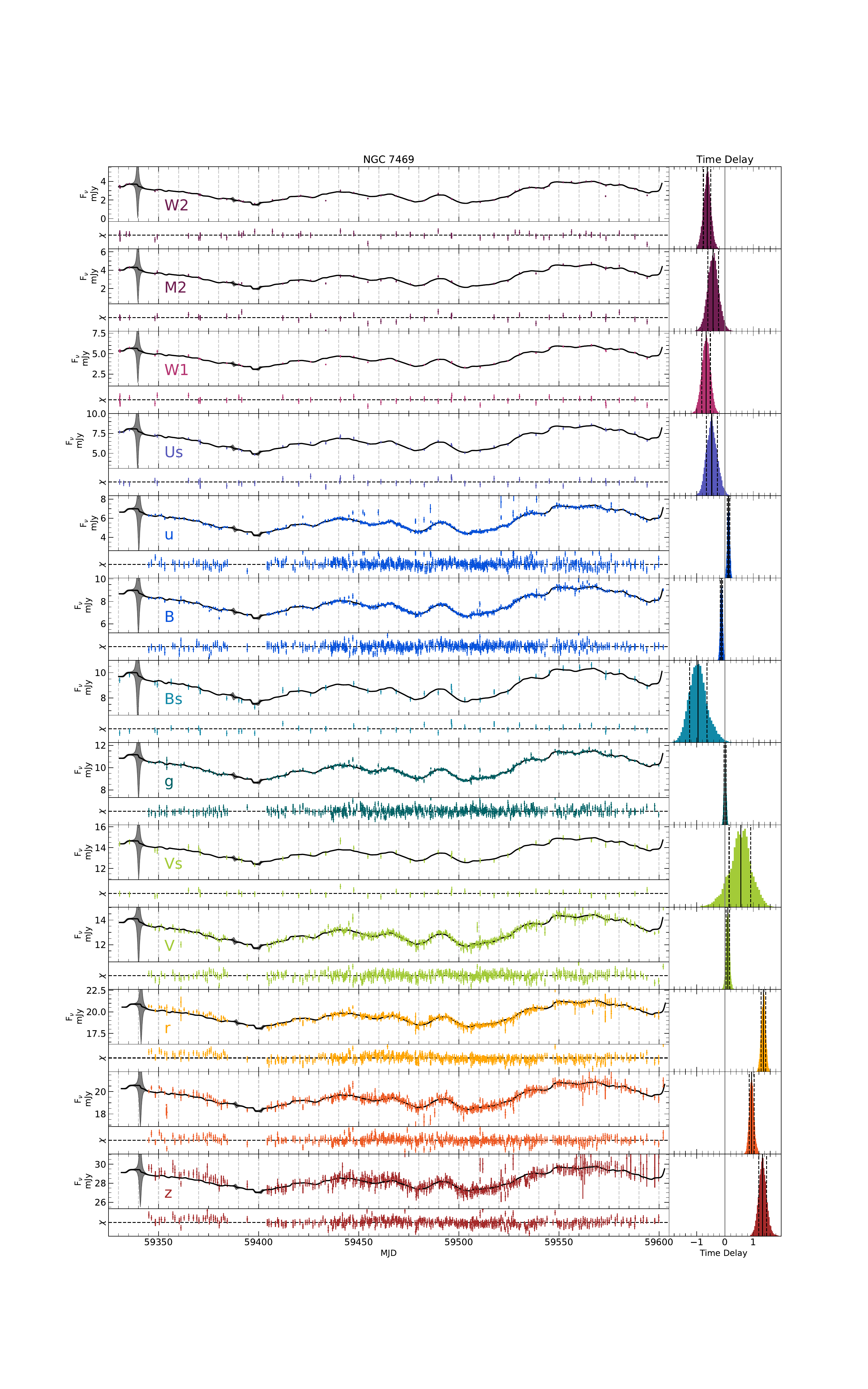}
    \caption{{\sc PyROA} fits to all the 13 light curves, the weekly cadence {\it Swift} light curves in 3 UV bands (W2, M2, W1) and 3 optical bands (Us, Bs, Vs) and the sub-day cadence 7-band LCO light curves ($u$, $B$, $g$, $V$, $r$, $i$, $z_s$). The panel format is the same as in Fig.~\ref{fig:LCOlags}.}
    \label{fig:lags}
\end{figure*}

\subsection{{\sc PyROA} fit to 6-band {\it Swift} and 7-band LCO light curves}
\label{sec:pyroa_swift}

As {\it Swift} was monitoring NGC~7469 with weekly cadence during our intensive sub-day cadence LCO monitoring, we performed additional {\sc PyROA} analyses fitting the {\it Swift} light curves on their own, with lags relative to UVW2, and fitting both {\it Swift} and LCO light curve datasets together, with lags relative to the $g$ band.
Fig.~\ref{fig:lags} shows the results of the {\sc PyROA} fit to all 13 light curves, with the data and model light curves in the left panels, and   
 the corresponding lags in the right panels. 
 Table~\ref{tab:lags} assembles the main parameters of this fit, including the mean and RMS fluxes, the rms residual parameter, and the lags relative to the $g$ band light curve.


 Including {\it Swift} data in this fit has almost no effect on the LCO lags or their uncertainties.
 Our fit to {\it Swift} data on their own did not define significant inter-band lags and this could be because of sparse sampling in the {\it Swift} data. However, 
 the fit combining both datasets yields {\it Swift} lags relative to LCO $g$ with uncertainties ranging from 0.14~d for W2 to 0.4~d for Vs. Two {\it Swift} bands, Bs and Vs, have response curves very similar to their LCO counterparts, $B$ and $V$.
 Comparing their lags relative to $g$,
 the {\it Swift} Vs lag $0.56\pm0.38$~d is consistent with the more accurate LCO $V$ lag $0.10\pm0.06$ day within 2$\sigma$,
but the {\it Swift} Bs lag $-0.96\pm0.30$~d is 2.8~$\sigma$ below the LCO B lag. 
The relatively large uncertainties in the {\it Swift} lags highlight the difficulty in measuring small lags with low-cadence monitoring data. The other issue with the {\it Swift} Bs and Vs bands is that they include a host galaxy contribution, which reduces the quality of data. Nevertheless, the {\it Swift} lags helpfully extend our optical LCO lag measurements to significantly shorter ultraviolet wavelengths.

We note that a somewhat longer lag occurs for the LCO $u$ band relative to neighboring bands at shorter and longer wavelengths. This feature is similar to the finding of excess $u$-band lags in most other well-studied AGN \citep{Hernandez20, Fausnaugh_2018} and is interpreted as due to Balmer continuum emission from the larger BLR mixing with the shorter lag from the accretion disc.
Also noteworthy here is an apparent excess lag in the LCO $r$ band, which has rarely if ever been reported in other AGN.  
We attribute this to H$\alpha$ emission from the BLR. 

\subsection{X-ray to optical lag}

We also used {\sc PyROA} to measure a
lag for the light curve of the full {\it Swift} X-ray band (0.3-10 keV) relative to the {\it g}-band light curve.
 We find a lag of $\tau_{\rm X-ray}=-1.8^{+1.2}_{-1.1}$~days. This low value is consistent with results from similar studies with joint {\it Swift} and LCO reverberation mapping experiments \citep[e.g.,][]{2019ApJ...870..123E}. We show this measurement in the lag spectrum of Fig.~\ref{fig:lagfits} and the lag spectrum fits in Sec.~\ref{sec:delay_spec}.



\begin{figure*}
    \centering
    \includegraphics[width=8.8cm]{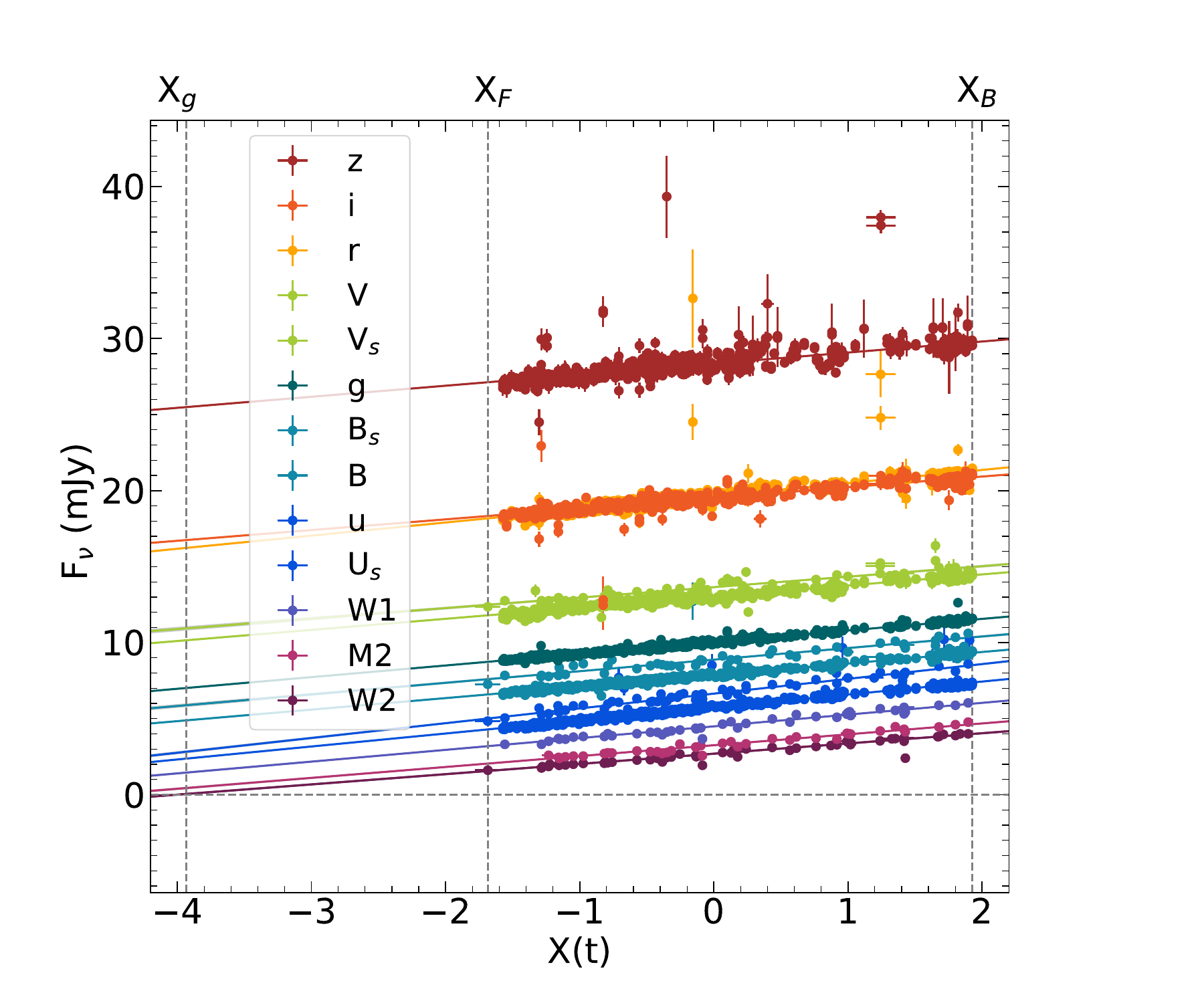}
    \includegraphics[width=8.8cm]{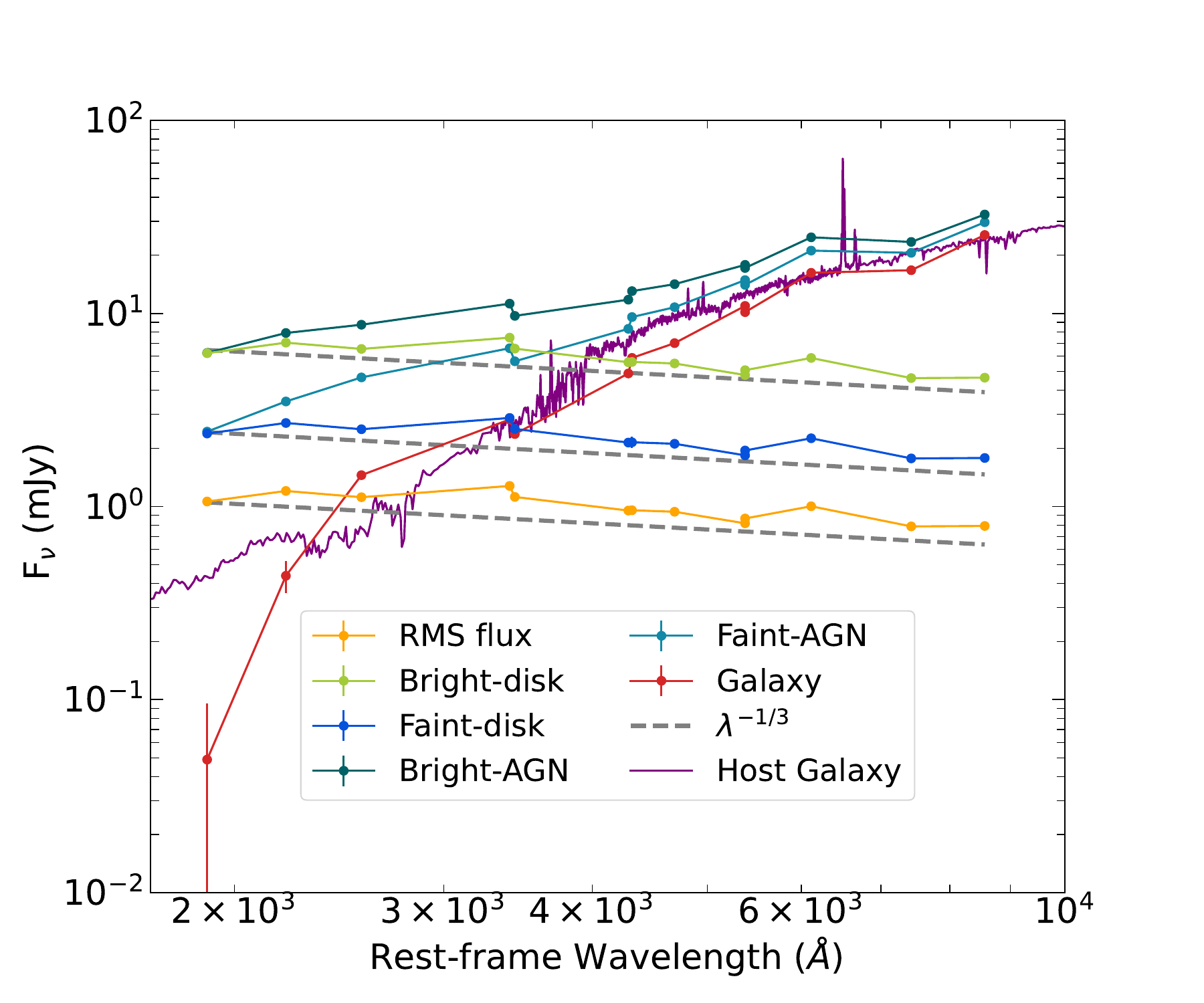}
    \caption{ Results of the flux-flux analysis. The left panel shows the observed fluxes as a function of the dimensionless light curve shape $X(t)$, which spans $\pm1$ for the mean$\pm$rms of the flux variations.
    The linear models fitted to the flux data adequately describe the variations in all bands, with the slopes $dF_\nu/dX$ giving the RMS of the light curve at each wavelength.
    Extrapolating to fainter levels, the W2 flux is 1$\sigma$ above 0 at $X_{\rm g}$.  
    Evaluating the model fluxes at $X_{\rm g}$ gives the constant (host galaxy) SED. Subtracting this leaves the variable (AGN disc) SED,
    ranging from $X_{\rm F}$ at the faintest to $X_{\rm B}$ at the brightest during the campaign.
     The right panel shows the SEDs obtained from the flux-flux plot, corrected for Galactic dust extinction and reddening. The variable disc SEDs, faint, bright, and RMS, are close to the
     theoretical prediction, $F_{\nu} \propto \lambda^{-1/3}$, shown by the dashed grey lines.
     The host galaxy SED (red) is much redder.
     A possible host contribution of NGC~7469 type galaxy (NGC 5953) is plotted in purple which has a similar shape as the estimated one.
    \label{fig:sed}}
\end{figure*}

\begin{table*}
    \centering
     \caption{Results of the flux-flux analysis. We present the static component (associated with the galaxy) and the variable component in its faintest and brightest state (associated with the AGN), before and after Galactic extinction correction E$(B-V)=0.069$ \citep{1998ApJ...500..525S}. These values are obtained from the joint analysis of the {\it Swift} and LCO data. { The average AGN flux is estimated for $X(t)$ = 0 from the flux-flux plot shown in Fig~\ref{fig:sed}.}}
    \begin{tabular}{ccccccc}
    \hline
     $\lambda$ & Galaxy flux (X$_g$) & AGN$_{\rm min}$(X$_F$)  & AGN$_{\rm max}$(X$_F$) & AGN$_{\rm min}$ & AGN$_{\rm average}$ & AGN$_{\rm max}$   \\
     & & & & (de-reddened)& (de-reddened)& (de-reddened) \\
     (\AA) & (mJy) & (mJy)  & (mJy) & (mJy)  & (mJy) & (mJy)  \\
     \hline
  1928    & $0.049\pm0.047$ & $1.514\pm0.055$ &  $3.946\pm0.056$& $2.388\pm0.086$& 4.178$\pm$0.080  & $6.223\pm0.088$ \\   
  2246     & $0.438\pm0.082$ & $1.611\pm0.094$ &  $4.198\pm0.096$& $2.708\pm0.158$& 4.737$\pm$0.146 & $7.057\pm0.161$\\
  2600    & $1.453\pm0.065$ & $1.739\pm0.074$ &  $4.532\pm0.076$& $2.513\pm0.107$ & 4.396$\pm$0.099 & $6.550\pm0.109$\\
  3465    & $2.822\pm0.101$ & $2.187\pm0.116$ &  $5.699\pm0.118$& $2.875\pm0.152$ & 5.029$\pm$0.140 &$7.493\pm0.155$\\
  3500     & $2.373\pm0.015$ & $1.921\pm0.018$ &  $5.007\pm0.018$& $2.522\pm0.023$ & 4.411$\pm$0.021 & $6.573\pm0.023$\\
 4361     & $4.887\pm0.015$ & $1.705\pm0.018$ &  $4.444\pm0.018$ & $2.144\pm0.022$ & 3.750$\pm$0.020 &$5.587\pm0.023$\\
  4392    & $5.885\pm0.124$ & $1.714\pm0.143$ &  $4.467\pm0.146$ & $2.151\pm0.179$ & 3.762$\pm$0.166 &$5.606\pm0.183$ \\
 4770     & $7.017\pm0.011$ & $1.723\pm0.014$ &  $4.491\pm0.014$ & $2.113\pm0.017$ & 3.696$\pm$0.016 &$5.508\pm0.017$\\
 5468     & $10.938\pm0.143$ & $1.550\pm0.165$ &  $4.039\pm0.168$& $1.841\pm0.195$ & 3.219$\pm$0.181  &$4.798\pm0.199$\\
 5468     & $10.156\pm0.020$ & $1.642\pm0.023$ &  $4.280\pm0.024$ & $1.950\pm0.028$ & 3.411$\pm$0.026 &$5.083\pm0.028$\\
 6215     & $16.230\pm0.030$ & $1.944\pm0.035$ &  $5.065\pm0.036$& $2.256\pm0.041$ & 3.946$\pm$0.038 &$5.880\pm0.042$\\
 7545     & $16.746\pm0.030$ & $1.584\pm0.035$ &  $4.127\pm0.036$& $1.774\pm0.039$ & 3.104$\pm$0.036 &$4.625\pm0.040$\\
 8700     & $25.490\pm0.051$ & $1.636\pm0.059$ &  $4.262\pm0.060$& $1.784\pm0.064$ & 3.120$\pm$0.059  &$4.649\pm0.065$\\
      \hline
    \end{tabular}
    \label{tab:fluxflux}
\end{table*}

\subsection{Flux-Flux analysis and Spectral Energy Distributions}

Our flux-flux analysis, illustrated in Fig.~\ref{fig:sed}, decomposes the observed light curves into constant (host galaxy) and variable (AGN disc) components, providing a separate spectral energy distribution (SED) for each component.
The left panel of Fig.~\ref{fig:sed} uses our {\sc PyROA} fit results 
to plot the observed fluxes, in mJy units, against the light curve shape $X(t-\tau)$, shifted in time to compensate for the time delay $\tau$. Note that while the mean fluxes $\bar{F}_\nu(\lambda)$ increase with wavelength, the trend lines are nearly parallel, with slopes $dF_\nu/dX=\Delta F_\nu(\lambda)$ that are nearly the same at all wavelengths.
The flux variations span a large range, up to a factor of $\sim2.5$ between the brightest and faintest UVW2 fluxes during the campaign. The left panel of Fig.~\ref{fig:sed} shows also that these variations are linear,
with no significant curvature that would be visible here if the AGN disc SED shape were to change as it brightens and fades. This validates the adequacy of our linear model, Eqn.~(\ref{eqn:pyroa}), which decomposes the observed variations into a constant component $\bar{F}_\nu(\lambda)$ and a variable component $\Delta F_\nu(\lambda)\, X(t-\tau)$.

From our {\sc PyROA} model, the mean fluxes $\bar{F}_\nu(\lambda)$ occur at $X=0$, and 
the slopes $\Delta F_\nu(\lambda)$ give the rms amplitude of the variations at each wavelength.
The faintest
and brightest states observed in the monitoring campaign occur at $X_{\rm F}\approx-1.7$ and $X_{\rm B}\approx+1.9$, respectively.
Anchored in the data between $X_{\rm F}$ and $X_{\rm B}$, 
the fitted linear trend line extrapolates
to fainter levels, effectively turning down the light of the AGN disc.
At $X_{\rm g}\approx-3.9$ the extrapolated UVW2 flux approaches zero, and the fluxes here define the SED of the host galaxy. By convention, we set $X_{\rm g}$ where the extrapolated UVW2 flux is $1~\sigma$ above 0. This is a lower limit because the host galaxy may have a more significant positive UVW2 flux from young high massive starts.

The right panel of Fig.~\ref{fig:sed} shows the resulting galaxy SED, $\bar{F}_\nu+X_g\, \Delta F_\nu$,
and the AGN disc SED, $\bar{F}_\nu+(X-X_g)\,\Delta F_\nu$,
at several brightness levels $X$, as well as their sum. The AGN disc SED is relatively blue, close to the expected $F_\nu \propto \nu^{1/3}$ power-law, with departures at the 0.1~dex level.
The relatively red host galaxy SED dominates in the optical spectrum and the AGN disc SED dominates in the UV.
The galaxy SED and the faint and bright AGN disc SEDs, before and after correction for Galactic dust extinction, are tabulated in Table~\ref{tab:fluxflux}.

NGC~7469 has a circumnuclear starburst ring, with a radius of $~1.8^{\prime\prime}$ \citep{DiazSantos2007,Armus2023}. As we see in Fig.~\ref{fig:hst}, the starburst ring is well within our 5" radius aperture and we therefore expect UV emission from these hot young stars.  This likely violates the convention of no significant UV flux that we use to set $X_g$. We should therefore slide $X_{\rm g}$ to higher values, shifting a fraction of the AGN disc SED to the host galaxy SED. This makes the galaxy SED bluer,
and scales the AGN disc SED down by a factor with little change in its spectral shape.

By modeling the HST image, we can estimate what fraction of the light within the 5" radius aperture arises from a point source vs the extended galaxy to obtain a better value of $X_g$.  We have not pursued this analysis and we expect the ambiguity in $X_g$ to have only a little effect on the interpretation of our results.
However, we plot in Fig.~\ref{fig:sed}
the spectrum of a galaxy with a mix of old and young stars, of similar type to NGC~7469. The shape is close but not a perfect match to the host galaxy SED derived from our flux-flux analysis, with a clear excess in the UVW2 band.

\section{Interpretation of the Results}\label{sec:interpretations}

\subsection{Fits to the Spectral Energy Distribution}


\begin{figure*}
      \includegraphics[width=0.45\textwidth]{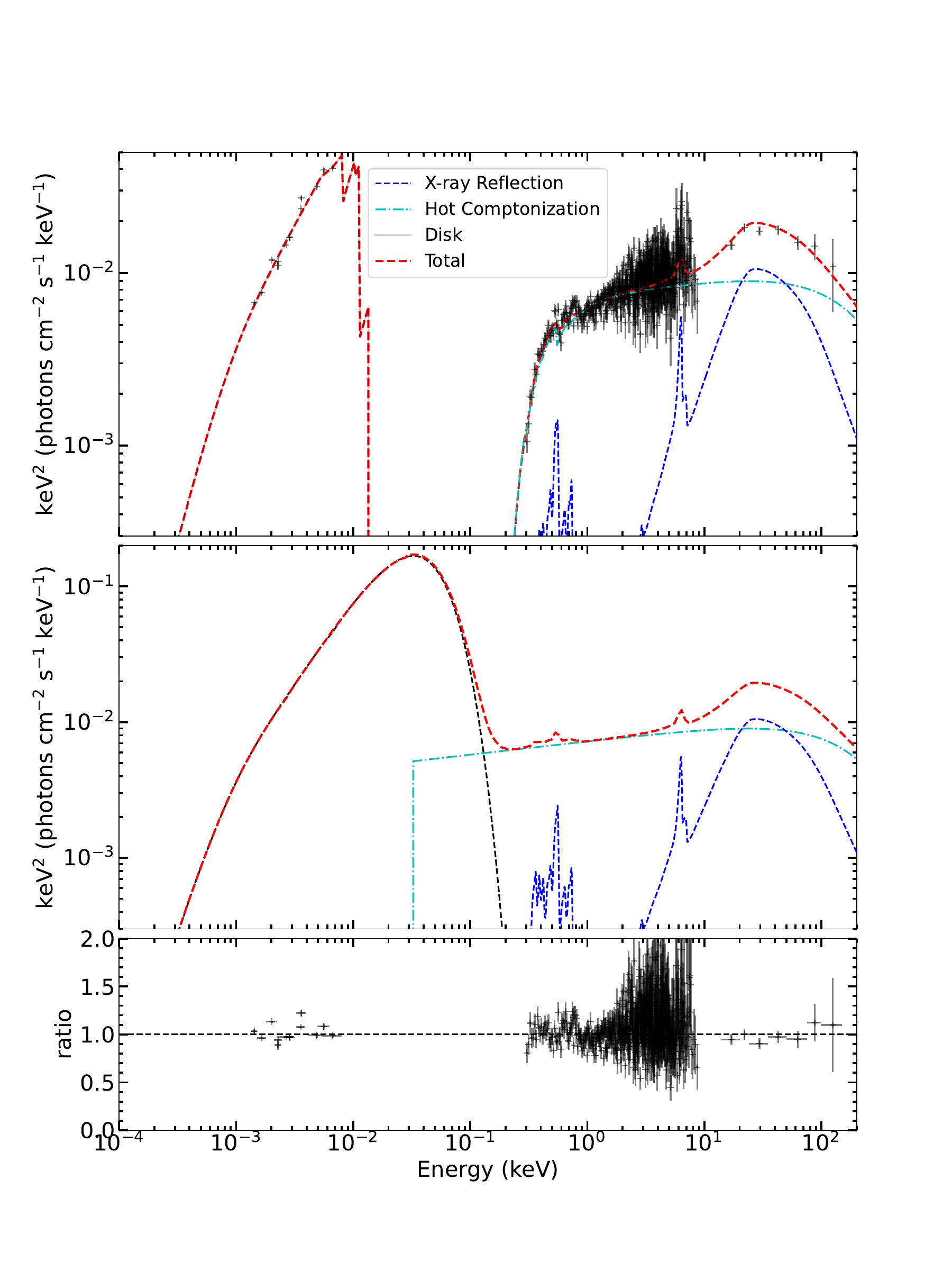}
      \includegraphics[width=0.45\textwidth]{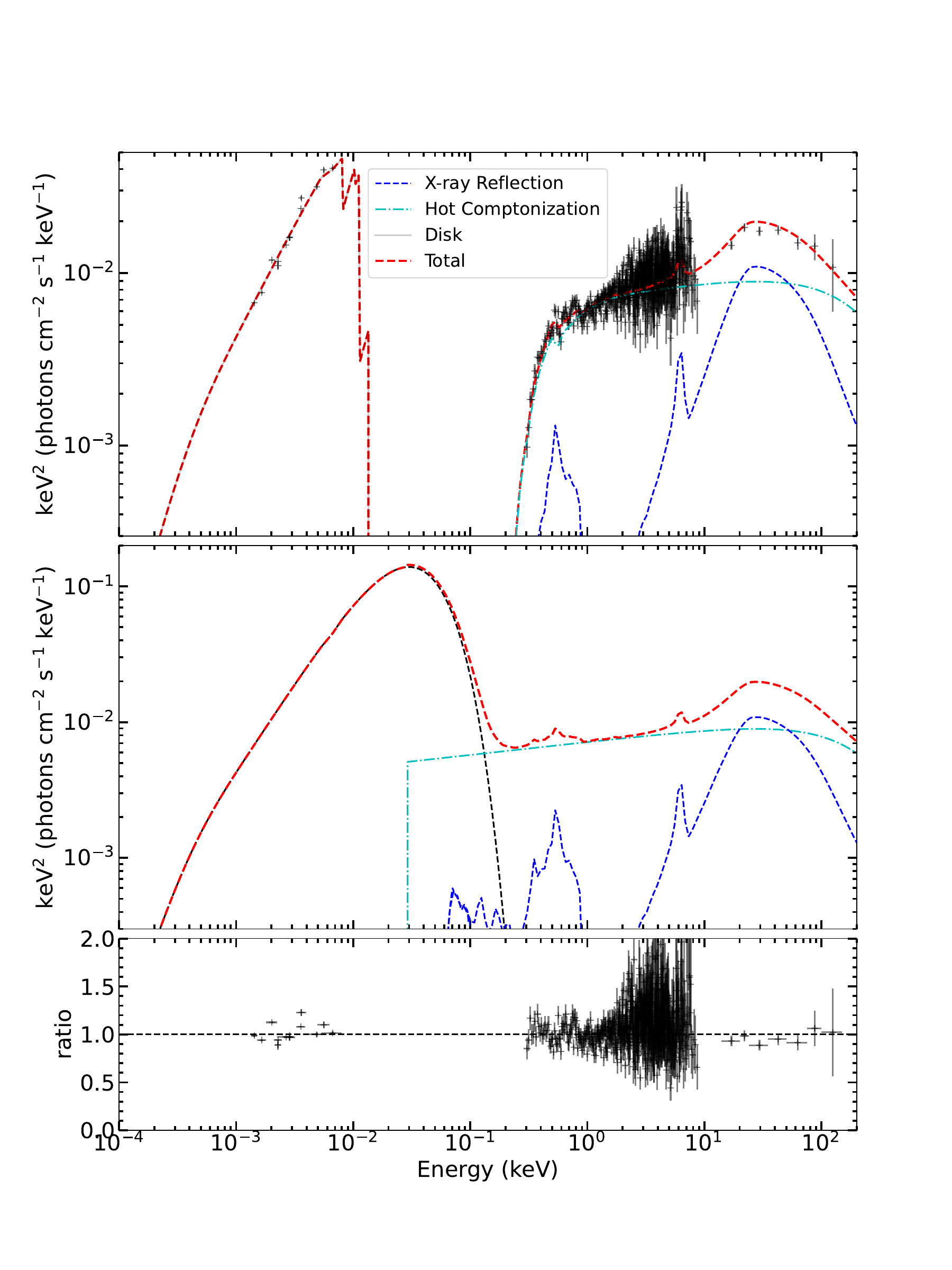}
      \caption{{ The broadband averaged SED of NGC 7469 is fitted with the \texttt{KYNSED} model in {\sc xspec}. The left panel shows the case when spin is fixed to a minimum value $\sim$0 and the right panel represents the case when spin is fixed to a maximum value 1. The optical data points are the averaged disc emission {from Table\,\ref{tab:fluxflux}}, the X-rays are from the {\it Swift} XRT between 0.3-10 keV and hard (15-200~keV) X-rays from the 70-month averaged {\it Swift} BAT spectra. The red line on the top and middle panels represents the total model with and without galactic absorption. The middle panel shows the decomposition of the total model in the disc (black), hard-comptonization (cyan), and reflection (blue) components. 
      The lower panel shows the residuals of the fit. The best-fit parameters are reported in Table~\ref{tab:total-sed}. 
      \label{fig:total-sed}}}
\end{figure*}

\begin{table}
    \centering
    \caption{{ SED parameters of the fitted \texttt{KYNSED} model for two values of black hole spin: Model-1 for spin, $a^\star=0$, and Model-2 for spin, $a^\star=1$. 
    The best-fit parameters and 68\% confidence intervals are estimated from the MCMC chain in {\sc xspec}. Parameters held fixed include the redshift ($z=0.016268$), luminosity distance ($D_{\rm L}=71.0$~Mpc), black hole mass ($\mbh=9\times10^6$~M$_{\odot}$; \citealt{2015PASP..127...67B}), disc density ($10^{15}$~cm$^{-3}$) and Fe abundance (solar) (see \citealt{2022A&A...661A.135D}) and $L_{\rm transfer}$ = 0.5.
   }}
    \begin{tabular}{lccc}
    \hline
     SED parameters & units & Model-1 & Model-2 \\ 
      &  & $a^\star=0$ & $a^\star=1$ \\ \hline 
     column density $N_{\rm H}$& $10^{20}$~cm$^{-2}$ & $5.7^{+0.6}_{-0.3}$
     & $6.2^{+0.6}_{-0.6}$ \\[1.2ex]
     accretion rate $\dot{m}_{\rm Edd}$& percent & 23$^{+1}_{-7}$ & 24$^{+2}_{-1}$ \\[1.2ex]
     $f_{\rm col}$ & - & 1.82$\pm$0.07 & 1.06$\pm$0.10  \\[1.2ex]
     inclination $i$ & deg & 14.9$^{+1.6}_{-1.2}$ & 15.0$^{+3}_{-2}$\\[1.2ex]
     outer radius $r_{\rm out}$ & $R_g$ & 10$^4$ & 10$^4$ \\[1.2ex]
     corona height $h$ & $R_g$ & 46$^{+4}_{-5}$ & 27$^{+10}_{-7}$ \\[1.2ex]
     photon index $\Gamma$ & -- & 1.90$^{+0.02}_{-0.03}$ &1.90$^{+0.09}_{-0.02}$ \\[1.2ex]
     high-energy cutoff $E_{\rm cut}$ & keV & 241$^{+16}_{-6}$ & 
     $284^{+31}_{-48}$ \\[1.2ex]

     $\chi^2/$dof. & --& 1169.2/526 &  1194.8/526\\
      \hline
    \end{tabular}
    \label{tab:total-sed}
\end{table}
AGN {emit} a very wide range of wavelengths and the complete information about the physical mechanism can only be accessed by modeling the broadband emission. This {also allows} us to estimate the accretion rate and the total energy of the system.
The broad energy range achieved by combining {\it Swift} and LCO light curves enables the study of the {SED} across the full X-ray, UV, and optical domains. 
To this end, we constructed the SED shown in Fig.~\ref{fig:total-sed}.
For the UV and optical SED, we use the average disc spectrum obtained from the flux-flux analysis (Table~\ref{tab:fluxflux}).
For the X-ray SED, we collected all the XRT observations from the {\it Swift} archive for the period concurrent with our LCO monitoring (see Fig.~\ref{fig:swift}) and constructed an average spectrum (covering the 0.3-10~keV range) using the online {\sc BUILD XRT PRODUCTS} tool (\citealt{Evans:2009}). 
To extend the SED to higher energies, we collected the 70 months of averaged spectra\footnote{\url{https://swift.gsfc.nasa.gov/results/bs70mon/}} from {\it Swift} BAT \citep{Baumgartner2013} which cover the hard X-ray band (15-200~keV). This region is particularly important as the corona's SED is expected to peak within this range \citep{Haardt1991}.

We model {with {\sc xspec}12.13.1 \citep{Arnaud1996}} the X-ray SED along with the {average} SED obtained from the flux-flux analysis, after subtracting the galaxy flux.
We used the \texttt{KYNSED} model \citep{2022A&A...661A.135D} to fit the broadband SED. Fig.~\ref{fig:total-sed} shows the best fit achieved for two spin values,
with the corresponding parameters and confidence intervals tabulated in Table~\ref{tab:total-sed}. 
Fits performed for spin-zero ($a^\star=0$) and maximally spinning ($a^\star=1$) black holes are referred to as Model-1 and Model-2, respectively. As shown in Fig.~\ref{fig:total-sed} the \texttt{KYNSED} model consists of 3 components:
the multi-temperature blackbody emission from the accretion disc, the coronal emission (a power-law continuum with high-energy cutoff), and the X-ray reflection from the disc.
The top and middle panels show the total and component SEDs with and without Galactic extinction, respectively, and the bottom panel shows normalized residuals.


{\texttt{KYNSED} computes the emission from an accretion disc that is illuminated by the X-ray corona (point-like) located on the axis above the central black hole (lamp-post geometry). The output spectrum includes the thermal disc emission, primary X-ray emission as well as the X-ray reflection in a self-consistent way.}
More details about the model can be found in \cite{2022A&A...661A.135D} and the model can be downloaded from the GitLab\footnote{\url{https://projects.asu.cas.cz/dovciak/kynsed}}. It considers a Keplerian, geometrically thin and optically thick accretion disc with an accretion rate $\dot{m}_{\rm Edd}$\footnote{the accretion rate normalized to Eddington accretion rate. i.e. $\dot{m}_{\rm Edd}\equiv\dot{M}/\dot{M}_{\rm Edd}$, where $\dot{M}$ is the accretion rate in physical units}, around a SMBH with a mass $\mbh$ and spin $a^\star$. A color correction, parameterized by $f_{\rm col}$, alters the blackbody spectrum emitted by each annulus of the accretion disc, we keep it as a free parameter for both the cases and best fit values were recovered as shown in Table~\ref{tab:total-sed}.
The X-ray corona is modeled in the lamp-post geometry as a point source situated at height $h$ above the black hole. The X-rays are emitted isotropically in the rest frame of the corona, with a power-law energy distribution parameterized by photon index $\Gamma$ and high-energy exponential cut-off at $E_{\rm cut}$. The corona is powered by the accretion flow below a transfer radius $r_{\rm transfer}$. Because the mechanism of transferring power from the inner accretion disc to the corona is not yet understood, the fraction of the accretion power transferred to the corona ($L_{\rm transfer}/L_{\rm disc}$) is fixed to 0.5.

{ The SED model parameters held fixed at typical values for this source include the redshift ($z=0.016268$), luminosity distance (71.0~Mpc), and black hole mass ($9\times10^6$~M$_\odot$; \citealt{2015PASP..127...67B}). We also fixed the parameters such as 
the accretion disc density (10$^{15}$~cm$^{-3}$), and the Fe abundance (solar) (for details see \citealt{2022A&A...661A.135D}). The remaining parameters are optimized for two cases of spin 0 and 1. The best-fit parameters and 68\% confidence intervals listed in Table~\ref{tab:total-sed}
are estimated using the MCMC package\footnote{\href{https://heasarc.gsfc.nasa.gov/xanadu/xspec/manual/node43.html}{https://heasarc.gsfc.nasa.gov/xanadu/xspec/manual/node43.html}} in {\sc xspec}. 
Both spin 0 and 1 give a similarly good fit to the data, with $\chi^2$/dof = 1169.2/526 and 1194.8/526. While $\Delta\chi^2=25.6$ nominally favors $a^\star=0$, we report results for both cases. }


{ For many of the SED parameters, the best-fit values are similar for the two spin cases.
The hydrogen column density ($N_{\rm H}$) accounts for absorption on the line of sight in our galaxy.  Consistent values are found for
$N_{\rm H}=(5.7,6.2)\times10^{20}$~cm$^{-2}$ in both cases of spin.
The accretion rate estimates, $\dot{m}_{\rm Edd}=23$\% for spin 0 and 24\% for spin 1,
 are similar and consistent within their uncertainties. 
The disc inclination is $i=14.9^\circ$ for spin 0 and $15^\circ$ for spin 1, almost equal.
The accretion disc outer radius is fixed at $r_{\rm out}=10^4\,R_g$ for both cases. 
The 
\texttt{KYNSED} model 
takes into account the height and luminosity of the hot corona. 
The corona height estimates are $h=46~R_g$ for $a^\star=0$ and 27~$R_g$ for $a^\star=1$, and the corona/disc luminosity ratio is fixed to 0.5.
The coronal X-ray photon index $\Gamma=(1.90,1.90)$ and cut-off energy $E_{\rm cut}=(241,284)$~keV are consistent within the errorbar. }

{For the \texttt{KYNSED} model we used the \texttt{XSET} command in {\sc xspec} to extract several additional physical parameters that are not directly
reported in the model.
For $a^\star=0$, the intrinsic and observed 2-10~keV X-ray luminosities are 0.065 and 0.058 in Eddington unit.
The corona radius is $32\,R_g$, the optical depth $\tau=0.99$ and 
the electron density in corona $n_{\rm e,c}= 1.7\times10^{10}$~cm$^{-3}$. 
Note this estimate for the coronal radius is comparable to its height.
The innermost stable circular orbit (ISCO)  radius and the inner edge of the disc are at 6~$R_g$. The transition (or transfer) radius is $\sim$31 $R_g$ indicating that a significant accretion flow below this radius is transferred to the corona to heat the electrons. The ratio of transferred disc power to the corona is, $L_{\rm transfer}/L_{\rm disc} = 0.5$. 
For spin 1, the ISCO radius and the inner edge of the disc are at $1.4\,R_g$. 
The transition radius at $\sim6\,R_g$ is much smaller than for $a^\star=0$ since a smaller inner part of the accretion disc suffices to power the corona. The power transferred to the corona from this part of the disc is $L_{\rm transfer}/L_{\rm disc} = 0.5$. The corona radius is estimated as 14$R_g$, the optical depth as 0.98, and the electron density in corona is $n_{\rm e,c}= 3.85\times10^{10}$~cm$^{-3}$.}

{As suggested in \citet{Mehdipour2018}, we also included the collisionally ionized component (\texttt{cie}) which is likely to be originated from the starburst activity in the source. The model has five parameters such as plasma temperature (which was kept free during the fitting), redshift, normalization, and abundance (fixed to solar), and the switch for the atomic database. Two cases for the redshift were chosen once it was fixed to the nominal value 0.01626 and in the second case it was as a free parameter. To estimate the normalization we used the value provided in \citet{Mehdipour2018}. In both cases, we do not see much improvement in the fit as the reduced $\chi^2$ does not change much suggesting the starburst activity does not have much effect on our results.}

To summarise, the \texttt{KYNSED} model tests a reverberation scenario in which the X-ray corona is powered by the inner disc, with the accretion-generated power that would normally be emitted as blackbody radiation from the inner disc transferred to and dissipated in the corona \citep{2022A&A...661A.135D}.
The corona then illuminates the accretion disc, elevating the disc temperatures and introducing reverberation and wavelength-dependent lags in the disc emission. An application of this model to the SED of NGC~5548, \citet{2022A&A...661A.135D} finds a corona height of 33-77~$R_g$ and X-ray luminosity 45-70\% of the accretion disc luminosity. 
{ In our SED modeling of NGC~7469, 
we find a somewhat lower 
coronal height 27-46~$R_g$ and 50\% of the accretion power transferred to the corona.}
 These findings indicate that the observed X-ray, UV, and optical SED can be generated by an accretion disc irradiated by an X-ray corona that is powered by the inner disc. {Note, however, that a substantial fraction of the accretion power (not just a few percent) must be transferred to the corona, and the corona height must be elevated to several tens of $R_g$ above the black hole to explain the reprocessing. }
We now consider whether this scenario, with rapid changes in the inner region causing reverberations on the disc that move out at the speed of light, can also account for the observed inter-band lags.

\begin{figure}
    \centering
    \includegraphics[width=8.0cm]{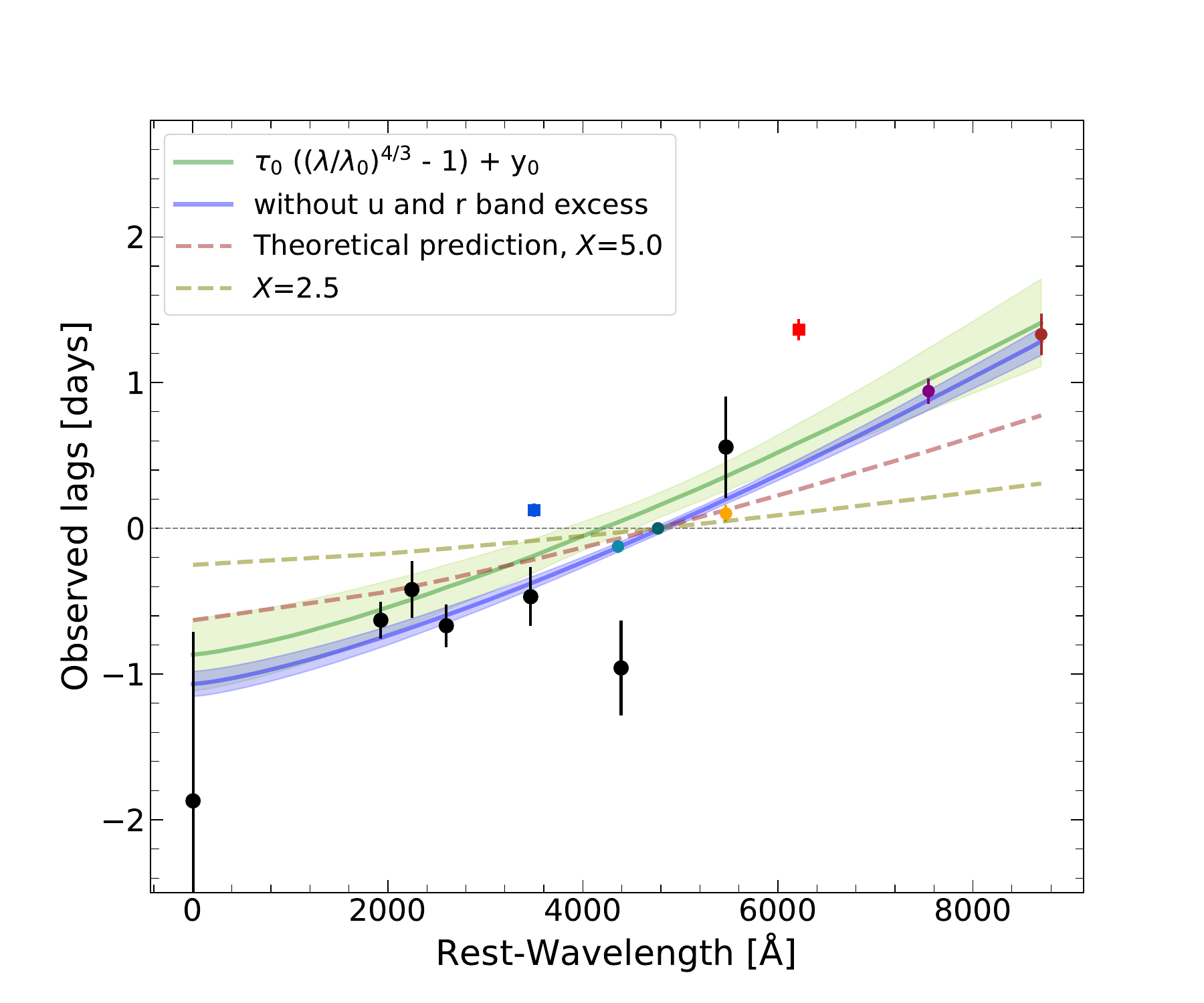}
    \caption{The lag spectrum $\tau(\lambda)$ derived from the measured time delays across the wavebands relative to the $g$ band. The black data points are the {\it Swift} observations and the colored ones are LCO's. The squares represent the $u$ and $r$ band excess.
    The time-lag data are fitted with $\tau$ = $\tau_0$ [ ($\lambda$/ $\lambda_{0}$)$^{4/3}$-1]+$y_0$ and the error envelope is estimated from the best-fit parameters and their uncertainties. The theoretical estimation considered the factors $X$=2.5 and 5.0.
    The best-fit parameters are mentioned in Table \ref{tab:tau_fit}. 
    \label{fig:lagfits}}
\end{figure}

\subsection{Disc size from the lag spectrum }\label{sec:delay_spec}

A powerful diagnostic of the size and temperature profile of the accretion disc arises from the scaling of the time delays as a function of wavelength, known as the delay spectrum. We fit the measured lags given in Table~\ref{tab:lags} with a power-law model for the lag spectrum:
\be \label{eq:taulam}
    \tau(\lambda) = \tau_0 \, \left( \, \left( \frac{ \lambda } { \lambda_0 } \right)^{\alpha} - 1 \, \right) + y_0 \ .
\ee
In this 3-parameter model,
$\tau_0$ is the light travel time delay corresponding to the disc size at the reference wavelength $\lambda_0$.
Within the accretion disc paradigm, $\alpha$ controls the shape of the disc radial temperature profile, $T\propto R^{-1/\alpha}$, and thus $\alpha=4/3$ is expected for a thin steady-state blackbody accretion disc ($T\propto R^{-3/4}$, \citealt{1973SS}). We fix $\alpha=4/3$ for the subsequent analysis. The parameter $y_0$ allows the model lag $\tau(\lambda_0)$ to depart from 0 since the best-fit model need not pass exactly through the measured lag at $\lambda_0$. We use a non-linear least-squares minimization procedure with {\sc lmfit}\footnote{\url{https://lmfit.github.io/lmfit-py/index.html}} to fit this power-law model to the observed delay spectrum. 
The results are summarised in Table~\ref{tab:tau_fit}. 

Our initial fit is shown as the green curve in Fig.~\ref{fig:lagfits}.
The 7 Swift lags (black) have relatively large uncertainties compared with the 7 LCO lags (in color). The $\tau\propto\lambda^{4/3}$ model fits the LCO lags well apart from the $u$ and $r$-band lags, which are significantly larger than the power-law model. The fit is formally rejected, with a reduced $\chi^2/{\rm dof}=221/(14-2)=18.4$.

The $u$-band lag excess seen here is similar to what is often seen in the lag spectra of other well-studied AGN \citep[e.g.,][]{Cackett_2018, 2019ApJ...870..123E,Hernandez20} and is interpreted as a bias due to Balmer continuum emission in the $u$ band.  In addition to broad emission lines, the BLR emits a diffuse continuum that can respond to changes in irradiation.  Given its larger size, the BLR responds with a larger lag than that of the more compact accretion disc
\citep{Korista2001,Korista2019}. 

The $r$-band lag excess seen here is even more significant given its small uncertainty and is rarely noted in other AGNs. A possible hint of $r$-band lag excess can be seen in two other AGNs, MCG+08-11-011 and NGC~2617, but is not discussed in detail (see the lag spectra of \citealt{Fausnaugh_2018}).
A plausible origin for this excess lag is the strong and variable H$\alpha$ line emission in the $r$ band. 

With the $u$ and $r$-band lag excesses having a physically motivated origin in the BLR response, we performed a second fit to the delay spectrum excluding the $u$ and $r$ lags. We retain the negative {\it Swift} B-band lag, an outlier with no physically motivated explanation, but note that it has little effect given its uncertainty.
With $u$ and $r$ on either side of the pivot wavelength, near the $g$ band, the main change is in the $y_0$ parameter, shifting the model lags by 0.17 days.
This second fit (blue curve in Fig.~\ref{fig:lagfits}), is formally acceptable, with reduced $\chi^2/{\rm dof}=14.6/(12-2)\approx1.46$. 
In both cases, the disc size at the reference wavelength is $\tau_0\approx$1~light day.

\begin{table}
    \centering
    \caption{Parameters of the power-law lag spectrum model (Eqn.~\ref{eq:taulam}) fitted to the inter-band lag data as shown in Fig.~\ref{fig:lagfits}. 
    The $1\sigma$ uncertainties are scaled to make 
    the reduced-$\chi^2 = 1$. { We also show the $p$-value here and the $p$-value $<$0.05 suggests the fit does not explain the parameters well. A $p$-value $<$0.05 suggests that the fit explains the data well and the fit is acceptable.}}
    \begin{tabular}{cccc}
    \hline
    parameter & units & LCO+Swift & omit $u$ and $r$
    \\ \hline \hline
    $\alpha$ & & 4/3 & 4/3
    \\
    $\tau_0$ & days & $1.02\pm0.23$ & $1.05\pm0.08$
    \\
    $y_0$ & days & $0.15\pm0.08$ & $-0.01\pm0.03$
    \\ $\lambda_0$ & \AA\ & 4749 & 4919
    \\ $\chi^2/{\rm dof}$ & & 221/(14-2)=18.4 & 14.6/(12-2)=1.46
    \\ p-value & & $< 0.0001$& 0.1473
   \\ \hline
    \end{tabular}
    \label{tab:tau_fit}
  \\  
\end{table}

\begin{figure*}
    \centering
        \includegraphics[width=0.45\textwidth]{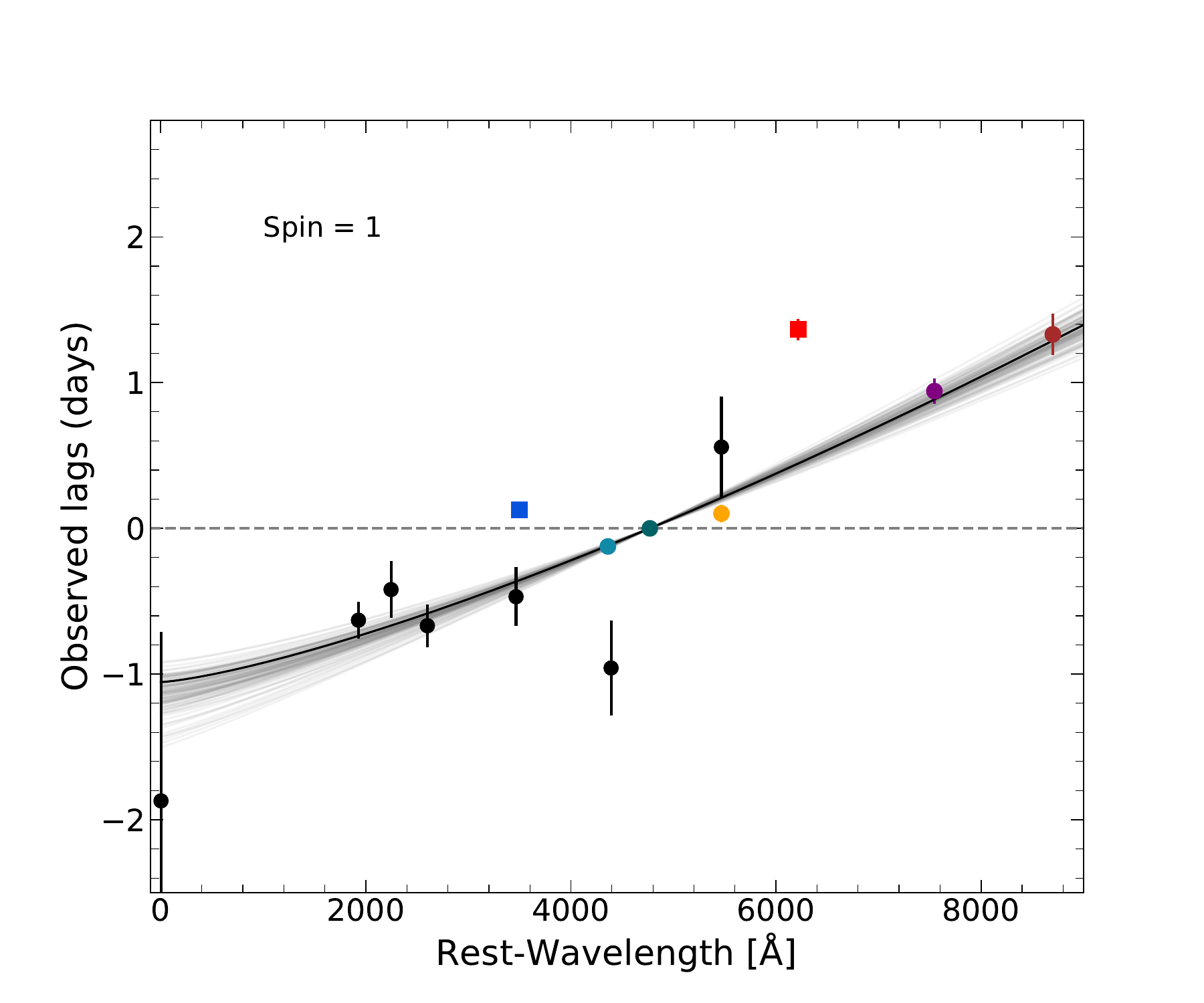}
        \includegraphics[width=0.36\textwidth]{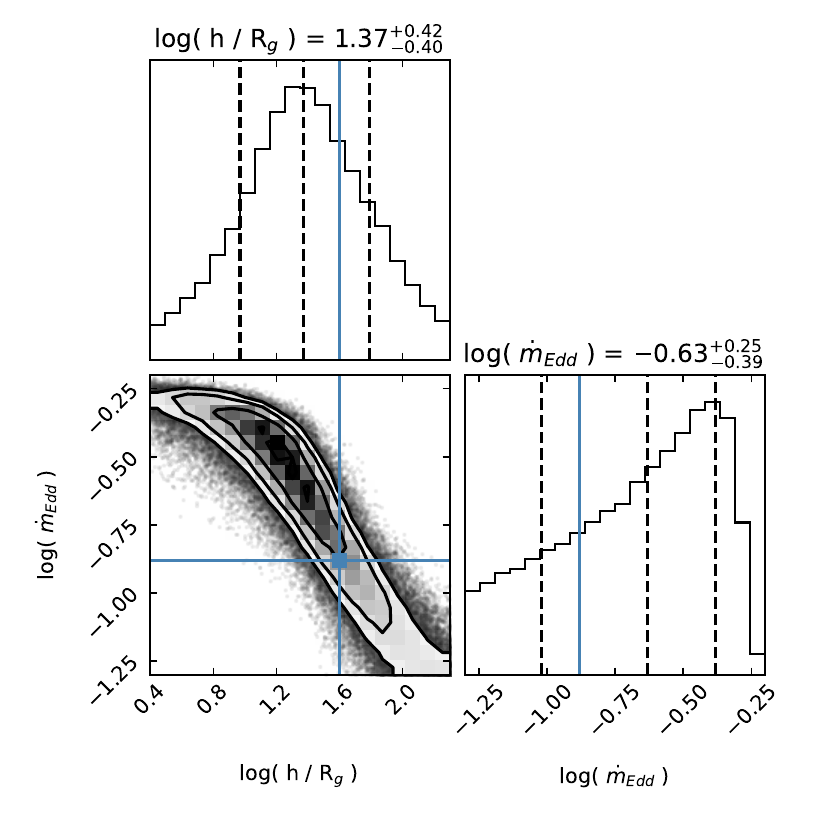}
    \caption{{\it Left:} The observed lag spectrum of NGC\,7469 is modeled with the \citet{10.1093/mnras/stab725, 10.1093/mnras/stad2701} parametrization of lag predictions for an isotropic lamp-post irradiating a thin accretion disc, with realistic X-ray reflection and full treatment of relativity effects in the Kerr geometry of a $9\times10^6$~M$_\odot$ black hole with maximum spin $a^\star=1$.
    The best-fit model (black line) and its 68\% uncertainty envelope (grey) are shown. The colors used for the lag data are the same as in Fig.~\ref{fig:LCOlags}. A horizontal dashed line $\tau = 0$ passes through the $g$-band lag at the reference wavelength (4770~\AA). {\it Right:} The joint posterior distribution and the marginalized probability distributions for the two parameters of interest, the accretion rate $\dot{m}$ in Eddington units, and the lamp-post height $h$ in units of $R_g$. Vertical dashed lines mark the 16, 50, and 84 percentiles of the distributions, which are reported in Table.\,\ref{tab:lag-fit}.
        This fit excludes the $u$ and $r$ lags, which are affected by Balmer continuum and H$\alpha$ emission line respectively, and
        the colour temperature boost factor for the reprocessed emission from the disc is held fixed at $f_{\rm col}=2.4$. The best-fit values of height and accretion rate from SED fit (Table ~\ref{tab:total-sed}) are marked in the right panel with blue colour.
    \label{fig:mcmc0}}
\end{figure*}

\section{Discussion}\label{sec:discussion}

Having presented the main results of our intensive continuum reverberation mapping campaign on NGC~7469, 
we now interpret them by comparing them with accretion disc theory predictions for the spectral energy distribution and the delay spectrum.

\subsection{Success and Failure of Standard Accretion Disc Theory}

We begin with a test of standard accretion disc theory by comparing the observed inter-band lags with theoretical predictions \citep{1973SS,2007MNRAS.380..669C}.
A simple approach assumes that the accretion disc temperature profile $T\propto R^{-3/4}$ corresponds to a power-law lag spectrum $\tau = \tau_0\, \left( \lambda/\lambda_0\right)^{4/3}$.
This power-law model assumes a steady-state thin blackbody accretion disc, ignoring lower fluxes in the UV and IR due to the finite temperature range between the inner and outer edges of the disc.
Fig.~\ref{fig:lagfits} presents our fits of this power-law model to the measured lags, and the resulting values of $\tau_0\approx1$~day are detailed in Table~\ref{tab:tau_fit}.

To compare the disc size parameter, $\tau_0\approx 1$~day, with the prediction of accretion disc theory, 
we use the following expression derived in \citet[][see their Eqn.~11 for details]{2016ApJ...821...56F}:
\begin{equation}
R_0 = 
\tau_0 \, c = 
\left( X\frac{k \,\lambda_0}{h\,c} \right)^{4/3}\left[\left( \frac{G\,\mbh}{8\,\pi\, \sigma}\right)\left(\frac{L_{\rm Edd}}{\eta \,c^2}\right)\left(3 + \kappa\right)\dot m_{\rm Edd}\right]^{1/3}.
\label{eq:tau0}
\end{equation}
Here $X\equiv h\,c/k\,T_0\, \lambda_0$ is a dimensionless parameter that connects the reference wavelength $\lambda_0$ and temperature $T_0$ at radius $R_0=c\,\tau_0$. $L_{\rm Edd}$ is the Eddington ratio, $\eta$ is the disc's radiative efficiency, $\kappa$ is the local ratio of external to internal heating assumed to be the same across all disc radii, and $\dot m_{\rm Edd}$ is the accretion rate in Eddington units. For the model lag spectra shown in Fig.~\ref{fig:lagfits}, we 
adopt $\eta= 0.1$ and $\kappa = 1$, following \cite{Fausnaugh_2018},
and use the value of $\dot m_{\rm Edd}$ = 0.15 (average of the spin 0 and 1 models) from Table~\ref{tab:total-sed}.

 One of the uncertainties in applying this model is the $X$ factor in Eqn.~\ref{eq:tau0}. The commonly adopted value of 2.5 corresponds to a flux-weighted radius at each wavelength (\citealt{2022arXiv220706432G}). However, a response-weighted radius is larger, perhaps as high as $X~\sim5$ \citep{2016ApJ...821...56F,10.1093/mnras/stx2348,2019ApJ...870..123E}. 
 The predicted lag spectra for $X = 2.5$ and 5.0,
  corresponding to $\tau_0=0.33$~d and 0.83~d,
 are shown as the dashed lines in Fig.~\ref{fig:lagfits}. 
For $X=2.5$ the observed lags exceed the predicted lags by a factor of 3, but for $X=5$ the discrepancy is just 
{50\%}.

\subsection{Relativistic disc model with realistic X-ray reflection} \label{sec:x-ray-reflection}

Several possible modifications to the standard thin steady-state blackbody accretion disc reprocessing model have been considered to explain why the disc sizes inferred from continuum echo mapping are larger than expected \citep{ 2019ApJ...879L..24K, 2023A&A...670A.147J}.
\citet{Gaskell17} has also suggested that the observed larger lags could be the effect of reddening.

 A means of testing the most sophisticated model currently available for continuum lag spectra is provided by
 \citet{10.1093/mnras/stab725, 10.1093/mnras/stad2701} in terms of a useful analytic parameterization of lag predictions for an isotropic lamp-post source irradiating a zero-thickness accretion disc, with realistic x-ray reflection and full treatment of relativity effects in the Kerr geometry. 
 In this model, the disc colour temperature is boosted above the effective temperature by a factor $f_{\rm col}$, which increases lags by shifting the reprocessed light to shorter wavelengths.
To apply this model to NGC\,7469,
we fix the black hole mass to $M_\bullet= 9\times$10$^6$ M$_{\odot}$ \citep{2015PASP..127...67B}. 
The analytic model requires an estimate for the
2-10~keV X-ray luminosity. 
For this we adopt $L_X=1.45\times10^{43}$~erg~s$^{-1}$, as
derived from our broadband SED model, corresponding to $L_{X, {\rm Edd}}= 0.012$ in Eddington units. 
We consider two cases of spin: $a^\star=0$ for a Schwarzshild black hole, and $a^\star=1$ for a maximally spinning Kerr black hole.
 The predicted lags then depend mainly on 3 parameters: the accretion rate ($\dot{m}$), the lamp-post height ($h$), and the colour temperature factor ($f_{\rm col}$).
 For a viable model, the constraints on $h$ and $\dot{m}$ from the lag data should overlap with those given in Table\,\ref{tab:total-sed} from the SED fit in Fig.\,\ref{fig:total-sed}. 

  We performed fits to the inter-band lag data both including and excluding the $u$ and $r$ lags, which may be affected by Balmer continuum and H$\alpha$ emission.
  We consider {initially} two fixed values $f_{\rm col}=1.5$ and 2.4 and employ the python package {\sc emcee} \citep{Foreman-Mackey_2013} to explore the $(h,\dot{m})$ parameter space, with uniform priors, and 32 MCMC walkers taking $5\times10^4$ steps. 
Fig.\,\ref{fig:mcmc0} shows our fit to the lag data, excluding the $u$ and $r$ lags, for spin\,1, and for $f_{\rm col}=2.4$.
This model provides a good fit for the lag data.
The model predicts a lag spectrum close to $\tau\propto\lambda^{4/3}$ since reprocessing occurs on the flat disc with $T\propto r^{-3/4}$ in regions far enough from the black hole that relativity effects can be neglected.
Similarly good fits to the lag data are achieved for all cases we considered but with different values of $h$ and $\dot{m}$.
The case shown in Fig.\,\ref{fig:mcmc0} has the best match between constraints from fitting the lags and those from fitting the SED.
Note the strong anti-correlation between $h$ and $\dot{m}$, which can be understood since increasing $h$ allows each disc annulus to intercept a larger solid angle of the lamp-post irradiation, and thus the accretion rate can be reduced to maintain the same effective temperature.

\begin{figure}
\begin{tabular}{c} 
 \includegraphics[trim=18mm 10mm 62mm 10mm,clip,width=0.45\textwidth]{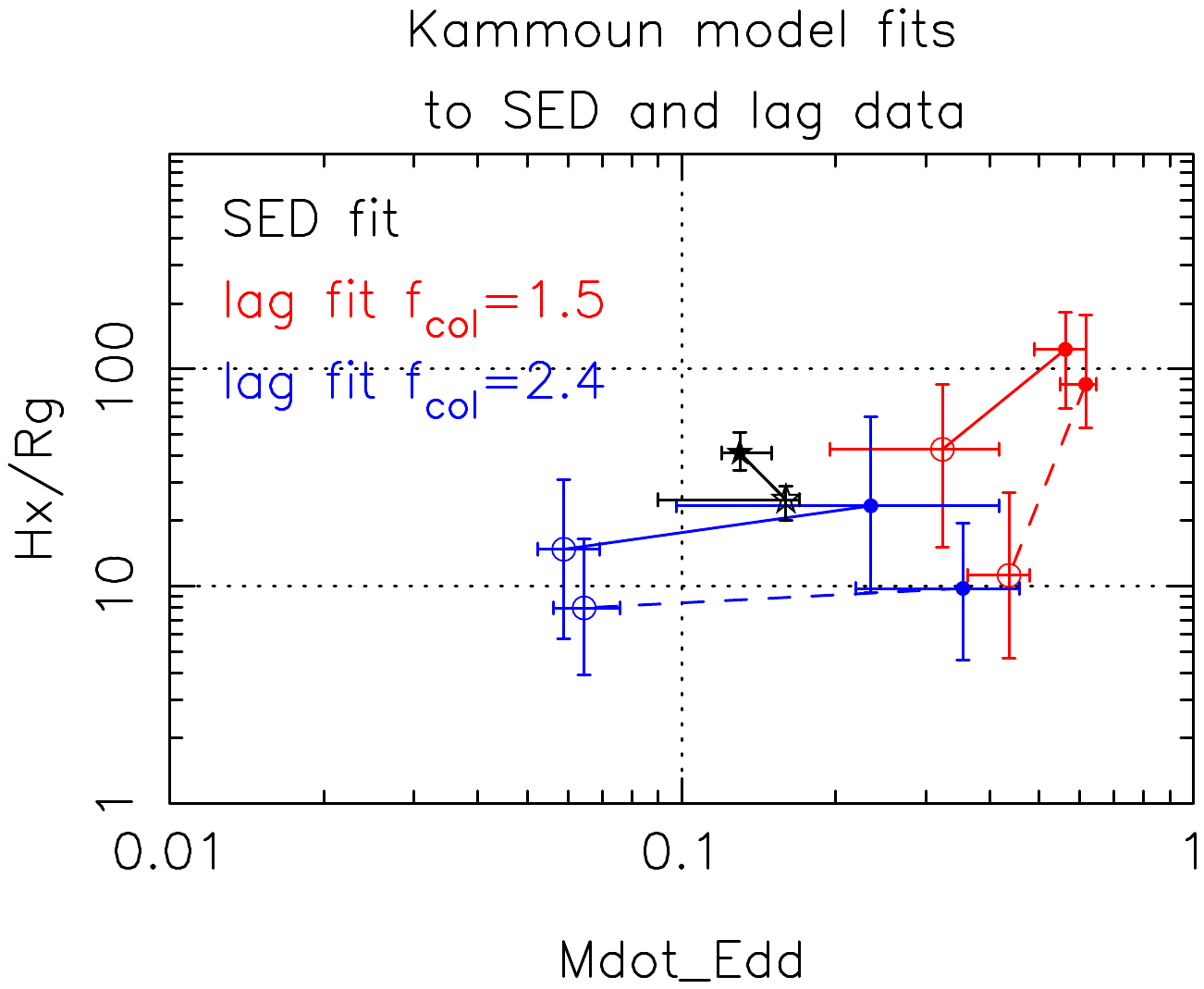} 
 \end{tabular}
 \caption{ Constraints on the lamp post height $H_x$ and accretion rate $\dot{m}_{\rm Edd}$ in Eddington units
 for the \citet{10.1093/mnras/stab725, 10.1093/mnras/stad2701} model, which features a compact isotropic lamp post model of the X-ray corona, fully relativistic light propagation near the black hole, and realistic treatment of the X-ray reflection from the disc.
 Open and filled symbols are for spin $0$ and $1$, respectively.  Black stars mark the tight constraints from the SED (Table~\ref{tab:total-sed}, Fig.~\ref{fig:total-sed}).   Circles mark constraints from fits to the inter-band lags (Table\,\ref{tab:lag-fit}) for $f_{\rm col}=1.5$ (red) and $2.4$ (blue). Lines connecting spin 0 and spin 1 constraints are dashed for lag fits that include the $u$ and $r$ bands.
 The SED and lag constraints overlap for $f_{\rm col}\sim2.4$.
 \label{fig:kammoun}}
\end{figure}

 Table~\ref{tab:lag-fit} summarises, for each of the 8 cases considered, the best-fit parameters ($h$, $\dot{m}$).
Fig.~\ref{fig:kammoun} plots these results to show that the relatively tight constraints from the SED fit overlap with those from the lag fit for $f_{\rm col}=2.4$.
For $f_{\rm col}=1.5$, and for lower values, the accretion rate required to fit the lags is well above that found in the SED fit, and can thus be ruled out.

For $f_{\rm col}=2.4$,
the accretion rate from the lags is lower than that from the SED for spin\,0, and vice versa for spin\,1.  An intermediate spin may thus provide an accretion rate consistent between the lag and SED data.
However, for both spins the lag fit for $f_{\rm col}=2.4$ gives $h\sim10\,R_g$, smaller by a factor of 2 to 4 than that required by the SED. The uncertainties from the lag fit are large enough that this discrepancy is only marginally significant.
When we omit the $u$ and $r$ lags the required height increases by about a factor 2, improving the consistency between the lag and SED constraints.
We conclude {provisionally} that a model with $f_{\rm col}\sim2.4$ can {probably} fit both the SED and the lag constraints.

  A detailed parameter exploration is done in \cite{10.1093/mnras/stad2701} for various reverberation-mapped AGNs. 
  They find and discuss a range of parameter degeneracies arising from fits to the lag spectra.  In particular, $f_{\rm col}>1$ is always needed to fit the lag data, but with different corona heights, accretion rates and black hole spin in each AGN. 
  The corona height and accretion rate estimated for NGC\,7469 in our work are 
  in the range provided for various objects in \cite{10.1093/mnras/stad2701}.

\begin{table}
    \centering
     \caption{Constraints on the lamp-post height $h$ and accretion rate $\dot{m}$ derived by fitting the inter-band lag data with Kammoun's analytic expression$^1$ for fixed values of the colour correction $f_{\rm col}=(1.5,2.4)$, black hole spin $a^\star=(0,1)$, and for fits including and excluding the $u$ and $r$ lags.}
    \begin{tabular}{c|cc|cc}
    \hline
     Parameters & \multicolumn{2}{c|}{{$f_{\rm col} = 2.4$}} &\multicolumn{2}{|c|}{ {$f_{\rm col}=1.5$}}\\
    \hline
   -& $a^\star=0$ & $a^\star=1$ & $a^\star=0$& $a^\star=1$  \\
    \hline
   log ( $h/R_g$ ) & 0.90$^{+0.32}_{-0.31}$  & 0.99$^{+0.30}_{-0.33}$& 1.05$^{+0.38}_{-0.38}$& 1.93$^{+0.32}_{-0.20}$\\
   log ( $\dot{m}_{\rm Edd}$ )& -1.19$^{+0.07}_{-0.06}$ & -0.45$^{+0.11}_{-0.21}$ &-0.36$^{+0.04}_{-0.08}$& -0.21$^{+0.02}_{-0.05}$ \\
    \hline
    &  excluding  & $u$ and $r$ band & & \\
    \hline
   log ( $h/R_g$ ) & 1.17$^{+0.32}_{-0.41}$ & 1.37$^{+0.41}_{-0.40}$ &1.63$^{+0.30}_{-0.45}$ & 2.09$^{+0.17}_{-0.27}$\\
   log ( $\dot{m}_{\rm Edd}$ )& -1.23$^{+0.07}_{-0.05}$& -0.63$^{+0.25}_{-0.38}$&-0.49$^{+0.11}_{-0.22}$& -0.25$^{+0.04}_{-0.06}$ \\
    \hline 
    \end{tabular}
    $^1$ Analytical expression is provided in \cite{10.1093/mnras/stab725, 10.1093/mnras/stad2701}
    \label{tab:lag-fit}
\end{table}

\begin{figure}
\centering
\includegraphics[trim=1.4cm 0.8cm 2.1cm 2.3cm,clip,width=0.45\textwidth]{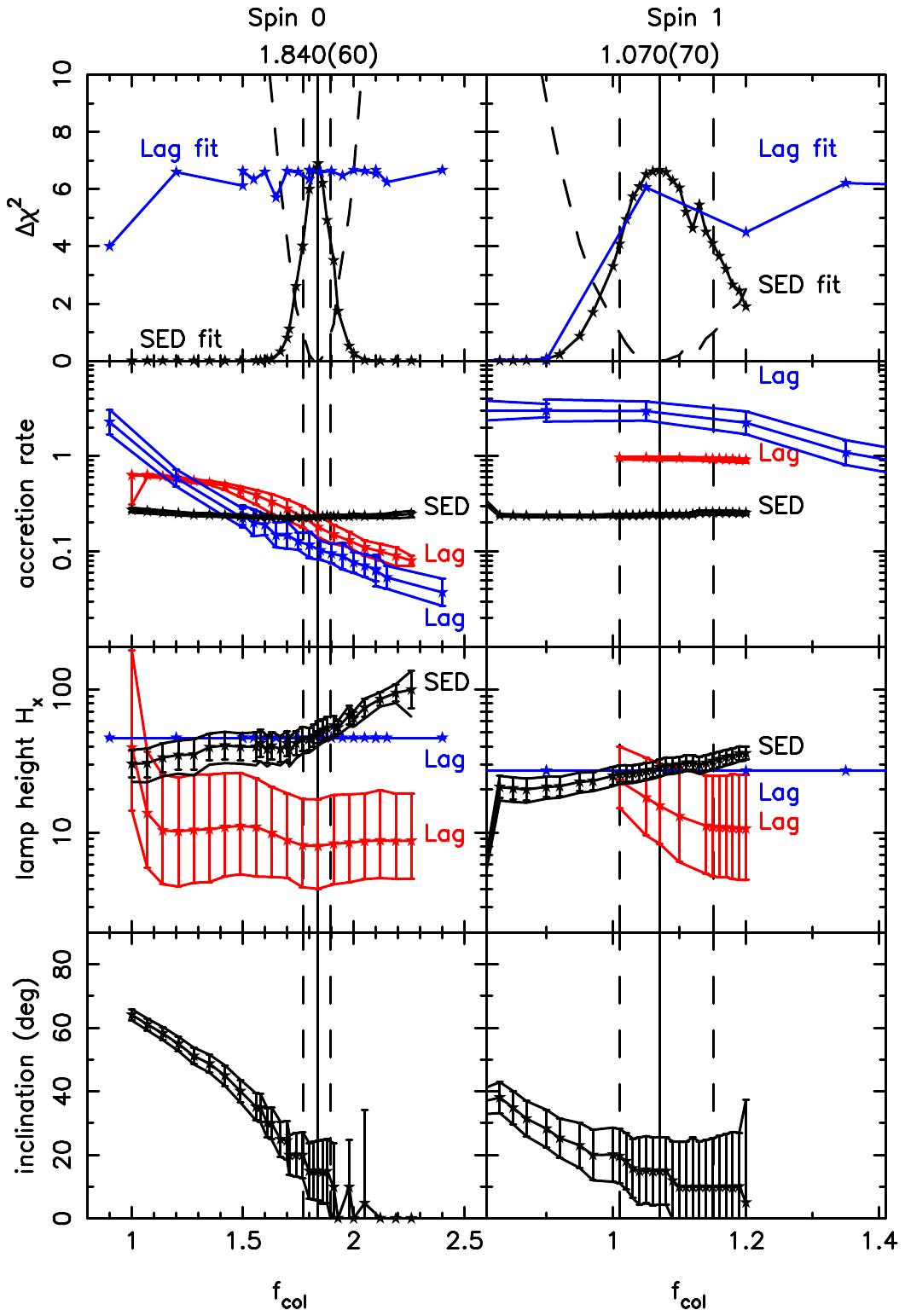}
\caption{{ Results of fitting the SED and lag spectra for spin\,0 (left column) and spin\,1 (right column). 
Three cases shown are the SED fit (black), and lag fits with lamp height free (red) and fixed (blue).
The results shown are for grids of fixed values of the disc colour temperature $f_{\rm col}$. 
Top row shows $\Delta \chi^2$ for the SED fit (dashed parabola) and relative probabilities
$P_{\rm rel}\propto \exp{\left(-\Delta\chi^2/2\right)}$ for the SED fit (black) and for the lag fit (blue).
The second row shows constraints on the accretion rate $\dot{m}$ in Eddington units.
The third row shows constraints on the
lamp height $H_x$ in units of $R_g$.
The bottom row shows constraints on the disc inclination.
The SED data provide tighter constraints than the lag data.
The best-fit $f_{\rm col}$ and its uncertainty based on $\Delta\,\chi^2=1$ are indicated at the top and by vertical lines.
The lag data provide a degenerate constraint of the form $\dot{m} \propto f_{\rm col}^{-4}$, arising from the disc temperature profile $T_{\rm eff} = (T/f_{\rm col}) \propto \dot{m}^{1/4}\, R^{-3/4}$.
} 
\label{fig:fcolgrid}
}
\end{figure}

{
To investigate the constraints in more detail,
Fig.\,\ref{fig:fcolgrid} summarizes for both spins 0 and 1, and for a range of $f_{\rm col}$, the constraints on the accretion rate $\dot{m}$,  lamp height $h$ and inclination $i$ arising from the fitting of the lag data compared to those of the fitting of the SED data.
To create this plot, we first ran a series of fits to the SED data to obtain for a grid of fixed values of $f_{\rm col}$, best-fit parameters and uncertainties for $\dot{m}$, $h$, and $i$ (black points with error bars in the lower 3 panels).
The top panel identifies
best-fit values of $f_{\rm col}$, $1.84\pm0.06$ for spin\,0 and $1.07\pm0.07$ for spin\,1, that minimise the $\chi^2$ and maximise the
relative probability $P_{\rm rel}\propto\exp{\left(-\Delta \chi^2/2\right)}$.
A similar series of fits to the lag data then yields the red points with error bars in the lower 3 panels.
For a consistent model, the lag constraints (red) and SED constraints (black) should agree.

The best-fit inclination is small $i<20^\circ$ for both spins.
The SED data fairly tightly constrain the accretion rate, $\dot{m}\approx 0.25$ in Eddington units, with little dependence on $f_{\rm col}$ or spin.
The lag data, however, admit a strong degeneracy of the form $\dot{m} \propto f_{\rm col}^{-4}$, which arises from the disc temperature profile $T_{\rm eff} = (T/f_{\rm col}) \propto \dot{m}^{1/4}\, R^{-3/4}$.  
In Fig.\,\ref{fig:fcolgrid} we show two lag fits, one (in red) with $h$ optimised for each $f_{\rm col}$, and another (in blue) with $h$ fixed at the best-fit value from the SED fit.
}
{
Note that the spin\,0 model can fit both the SED and the lags with a consistent $\dot{m}\approx 0.25$, but with a high $f_{\rm col} \approx 1.8$, and with somewhat discrepant lamp heights, $h \sim50$ for the SED fit and $h\sim10$ for the lag fit
 a 2$\sigma$ discrepancy
given the 0.4\,dex uncertainty in $h$ from the lag fit.
With $h=46$ $R_g$ to match the SED fit, the lag fit (blue) reduces $\dot{m}$ by a factor of 2, again
2\,$\sigma$ below the SED fit.
For spin\,1 the SED fit again finds $\dot{m} \approx 0.25$ with a very plausible $f_{\rm col} = 1.1$. Unfortunately, this combination fails to fit the lags, which require $\dot{m}\sim 0.25 \,\left( 1.8 / f_{\rm col} \right)^4$, 
an order of magnitude higher than from the SED fit.
The $\chi^2$ favours the spin\,0 model, but the required boosting of disc temperatures by a factor of 1.8 is quite high. However, some of the earlier studies modelling the EUV and Soft X-ray emission from
AGN have used $f_{\rm col}=2.4$ \citep{10.1111/j.1365-2966.2011.19779.x} and also for X-ray binaries it has been shown that the $f_{\rm col}$ can be around 1.7 \citep{1995ApJ...445..780S, Mirzaev_2024}.
}

Note that  
{ $f_{\rm col}\sim1.8$ } implies that the colour temperature of the reprocessed emission from the outer disc is boosted { considerably above} 
the effective temperature.
This shifts the reprocessed emission to shorter wavelengths and thus increases the lags at a given wavelength.
The physical motivation for introducing $f_{\rm col}>1$ is to approximately describe the effect of electron scattering in the disc atmosphere \citep{2018MNRAS.480.1247K}.
This is appropriate for the hotter regions of the disc ($T>10^5$\,K) where electron scattering dominates the opacity, but not for the cooler disc regions ( $T<10^4$\,K) where $f_{\rm col}=1$ should apply.
For these flat-disc models, blackbody reprocessing ($f_{\rm col}=1$) fails to fit the lags, and the finding that 
{$f_{\rm col}\sim1.8$}
is needed to fit the lags is another way of quantifying the problem of disc sizes being 
larger, { and thus hotter at each radius}, than expected.

\begin{figure*}
\begin{tabular}{cc} 
\\ {\bf \Large (a)}
 \includegraphics[trim=15mm 1cm 60mm 1cm,clip,width=0.43\textwidth]{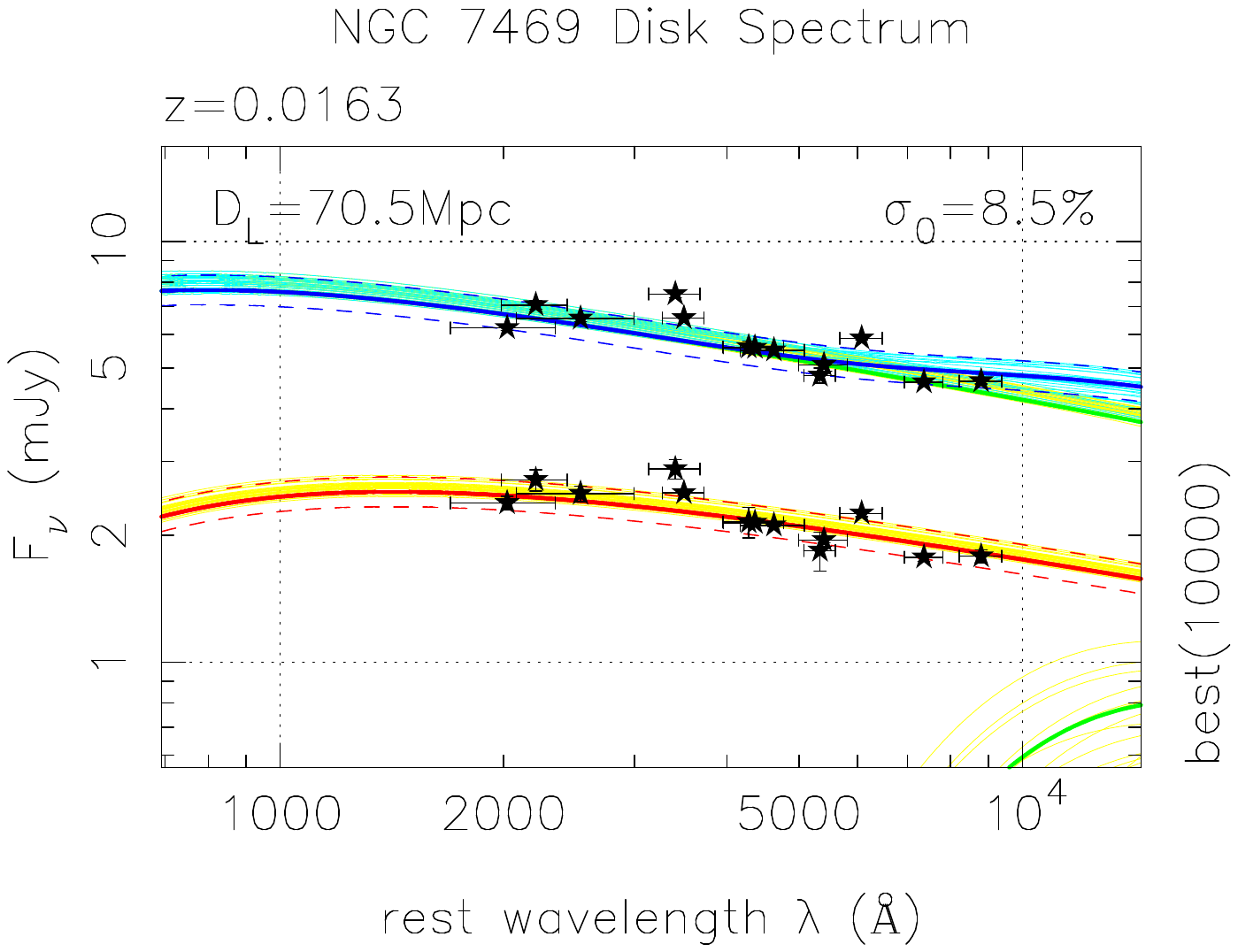} 
 & {\bf \Large (b)}
 \includegraphics[trim=15mm 1cm 60mm 1cm,clip,width=0.43\textwidth]{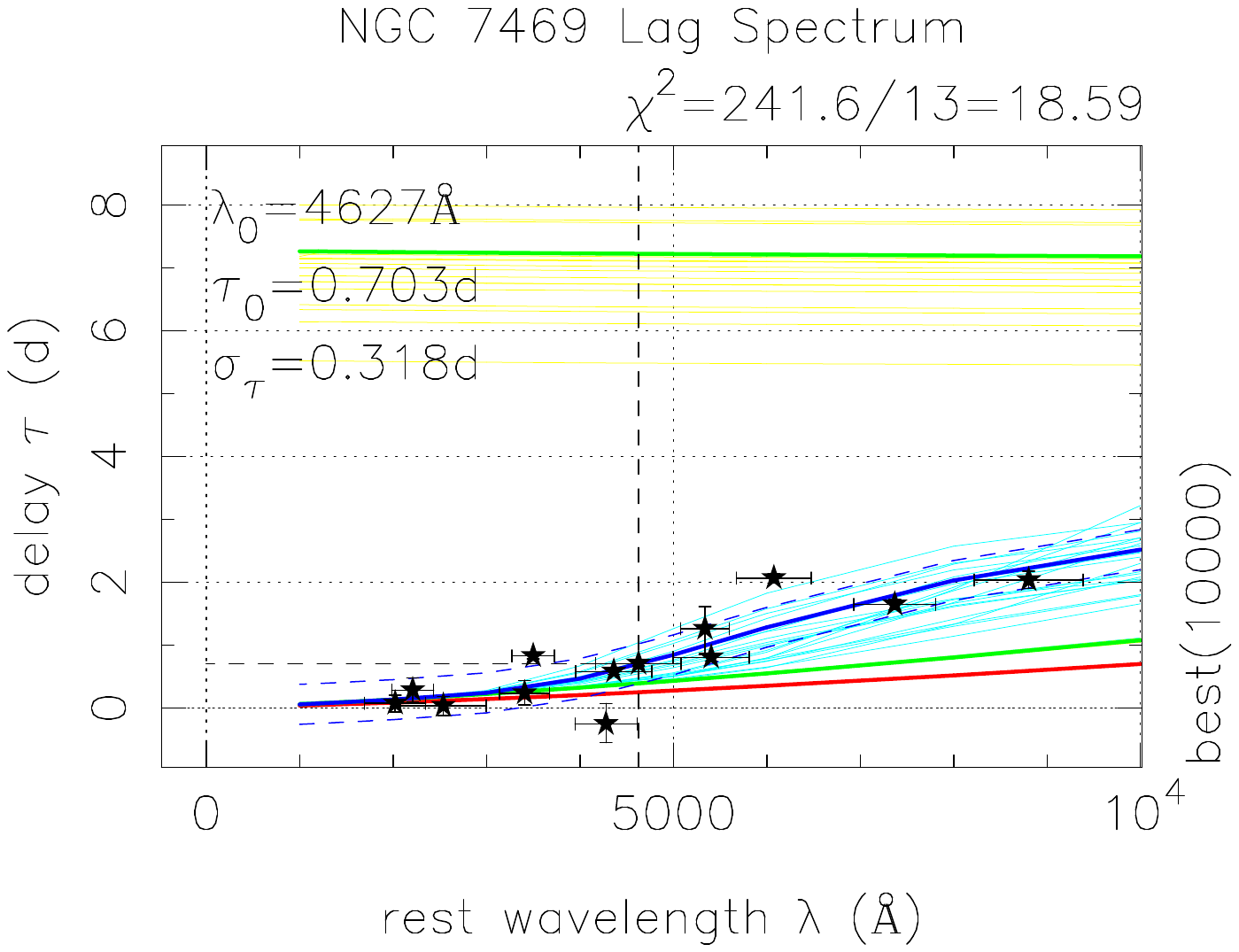}
 \\ {\bf \Large (c) }
  \includegraphics[trim=15mm 1cm 60mm 1cm,clip,width=0.43\textwidth]{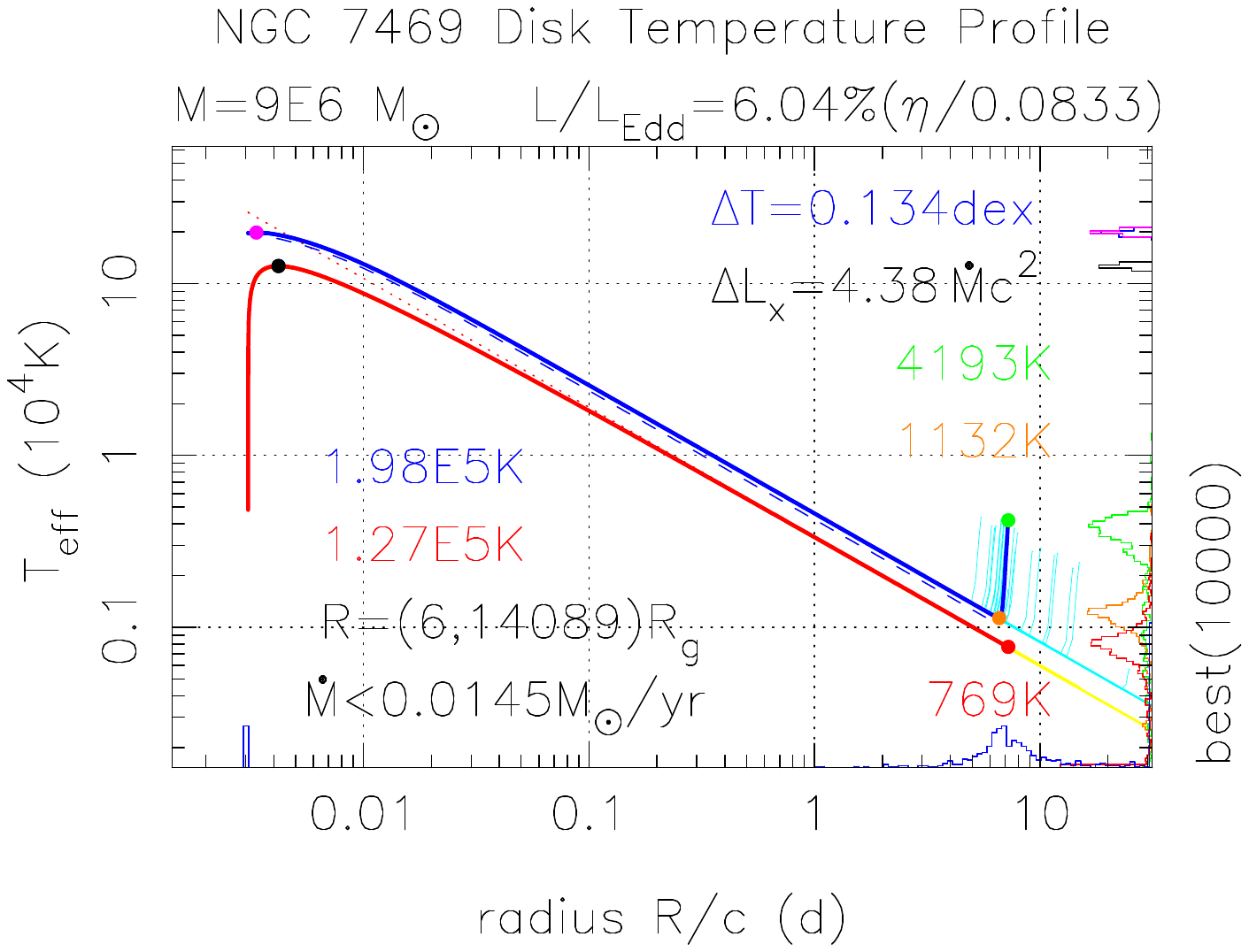} 
 & { \bf \Large (d)}
 \includegraphics[trim=15mm 1cm 60mm 1cm,clip,width=0.43\textwidth]{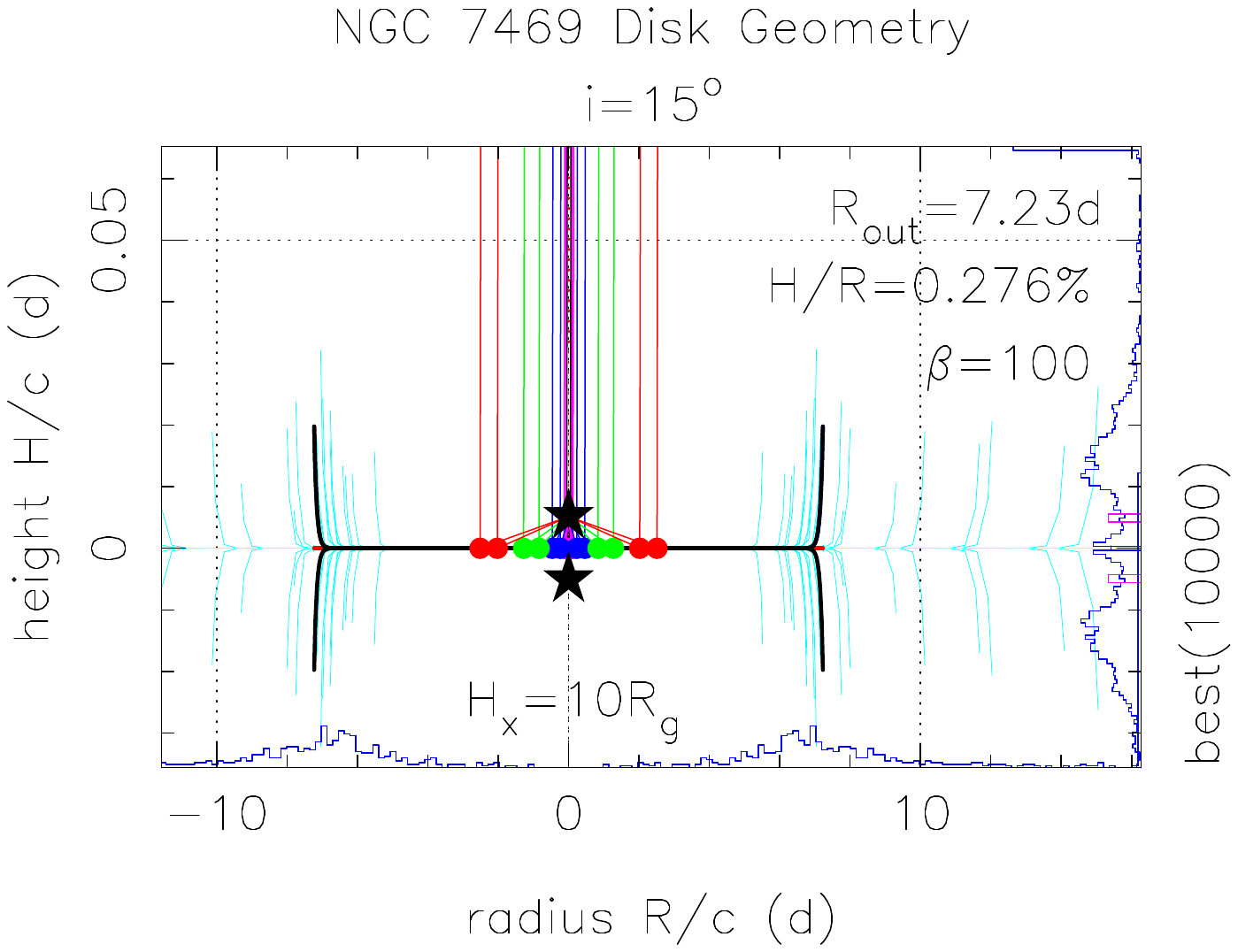}
 \end{tabular}
 \caption{
 Bowl model fits to the disc SED and inter-band lag data for NGC\,7469. 
 The faint and bright disc SED data (a) and the inter-band lags (b) are compared with a blackbody reprocessing model for a thin disc,
 with a bowl-shaped power-law $H(R)\propto R^\beta$ profile, irradiated by a lamp post just above the black hole.
 The disc geometry (d) has a flat disc with a steep rim at $R_{\rm out} \sim 5-10$ light days. 
 The temperature profile (c) decreases as $T\propto R^{-3/4}$ on the flat disc, reaching a minimum 
 and then rising
 to {$\sim4000$\,K}
 on the inward-tilted face of the steep rim.
 With the lamp off (red curves) the model fits the
 faint-disc SED by adjusting the accretion rate $\dot{M}$
 to fit the red end and the inner radius $R_{\rm in}$,
 hence maximum temperature, to fit the UV end.
 With the lamp on (blue curves), the disc temperatures rise and the SED flux increases to match the bright disc SED.
 In the upper panels, green curves show the bright state model SED and lag for the flat disc inside and the steep rim outside the temperature minimum.
Time lags are dominated by the flat disc in the UV and are pulled toward the steep rim's lag in the optical. This transition from disc to rim reprocessing increases the optical lags to fit the lag data. 
Yellow and cyan curves show
30 random MCMC samples to indicate uncertainties in the constrained Bowl model.
In the lower panels, histograms on the lower and right edges give uncertainties in fiducial temperatures, radii, and heights.
{Best-fit parameters are collected in Table.\,\ref{tab:bowl}}
 \label{fig:bowl}}
\end{figure*}

\begin{figure}
\begin{tabular}{c} 
 \includegraphics[trim=15mm 1cm 62mm 1cm,clip,width=0.45\textwidth]{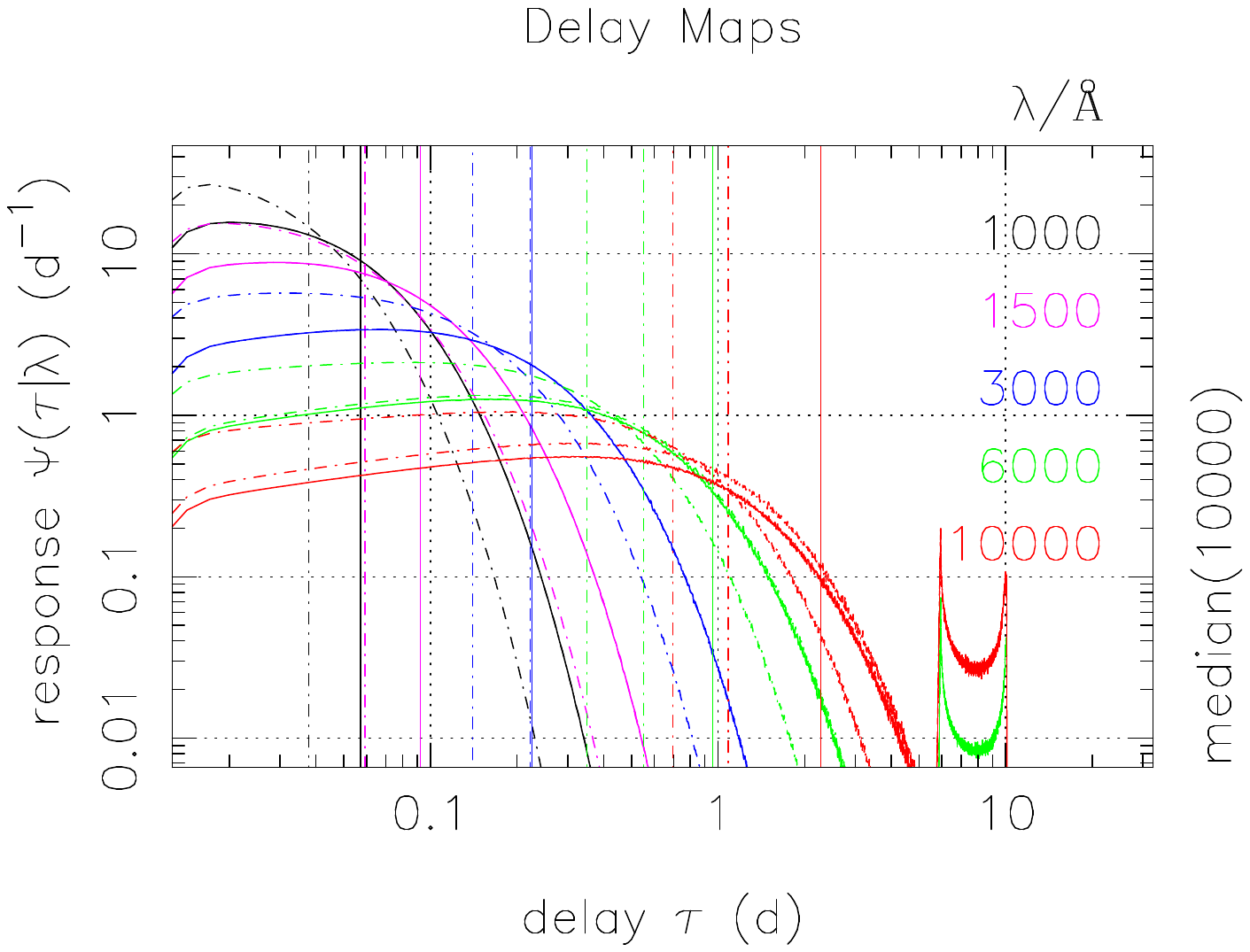} 
 \end{tabular}
 \caption{Delay maps $\Psi(\tau|\lambda)$,
showing the time delay distribution
of the response to changes in
lamp-post luminosity,
 for the Bowl model 
 in Fig.~\ref{fig:bowl}.
 The delay maps are plotted on a log-log scale for 5 wavelengths indicated by colors.
 The solid and dot-dash curves give the marginal response for the faint and bright states (red and blue curves in Fig.~\ref{fig:bowl}), respectively.
 At UV wavelengths (black, purple, blue) the response is dominated by reprocessing on the flat disc. These UV delay maps rise to a peak and then decline, giving mean delays
 (vertical lines) scaling as
 $\tau_0\,\left(\lambda/\lambda_0\right)^{4/3}$.
 At optical wavelengths (green, red)
 a U-shaped feature appears due to reprocessing on
 the steep rim, spanning the range
 $\tau=\left(1\pm\sin{i}\right)\, R_{\rm out}/c$.
 The steep rim faces away from the observer on the near side, and is thus more visible on the far side of the inclined disc, significantly increasing the optical lags.
 \label{fig:psi}}
\end{figure}

\subsection{ Bowl model with power-law disc thickness profile}


Another option that may help to explain why disc sizes inferred from continuum reverberation mapping are larger than expected 
is to
{retain blackbody reprocessing with $f_{\rm col}=1$ but}
relax the assumption of a flat disc thickness profile $H(R)$.
\cite{Starkey2023} investigate how
vertical structure on the disc surface affects the lag spectrum. They find that
steeply sloped rims or wave-like ripples can have the desired effect of increasing the lags.
The inwardly tilted slopes intercept more flux from the central lamp-post than a flat disc surface. This elevates the local temperature and thus 
shifts the lagged blackbody response to shorter wavelengths.
Applied to the 2014 STORM campaign data on NGC~5548, these models succeed in fitting the larger-than-expected lags simultaneously with the
faint and bright disc SEDs. 
In particular, an increase in disc thickness \citep{BaskinLaor2018}, or launching of a failed radiatively-accelerated dusty wind at the inner edge of the BLR \citep{CzernyHryniewicz2011}
is expected at the radius where the disc temperature profile crosses the dust sublimation threshold at $\sim1500$\,K.
This promising model works for NGC~5548 and, given the connection with upward radiation pressure on dust in the disc atmosphere, it may apply more generally to all AGN discs.
 
We now apply to NGC~7469 this blackbody reprocessing model on a disc with a concave power-law $H(R)$ thickness profile.  We refer to this as the Bowl model, since the concave $H(R)$ profile can resemble the interior surface of a bowl.
A pair of lamp-posts, isotropic point sources with bolometric luminosity $L_x$ located at $\pm H_x$ above and below the central black hole, irradiate the disc surface at radius $R$ and height $\pm H(R)$ above and below the disc plane.
Balancing heating and radiative cooling,
the effective temperature $T(R)$ is
given by
\be   \label{eqn:tfour}
    \sigma \, T^4 (R) = \frac{ 3\, G\, \mbh \, \dot{M} } 
    { 8 \, \pi \, R^3 }
    \, \left( 1 - \left( \frac{R_{\rm in}}{R} \right)^{1/2} \right) 
  ~  + ~ \frac{ L_x \, \left( 1-A \right) \cos{\theta} }
    { 4 \, \pi \, \left( R^2 + \left(H(R)-H_x\right)^2 \right) }
    \ ,
\ee
Here $G$ is Newton's gravitational constant,
$\sigma$ is the Stefan-Boltzmann constant, and
$\mbh$ is the black hole mass.
The first term represents heating by the
viscous dissipation and torques associated with the accretion rate $\dot{M}$.
For zero torque between the disc and the black hole, the viscous heating vanishes at the inner radius
$R_{\rm in}$, which we assume is at the
ISCO.
The second term accounts for irradiative heating from the lamp-posts. 
The disc albedo $A$ is assumed to be independent of $R$.
For a thin disc the grazing incidence angle
$\theta$ is close to $90^\circ$.
For a thin flat disc, $\cos{\theta}\approx (H_x-H)/R$ so that both terms scale as $R^{-3}$ at large $R$.
This introduces a degeneracy between accretion power and irradiation, i.e., between $\mbh\,\dot{M}$ and $L_x\,(H_x-H)\,(1-A)$.

The Bowl model {as currently implemented} makes simplifying assumptions such as isotropic point source lamps, radius-independent disc albedo, blackbody reprocessing, and Euclidean geometry neglecting
relativity effects (Doppler boost, gravitational redshift, light trajectory bending).  
{A more detailed modelling of the inner disc region would be needed to enable the Bowl model to fit the observable X-ray spectrum, 
and predict the unobservable EUV spectrum, but this is beyond the scope of the paper.}
As a consequence, 
parameters associated with the region near the black hole are unreliable and should be regarded as a useful approximate description of the irradiation that emerges to reach the outer disc at $R\gg R_g=G\,\mbh/c^2$ where neglecting relativity is a good approximation.
Nevertheless, 
{since the height and bolometric luminosity of the lamps determine the temperature structure on the irradiated disc surface, and thus the UV/optical lag spectrum and SED,}
the Bowl model provides useful parameters for interpreting disc lags and SEDs from reverberation mapping experiments,
especially if AGN accretion discs are flared and dusty like protoplanetary discs \citep{2023FrASS..1056088L}.

A Bowl model fit is presented in Fig.~\ref{fig:bowl}, where Panel~(a) shows the fit to the faint and bright disc SEDs, Panel~(b) shows the fit to the inter-band lags, Panel~(c) shows the temperature-radius profile,
and Panel~(d) shows the disc geometry.
We employ MCMC methods to explore the parameter space, fitting {typically 6-10} 
Bowl parameters with 26 constraints from the observed disc SEDs at minimum and maximum light, and 13 inter-band lags.
In fitting the Bowl model to NGC~7469, we hold several parameters fixed.
We adopt $\mbh=9\times10^6$~\Msun\ for the black hole mass
\citep{2015PASP..127...67B}. 
The luminosity distance $D_{\rm L}=70.5$~Mpc is from the redshift.
Given that broad emission lines are present in the spectrum, we may expect $i<60^\circ$, but the disc inclination is not well determined. 
 {There is} a strong degeneracy because the disc SED scales as $(\mbh\,\dot{M})^{2/3}\,\cos{i}$.
{ After running a variety of fits with different parameter values free and fixed, we opt to present in Fig.\,\ref{fig:bowl} the fit for }
{$i=15^\circ$, matching the SED-fit results in Fig\,\ref{fig:fcolgrid}}.
{ Fig\,\ref{fig:bowl}
presents the fit for $R_{\rm in}=6\,R_g$ fixed at the ISCO radius of a spin-0 Schwarzshild black hole,
and with the lamp height fixed at $H_x=10\,R_g$, which resolves its degeneracy with the change in irradiating luminosity $\Delta L_x$,
since the flat disc irradiation scales approximately as $H_x\, \Delta L_x$.
Finally, we fix the $H(r)$ power-law index $\beta=100$, which produces a flat disc with a steep outer rim.}

As discussed in \cite{Starkey2023}, the 
Bowl model parameters can be optimized in 3 stages as follows.
First, starting with a flat disc model (zero thickness), adjust $\dot{M}$ and $R_{\rm in}$ to fit the faint disc SED. 
Turning up the lamp irradiation then increases the flat disc temperature by
$\Delta T$ to match the change between the faint and bright disc SED. 
Such models fit the SEDs but underpredict the lags.
Finally, adjust the size and shape of the disc $H(R)$ profile to fit the lags. These adjustments can be done by hand to find rough initial parameter estimates that are subsequently refined by the MCMC algorithm.
However, the burn-in stage of the MCMC algorithm succeeds in discovering the same result even when starting from very distant parameter values.

The faint disc $T(R)$ profile (red curve in Fig.~\ref{fig:bowl}c) requires the blackbody emission from both sides of the disc to balance
heating by viscous dissipation and torques associated with the accretion rate $\dot{M}$, thus using only the first term in Eqn.~(\ref{eqn:tfour}).
The fit parameters at this stage are $\dot{M}$ and $R_{\rm in}$.
Adjusting $\dot{M}$ scales the $F_\nu \propto \nu^{1/3}$ segment of the flat-disc SED model to match the faint-disc SED data.
This fit defines a lower limit, 
$\dot{M}<{0.0134}$~\Msun~yr$^{-1}$because of course some irradiation elevates disc temperatures even at the minimum flux level.

The Bowl model places $R_{\rm in}$ at the ISCO radius, and thus $1 < R_{\rm in}/R_g < 9$, depending on the black hole spin.
With $\dot{M}$ constrained as noted above, increasing $R_{\rm in}$ lowers the maximum temperature and thus the UV end of the SED. 
The MCMC samples { confirm the expected positive correlation} 
 {between $R_{\rm in}$ and} $\dot{M}$ that keeps $T_{\rm max}>2\times10^5$~K.
The disc luminosity is $L_{\rm disc}=\eta\,\dot{M}\,c^2$,
and with a radiative efficiency $\eta=\left(R_g/2\,R_{\rm in}\right)=1/12$  the faint disc model has an Eddington ratio 
{ $L/L_{\rm Edd}\sim6\%$} .

The maximum temperature needs to exceed $T_{\rm max}>2\times10^5$\,K to avoid a significant decrease on the UV end of the model SED.
Turning up by $\Delta L_x$ the bolometric luminosity of the lamp-posts above and below the disc plane, the additional irradiation increases the disc temperature profile, 
as shown by comparing the blue and red curves in Fig.~\ref{fig:bowl}(c).
 The disc temperatures rise by 
 {$\Delta T\approx0.134$~dex}, a factor {1.36}, and the model SED rises by a corresponding factor 
 
 {$1.36^{8/3}\approx2.3$} to match the increase between the faint and bright disc SED data, as shown in Fig.~\ref{fig:bowl}(a).
 Increased irradiation may arise from changes in the lamp height $H_x$ and/or changes in its luminosity $\Delta L_x$.
 The MCMC samples confirm the expected degeneracy between $H_x$ and $L_x$, with the product $L_x\,H_x$ being tightly constrained by the required rise in temperature needed to match the observed rise in flux.


With Bowl parameters $\dot{M}$, $R_{\rm in}$ and $\Delta T$ constrained to match the faint and bright disc SED, the flat-disc model predicts time delays that increase by about 1 day between $10^3$ and $10^4$~\AA, as indicated by the red curve in Fig.~\ref{fig:bowl}(b).
In contrast,
the observed lags rise by about 2 days,
a significant discrepancy.
The model lags increase if we consider a concave 
bowl-shaped power-law thickness profile $H\propto R^\beta$. 
This adds 3 parameters,
the outer disc radius $R_{\rm out}$, the aspect ratio $H/R$,
and the power-law index $\beta$. 
Irradiation of the concave disc increases temperatures in the outer disc more than the inner disc. This helpfully increases the model lags but also increases the red end of the SED and can upset the fit to SED data.  The MCMC chain discovers that both constraints can be met with a high value of $\beta$ to produce a flat inner disc with a steep outer rim.
The resulting geometry,
shown in Fig.~\ref{fig:bowl}(d),
has a thin flat inner disc 
encircled by a thicker flared outer rim.
The data constrain the rim radius 
to the range $R_{\rm out}\approx5-10$ light days, with steeper rims ($\beta>100$) at larger radii.
The rim is steep but not tall, $H/R<1$\%.
{This allows irradiation to pass over the top of the rim to illuminate structures such as the BLR and dusty torus at larger radii.}


In fitting the Bowl model simultaneously to the lag and SED data, we include a noise model parameter, the fractional uncertainty $\sigma_0$, added in quadrature with the individual flux uncertainties $\sigma_i$.
The standard maximum likelihood criterion
implements the trade-off between lowering $\chi^2$ to fit the data and lowering $\sigma_0$ to decrease the model SED uncertainties.
A relatively large $\sigma_0$ softens the SED constraints and allows the MCMC chain to attend primarily to fitting the lags. 
The MCMC chain then lowers $\sigma_0$, strengthening the SED constraints, and settles around $\sigma_0\sim9\%$.
A similar parameter $\sigma_\tau$ applies to the lags, 
{and the best fit is for $\sigma_\tau\sim0.3$\,d.}
{These noise model parameters quantify model uncertainty, since blackbody reprocessing neglects wavelength-dependent opacities that can produce Balmer edge and H$\alpha$ features increasing the lag and flux in the $u$ and $r$ bands.}

\begin{table}
\begin{center}
\caption{ Parameters of the BOWL model 
shown
in Fig.~\ref{fig:bowl}.
\label{tab:bowl}}
\begin{tabular}{l|c|r|l}
\hline
 parameter & symbol & value & units
 \\ \hline \hline
redshift    
    & $z$ & 0.016268 & --
\\ luminosity distance 
    & $D_{\rm L}$ & 70.5 & Mpc
\\ inclination
    & $i$ &  $15^\circ$ 
    & degrees
\\  black hole mass
    & $M$ & $9\times10^6$ & \Msun
\\  accretion rate
    & $\dot{M}$ & $<0.0134$ 
    & \Msun~yr$^{-1}$
\\  inner radius
    & $R_{\rm in}$ & 6 
    & $G\,M/c^2$
\\  outer radius
    & $R_{\rm out}$ & 7.23 
    & light days
\\  rim height
    & $H_{\rm out}$ & 0.0276 
    & $R_{\rm out}$
\\ rim shape
    & $\beta\equiv d\ln{H}/d\ln{R}$ 
    & 100 
    & --
\\ irradiation
    & $\Delta T$ & 0.134 
    & dex
\\ lamp height
    & $H_x$ & 10 
    & $G\,M/c^2$
\\  lamp luminosity
    & $\Delta L_x$ & 4.38 
    & $\dot{M}\, c^2$
\\  disc efficiency
    & $\eta = L_{\rm disc}/\dot{M}\,c^2$ & 0.083 
    & --
\\  disc luminosity & 
$L_{\rm disc}=\eta\,\dot{M}\,c^2$ & 0.060 
& \Ledd
\\ max temperature 
    & $T_{\rm max}$ & >2.0 
    & $10^5$~K
\\ rim temperature & $T_{\rm min}\rightarrow T_{\rm rim}$ & $1.1\rightarrow4.2$ 
& $10^3$~K
\\ SED uncertainty & $\sigma_0$ 
& 8.5 
& percent
\\ lag uncertainty & $\sigma_\tau$
& 0.32 & day
\\ \hline
\end{tabular}
\end{center}
\end{table}


To show more clearly how the disc rim increases the time lags,
Fig.~\ref{fig:psi} shows the delay maps $\Psi(\tau|\lambda)$
for the Bowl model of Fig.~\ref{fig:bowl}.
At UV wavelengths (black, purple, blue) the delay maps,
 are dominated by reprocessing on the flat disc.
 They rise to a peak and then decline, giving mean delays
 (vertical lines) that scale as
 $\tau_0\,\left(\lambda/\lambda_0\right)^{4/3}$.
 At optical wavelengths (green, red)
 a U-shaped feature appears at longer lags, due to response on the steep rim where irradiation elevates the temperature to $\sim3000$~K, emitting optical but very little UV light.
 The time lag at radius $R$ spans the range
 $\tau=\left(1\pm\sin{i}\right)\, R/c$.
 The response peak at $\tau=(1-\cos{i})\,R/c$, from the near side of the rim, is diminished because the rim here is tilted
 away from the observer, and vice versa for the peak at $\tau=(1+\cos{i})\,R/c$ from the far side tilted toward the observer.
 Thus the mean lag for the disc + rim geometry is a $\lambda$-dependent mix interpolating
 between small lags that scale as $\lambda^{4/3}$ in the UV, and larger lags
 from the steep rim, which dominates at longer optical and near infra-red wavelengths.


Why should the disc have a steep rim at this radius?
Note in Fig.~\ref{fig:bowl}(c) that the Bowl model's faint disc $T\propto R^{-3/4}$ profile (red)
drops to $\sim1000$~K at the rim, 
and the irradiated disc (blue) drops to a minimum temperature 
before rising to $\sim3000$~K.
The 1000-1500~K temperature range correspond to expected dust sublimation temperatures in the disc atmosphere.
A similar rim temperature was found for a Bowl model fit to lag and SED data on NGC~5548 \citep{Starkey2023} and Mrk 817 \citep{Cackett_2023}. 
An increase in the disc thickness is expected outside this radius due to radiation pressure on dust grains in the disc atmosphere.
Our findings are compatible with and lend support to the FRADO (Failed Radiatively Accelerated Dusty Outflow) wind model for the inner edge of the broad emission-line region \citep{CzernyHryniewicz2011}.
In this context, the disc rim is the base of the dusty outflow at the inner edge of the low-ionization BLR and should occur at the same temperature in all AGN.

Finally, note in Fig.~\ref{fig:bowl}(a) that the SED data are above the model in the $u$ and $r$ bands near $3600\AA$ and $6500\AA$, respectively. 
Similarly, in Fig.~\ref{fig:bowl}(b), the lag data are above the model in both $u$ and $r$.
These lag and SED excesses are not straightforward to produce in a blackbody reprocessing model and may be interpreted as contributions of 10-20~\% from the Balmer continuum in the $u$ band and H$\alpha$ emission in the $r$ band.
Increased lags in the $u$ band are seen in most if not all cases when intensive monitoring has succeeded in measuring reliable inter-band continuum lags.
In the Bowl model, these increased lags and fluxes associated with Balmer emission may arise naturally if the steep rim encircling the disc has a region of optically thin gas near its crest, rather than the current model's sharp upper edge.
An opacity-dependent rim height, taller at high-opacity wavelengths, should produce bound-free edges and emission line features in the SED and lag spectra.
The Bowl model might be elaborated to include these effects, but that is beyond the scope of this paper.

\subsection{Diffuse Continuum Emission}


 The observed inter-band time lags, typically larger by a factor of 2 to 4 compared to theoretically predicted values, can arise 
 in several ways. 
 An important clue is the excess lag seen in the $u$ band in many sources, and now also in the $r$ band in our study of NGC~7469. These excess lags suggest that small time lags from the relatively compact accretion disc are increased by reverberations in the diffuse continuum and line emissions from the larger broad-line region \citep{10.1093/mnras/sty2242, Korista2019, 2019NatAs...3..251C}. 
   \citet{10.1093/mnras/stab3133} argues that the observed lags could be dominated by the BLR reverberations, diluted by flux from a compact zero-lag disc. An important point is that the u-band excess indicates that there is a Balmer continuum.  However, the Balmer continuum lags should be at all wavelengths, not just the u band.  So, simply ignoring the u band does not fully take into account the Balmer continuum lags and parameters determined from fitting lag vs lambda will not be correct.
   
  Competing models currently ignore the BLR contribution but seek to increase the lags by modifying the standard thin-disc blackbody reverberations. In the relativistic disc+corona model, the lags are increased by elevating the height of the irradiating source, i.e. the corona, to several tens of $R_g$, by realistic treatment of X-ray reflections from the disc atmosphere, and by modifying the thermal emission from the irradiated disc by elevating the colour temperature by a factor $f_{\rm col}=2.4$ above the effective temperature \cite{2019ApJ...879L..24K, 10.1093/mnras/stab725}.

Simulations by \citet{2023A&A...670A.147J} consider both the height of the irradiating source and the contribution from the diffuse continuum emission from the BLR. They find a degeneracy with both contributions being able to explain the observed lag spectra. However, they noticed also that for high accretion rate sources, this degeneracy could be resolved.
This kind of modeling will be important to apply to real data to examine the exact contribution of diffuse BLR continuum and containing the corona height. It is also important to note that in their modeling the diffuse BLR contribution comes from the simple Thomson scattering but in reality, a more rigorous treatment of the photo-ionization physics, e.g. using {\sc cloudy}, would be important.

\section{Summary and Conclusions}\label{sec:conclusions}

We report the results of an intensive reverberation mapping campaign targeting NGC~7469, a Seyfert~I galaxy with a circum-nuclear starburst ring (Fig.~\ref{fig:hst}).  We used the Las Cumbres Observatory robotic telescope network to
monitor variations in 7 optical bands with sub-day cadence over 257 days (Fig.~\ref{fig:lco}). 
Parallel monitoring at weekly intervals with {\it Swift} sampled variations in hard and soft X-rays and in 3 UV and 3 optical bands (Fig.~\ref{fig:swift}). 

We model the light curves, using the {\sc PyROA} methodology, to derive inter-band time lags for 13 UV and optical bands (Fig.~\ref{fig:lags}) and to
decompose the observed variations into the constant spectral energy distribution (SED) of the host galaxy starlight and the variable SED from the AGN accretion disc (Fig.~\ref{fig:sed}). The variable SED is close to a standard $F_\nu \propto \nu^{1/3}$ disc spectrum
and maintains this SED shape
while varying over a factor of 2.6 between the faintest and brightest states seen in the campaign.
The lags collected in Table~\ref{tab:lags} and SEDs in Table~\ref{tab:fluxflux} can be used to test models of reverberating accretion discs.

We fit the broad X-ray, UV, and optical SED with the \texttt{KYNSED} model in {\sc xspec} to test and infer parameters of a relativistic disc+corona model in the Kerr geometry, treating the x-ray corona as an isotropic point source emitting a power-law X-ray spectrum powered from the inner disc and partially reflected from the irradiated accretion disc.
The resulting SED fit achieved by MCMC methods is similar for black hole spin~0 and spin~1 (Fig.~\ref{fig:total-sed}), with corresponding parameters (Table~\ref{tab:total-sed}) including 
the inclination (65 or 41 degrees), the
accretion rate (16 or 13\% in Eddington units),
corona height (25 or $41~R_g$) 
and coronal power (75 or 90\% of the accretion power) transferred from the inner disc to the corona.

As in many previously studied cases, we find that the inter-band lags in NGC~7469 are consistent with $\tau \propto \lambda^{4/3}$, as predicted for a geometrically-thin optically-thick steady-state accretion disc with
$T\propto R^{-3/4}$, but with lags that are
up to 3 times larger than expected (Fig.~\ref{fig:lagfits}).
This suggests that the disc is larger at each wavelength and thus hotter at each radius than predicted. 

 We combined the SED and lag fits together to achieve the best fit values for the corona height and the accretion rate. Both the fits were explored for the grid of $f_{\rm col}$, and a plot summarizing fits is shown in Figure~\ref{fig:fcolgrid}. Based on the $\chi^2$ distribution, we found that the best fit value for the  $f_{\rm col}$ is 1.8, which seems to be quite high.
We note that the irradiation model generates larger disc lags primarily by assuming that at every radius the disc colour temperature is a factor $f_{\rm col}=1.8$ larger than the effective temperature.
This $f_{\rm col}$ parameter shifts the reverberation response at each radius to shorter wavelengths and lowers the surface brightness to maintain the effective temperature.
That is plausible for the hot inner disc where electron scattering opacity dominates, but not for the lower temperature at which the optical emission is produced. As such $f_{\rm col}=1.8$ merely quantifies the problem of discs being hotter than expected in the standard model of blackbody reprocessing with $f_{\rm col}=1$.

We also consider a Bowl model \citep{Starkey2023} that assumes blackbody reprocessing ($f_{\rm col}=1$) on a disc with a power-law thickness profile $H\propto R^\beta$.  
This Bowl model matches both the SED and lag constraints
(Fig.\ref{fig:bowl}) provided $\beta$ is large enough to produce a flat inner disc geometry that determines the UV-optical SED and the UV lags, encircled by a steep rim that increases the optical lags. 
The fitted model places the rim near the radius where the disc temperature is similar to dust sublimation temperatures. 
This result supports models that invoke dust opacity to thicken the disc \citep{BaskinLaor2018} and/or to launch a failed radiatively accelerated dusty outflow (FRADO) outside this radius \citep{CzernyHryniewicz2011}, as is proposed to occur at the inner edge of the low-ionization BLR.

 We note that the observed lags and SED fluxes are larger in the $u$ and $r$ bands relative to the adjacent bands and relative to the fitted models discussed above.
The SED excesses imply a 10-20\% contribution of
Balmer continuum emission in the $u$ band and H$\alpha$ emission in the $r$ band. However, rigorous modeling of the Balmer continuum at all wavelengths is required for a fair conclusion.
The lag and SED constraints reported in our paper can now be used to test more detailed models that combine reverberations from the compact accretion disc and from the larger photo-ionized broad emission-line region, to assess the impact on the inferred parameters.

Finally, we note that monitoring of NGC~7469 is continuing with ground-based robotic telescopes in parallel with a more intensive sub-day cadence monitoring with {\it Swift} and {\it NICER} for analysis in future work. 

\section*{Acknowledgements}
We thank the anonymous referee for the constructive suggestions.
This work makes use of observations from the Las Cumbres Observatory global telescope network. We acknowledge the use of public data from the {\it Swift} data archive.
R.~Prince is grateful for a visiting fellowship funded by the Scottish Universities Physics Alliance (SUPA). The project is partially supported by the Polish Funding Agency National Science Centre, project 2017/26/A/ST9/-00756 (MAESTRO 9) and the European Research Council (ERC) under the European Union’s Horizon 2020 research and innovation program (grant agreement No. [951549]) and BHU IoE scheme.
HL acknowledges a Daphne Jackson Fellowship sponsored by the Science and Technology Facilities Council (STFC), UK. We also thank Tingting Liu and Marcin Marculewicz for their helpful comments.

\section*{Data Availability}
The raw datasets were derived from sources in the public domain: LCO archive \url{https://archive.lco.global} and {\it Swift} archive \url{https://www.swift.ac.uk/swift_live}. The inter-calibrated light curves are available on request.
This research made extensive use of {\sc astropy}, a community-developed core Python package for Astronomy \citep{Astropy-Collaboration:2013aa}, {\sc matplotlib} \citep{Hunter:2007aa} and {\sc corner} to visualize MCMC posterior distributions \citep{corner2016}.



\bibliographystyle{mnras}
\bibliography{example} 

\begin{thebibliography}{}
\makeatletter
\relax
\def\mn@urlcharsother{\let\do\@makeother \do\$\do\&\do\#\do\^\do\_\do\%\do\~}
\def\mn@doi{\begingroup\mn@urlcharsother \@ifnextchar [ {\mn@doi@}
  {\mn@doi@[]}}
\def\mn@doi@[#1]#2{\def\@tempa{#1}\ifx\@tempa\@empty \href
  {http://dx.doi.org/#2} {doi:#2}\else \href {http://dx.doi.org/#2} {#1}\fi
  \endgroup}
\def\mn@eprint#1#2{\mn@eprint@#1:#2::\@nil}
\def\mn@eprint@arXiv#1{\href {http://arxiv.org/abs/#1} {{\tt arXiv:#1}}}
\def\mn@eprint@dblp#1{\href {http://dblp.uni-trier.de/rec/bibtex/#1.xml}
  {dblp:#1}}
\def\mn@eprint@#1:#2:#3:#4\@nil{\def\@tempa {#1}\def\@tempb {#2}\def\@tempc
  {#3}\ifx \@tempc \@empty \let \@tempc \@tempb \let \@tempb \@tempa \fi \ifx
  \@tempb \@empty \def\@tempb {arXiv}\fi \@ifundefined
  {mn@eprint@\@tempb}{\@tempb:\@tempc}{\expandafter \expandafter \csname
  mn@eprint@\@tempb\endcsname \expandafter{\@tempc}}}

\bibitem[\protect\citeauthoryear{{Armus} et~al.,}{{Armus}
  et~al.}{2023}]{Armus2023}
{Armus} L.,  et~al., 2023, \mn@doi [\apjl] {10.3847/2041-8213/acac66}, \href
  {https://ui.adsabs.harvard.edu/abs/2023ApJ...942L..37A} {942, L37}

\bibitem[\protect\citeauthoryear{{Arnaud}}{{Arnaud}}{1996}]{Arnaud1996}
{Arnaud} K.~A.,  1996, in {Jacoby} G.~H.,  {Barnes} J.,  eds,  Astronomical
  Society of the Pacific Conference Series Vol. 101, Astronomical Data Analysis
  Software and Systems V. p.~17

\bibitem[\protect\citeauthoryear{{Astropy Collaboration} et~al.,}{{Astropy
  Collaboration} et~al.}{2013}]{Astropy-Collaboration:2013aa}
{Astropy Collaboration} et~al., 2013, \aap, 558, A33

\bibitem[\protect\citeauthoryear{{Baskin} \& {Laor}}{{Baskin} \&
  {Laor}}{2018}]{BaskinLaor2018}
{Baskin} A.,  {Laor} A.,  2018, \mn@doi [\mnras] {10.1093/mnras/stx2850}, \href
  {https://ui.adsabs.harvard.edu/abs/2018MNRAS.474.1970B} {474, 1970}

\bibitem[\protect\citeauthoryear{{Baumgartner}, {Tueller}, {Markwardt},
  {Skinner}, {Barthelmy}, {Mushotzky}, {Evans}  \& {Gehrels}}{{Baumgartner}
  et~al.}{2013}]{Baumgartner2013}
{Baumgartner} W.~H.,  {Tueller} J.,  {Markwardt} C.~B.,  {Skinner} G.~K.,
  {Barthelmy} S.,  {Mushotzky} R.~F.,  {Evans} P.~A.,   {Gehrels} N.,  2013,
  \mn@doi [\apjs] {10.1088/0067-0049/207/2/19}, \href
  {https://ui.adsabs.harvard.edu/abs/2013ApJS..207...19B} {207, 19}

\bibitem[\protect\citeauthoryear{{Bentz} \& {Katz}}{{Bentz} \&
  {Katz}}{2015}]{2015PASP..127...67B}
{Bentz} M.~C.,  {Katz} S.,  2015, \mn@doi [\pasp] {10.1086/679601}, \href
  {https://ui.adsabs.harvard.edu/abs/2015PASP..127...67B} {127, 67}

\bibitem[\protect\citeauthoryear{{Bertin} \& {Arnouts}}{{Bertin} \&
  {Arnouts}}{1996}]{bertin:1996}
{Bertin} E.,  {Arnouts} S.,  1996, \mn@doi [\aaps] {10.1051/aas:1996164}, \href
  {https://ui.adsabs.harvard.edu/abs/1996A&AS..117..393B} {117, 393}

\bibitem[\protect\citeauthoryear{{Blandford} \& {McKee}}{{Blandford} \&
  {McKee}}{1982}]{1982ApJ...255..419B}
{Blandford} R.~D.,  {McKee} C.~F.,  1982, \mn@doi [\apj] {10.1086/159843},
  \href {https://ui.adsabs.harvard.edu/abs/1982ApJ...255..419B} {255, 419}

\bibitem[\protect\citeauthoryear{{Brown} et~al.,}{{Brown}
  et~al.}{2013}]{Brown:2013}
{Brown} T.~M.,  et~al., 2013, \mn@doi [\pasp] {10.1086/673168}, \href
  {https://ui.adsabs.harvard.edu/abs/2013PASP..125.1031B} {125, 1031}

\bibitem[\protect\citeauthoryear{{Burrows} et~al.,}{{Burrows}
  et~al.}{2005}]{2005SSRv..120..165B}
{Burrows} D.~N.,  et~al., 2005, \mn@doi [\ssr] {10.1007/s11214-005-5097-2},
  \href {https://ui.adsabs.harvard.edu/abs/2005SSRv..120..165B} {120, 165}

\bibitem[\protect\citeauthoryear{{Cackett}, {Horne}  \& {Winkler}}{{Cackett}
  et~al.}{2007}]{2007MNRAS.380..669C}
{Cackett} E.~M.,  {Horne} K.,   {Winkler} H.,  2007, \mn@doi [\mnras]
  {10.1111/j.1365-2966.2007.12098.x}, \href
  {https://ui.adsabs.harvard.edu/abs/2007MNRAS.380..669C} {380, 669}

\bibitem[\protect\citeauthoryear{Cackett, Chiang, McHardy, Edelson, Goad, Horne
   \& Korista}{Cackett et~al.}{2018}]{Cackett_2018}
Cackett E.~M.,  Chiang C.-Y.,  McHardy I.,  Edelson R.,  Goad M.~R.,  Horne K.,
    Korista K.~T.,  2018, \mn@doi [ApJ] {10.3847/1538-4357/aab4f7}, 857, 53

\bibitem[\protect\citeauthoryear{Cackett et~al.,}{Cackett
  et~al.}{2020}]{Cackett_2020}
Cackett E.~M.,  et~al., 2020, \mn@doi [ApJ] {10.3847/1538-4357/ab91b5}, 896, 1

\bibitem[\protect\citeauthoryear{Cackett et~al.,}{Cackett
  et~al.}{2023}]{Cackett_2023}
Cackett E.~M.,  et~al., 2023, \mn@doi [ApJ] {10.3847/1538-4357/acfdac}, 958,
  195

\bibitem[\protect\citeauthoryear{{Chelouche}, {Pozo Nu{\~n}ez}  \&
  {Kaspi}}{{Chelouche} et~al.}{2019}]{2019NatAs...3..251C}
{Chelouche} D.,  {Pozo Nu{\~n}ez} F.,   {Kaspi} S.,  2019, \mn@doi [Nature
  Astronomy] {10.1038/s41550-018-0659-x}, \href
  {https://ui.adsabs.harvard.edu/abs/2019NatAs...3..251C} {3, 251}

\bibitem[\protect\citeauthoryear{{Collier} et~al.,}{{Collier}
  et~al.}{1998}]{1998ApJ...500..162C}
{Collier} S.~J.,  et~al., 1998, \mn@doi [\apj] {10.1086/305720}, \href
  {https://ui.adsabs.harvard.edu/abs/1998ApJ...500..162C} {500, 162}

\bibitem[\protect\citeauthoryear{{Combes}}{{Combes}}{2023}]{Combes2023}
{Combes} F.,  2023, \mn@doi [arXiv e-prints] {10.48550/arXiv.2302.12917}, \href
  {https://ui.adsabs.harvard.edu/abs/2023arXiv230212917C} {p. arXiv:2302.12917}

\bibitem[\protect\citeauthoryear{{Czerny} \& {Hryniewicz}}{{Czerny} \&
  {Hryniewicz}}{2011}]{CzernyHryniewicz2011}
{Czerny} B.,  {Hryniewicz} K.,  2011, \mn@doi [\aap]
  {10.1051/0004-6361/201016025}, \href
  {https://ui.adsabs.harvard.edu/abs/2011A&A...525L...8C} {525, L8}

\bibitem[\protect\citeauthoryear{{D{\'\i}az-Santos}, {Alonso-Herrero},
  {Colina}, {Ryder}  \& {Knapen}}{{D{\'\i}az-Santos}
  et~al.}{2007}]{DiazSantos2007}
{D{\'\i}az-Santos} T.,  {Alonso-Herrero} A.,  {Colina} L.,  {Ryder} S.~D.,
  {Knapen} J.~H.,  2007, \mn@doi [\apj] {10.1086/513089}, \href
  {https://ui.adsabs.harvard.edu/abs/2007ApJ...661..149D} {661, 149}

\bibitem[\protect\citeauthoryear{Done, Davis, Jin, Blaes  \& Ward}{Done
  et~al.}{2012}]{10.1111/j.1365-2966.2011.19779.x}
Done C.,  Davis S.~W.,  Jin C.,  Blaes O.,   Ward M.,  2012, \mn@doi [MNRAS]
  {10.1111/j.1365-2966.2011.19779.x}, 420, 1848

\bibitem[\protect\citeauthoryear{{Donnan}, {Horne}  \& {Hern{\'a}ndez
  Santisteban}}{{Donnan} et~al.}{2021}]{Donnan2021}
{Donnan} F.~R.,  {Horne} K.,   {Hern{\'a}ndez Santisteban} J.~V.,  2021,
  \mn@doi [\mnras] {10.1093/mnras/stab2832}, \href
  {https://ui.adsabs.harvard.edu/abs/2021MNRAS.508.5449D} {508, 5449}

\bibitem[\protect\citeauthoryear{{Donnan} et~al.,}{{Donnan}
  et~al.}{2023}]{Donnan2023}
{Donnan} F.~R.,  et~al., 2023, \mn@doi [\mnras] {10.1093/mnras/stad1409}, \href
  {https://ui.adsabs.harvard.edu/abs/2023MNRAS.523..545D} {523, 545}

\bibitem[\protect\citeauthoryear{{Dov{\v{c}}iak}, {Papadakis}, {Kammoun}  \&
  {Zhang}}{{Dov{\v{c}}iak} et~al.}{2022}]{2022A&A...661A.135D}
{Dov{\v{c}}iak} M.,  {Papadakis} I.~E.,  {Kammoun} E.~S.,   {Zhang} W.,  2022,
  \mn@doi [\aap] {10.1051/0004-6361/202142358}, \href
  {https://ui.adsabs.harvard.edu/abs/2022A&A...661A.135D} {661, A135}

\bibitem[\protect\citeauthoryear{{Edelson} et~al.,}{{Edelson}
  et~al.}{2015}]{2015ApJ...806..129E}
{Edelson} R.,  et~al., 2015, \mn@doi [\apj] {10.1088/0004-637X/806/1/129},
  \href {https://ui.adsabs.harvard.edu/abs/2015ApJ...806..129E} {806, 129}

\bibitem[\protect\citeauthoryear{{Edelson} et~al.,}{{Edelson}
  et~al.}{2017}]{2017ApJ...840...41E}
{Edelson} R.,  et~al., 2017, \mn@doi [\apj] {10.3847/1538-4357/aa6890}, \href
  {https://ui.adsabs.harvard.edu/abs/2017ApJ...840...41E} {840, 41}

\bibitem[\protect\citeauthoryear{{Edelson} et~al.,}{{Edelson}
  et~al.}{2019}]{2019ApJ...870..123E}
{Edelson} R.,  et~al., 2019, \mn@doi [\apj] {10.3847/1538-4357/aaf3b4}, \href
  {https://ui.adsabs.harvard.edu/abs/2019ApJ...870..123E} {870, 123}

\bibitem[\protect\citeauthoryear{{Evans} et~al.,}{{Evans}
  et~al.}{2009}]{Evans:2009}
{Evans} P.~A.,  et~al., 2009, \mn@doi [\mnras]
  {10.1111/j.1365-2966.2009.14913.x}, \href
  {https://ui.adsabs.harvard.edu/abs/2009MNRAS.397.1177E} {397, 1177}

\bibitem[\protect\citeauthoryear{{Event Horizon Telescope Collaboration},
  {Akiyama}, {Alberdi}, {Alef}, {Asada}  \& {Azulay}}{{Event Horizon Telescope
  Collaboration} et~al.}{2019}]{2019ApJ...875L...1E}
{Event Horizon Telescope Collaboration} {Akiyama} K.,  {Alberdi} A.,  {Alef}
  W.,  {Asada} K.,   {Azulay} R.,  2019, \mn@doi [\apjl]
  {10.3847/2041-8213/ab0ec7}, \href
  {https://ui.adsabs.harvard.edu/abs/2019ApJ...875L...1E} {875, L1}

\bibitem[\protect\citeauthoryear{{Event Horizon Telescope Collaboration}
  et~al.,}{{Event Horizon Telescope Collaboration} et~al.}{2022}]{EHT2022}
{Event Horizon Telescope Collaboration} et~al., 2022, \mn@doi [\apjl]
  {10.3847/2041-8213/ac6674}, \href
  {https://ui.adsabs.harvard.edu/abs/2022ApJ...930L..12E} {930, L12}

\bibitem[\protect\citeauthoryear{{Fausnaugh} et~al.,}{{Fausnaugh}
  et~al.}{2016}]{2016ApJ...821...56F}
{Fausnaugh} M.~M.,  et~al., 2016, \mn@doi [\apj] {10.3847/0004-637X/821/1/56},
  \href {https://ui.adsabs.harvard.edu/abs/2016ApJ...821...56F} {821, 56}

\bibitem[\protect\citeauthoryear{Fausnaugh et~al.,}{Fausnaugh
  et~al.}{2018}]{Fausnaugh_2018}
Fausnaugh M.~M.,  et~al., 2018, \mn@doi [ApJ] {10.3847/1538-4357/aaaa2b}, 854,
  107

\bibitem[\protect\citeauthoryear{{Flewelling} et~al.,}{{Flewelling}
  et~al.}{2020}]{Flewelling2020}
{Flewelling} H.~A.,  et~al., 2020, \mn@doi [\apjs] {10.3847/1538-4365/abb82d},
  \href {https://ui.adsabs.harvard.edu/abs/2020ApJS..251....7F} {251, 7}

\bibitem[\protect\citeauthoryear{{Foreman-Mackey}}{{Foreman-Mackey}}{2016}]{corner2016}
{Foreman-Mackey} D.,  2016, \mn@doi [The Journal of Open Source Software]
  {10.21105/joss.00024}, \href
  {https://ui.adsabs.harvard.edu/abs/2016JOSS....1...24F} {1, 24}

\bibitem[\protect\citeauthoryear{Foreman-Mackey, Hogg, Lang  \&
  Goodman}{Foreman-Mackey et~al.}{2013}]{Foreman-Mackey_2013}
Foreman-Mackey D.,  Hogg D.~W.,  Lang D.,   Goodman J.,  2013, \mn@doi [PASP]
  {10.1086/670067}, 125, 306

\bibitem[\protect\citeauthoryear{{Gardner} \& {Done}}{{Gardner} \&
  {Done}}{2017}]{Gardner2017}
{Gardner} E.,  {Done} C.,  2017, \mn@doi [\mnras] {10.1093/mnras/stx946}, \href
  {https://ui.adsabs.harvard.edu/abs/2017MNRAS.470.3591G} {470, 3591}

\bibitem[\protect\citeauthoryear{Gaskell}{Gaskell}{2017}]{Gaskell17}
Gaskell C.~M.,  2017, \mn@doi [MNRAS] {10.1093/mnras/stx094}, 467, 226

\bibitem[\protect\citeauthoryear{{Gehrels} et~al.,}{{Gehrels}
  et~al.}{2004}]{Gehrels:2004}
{Gehrels} N.,  et~al., 2004, \mn@doi [\apj] {10.1086/422091}, \href
  {https://ui.adsabs.harvard.edu/abs/2004ApJ...611.1005G} {611, 1005}

\bibitem[\protect\citeauthoryear{{Guo}, {Barth}  \& {Wang}}{{Guo}
  et~al.}{2022}]{2022arXiv220706432G}
{Guo} H.,  {Barth} A.~J.,   {Wang} S.,  2022, arXiv e-prints, \href
  {https://ui.adsabs.harvard.edu/abs/2022arXiv220706432G} {p. arXiv:2207.06432}

\bibitem[\protect\citeauthoryear{{Haardt} \& {Maraschi}}{{Haardt} \&
  {Maraschi}}{1991}]{Haardt1991}
{Haardt} F.,  {Maraschi} L.,  1991, \mn@doi [\apjl] {10.1086/186171}, \href
  {https://ui.adsabs.harvard.edu/abs/1991ApJ...380L..51H} {380, L51}

\bibitem[\protect\citeauthoryear{{Henden}, {Levine}, {Terrell}, {Welch},
  {Munari}  \& {Kloppenborg}}{{Henden} et~al.}{2018}]{henden:2018}
{Henden} A.~A.,  {Levine} S.,  {Terrell} D.,  {Welch} D.~L.,  {Munari} U.,
  {Kloppenborg} B.~K.,  2018, in American Astronomical Society Meeting
  Abstracts \#232. p. 223.06

\bibitem[\protect\citeauthoryear{{Hern{\'a}ndez Santisteban}
  et~al.,}{{Hern{\'a}ndez Santisteban} et~al.}{2020}]{Hernandez20}
{Hern{\'a}ndez Santisteban} J.~V.,  et~al., 2020, \mn@doi [\mnras]
  {10.1093/mnras/staa2365}, \href
  {https://ui.adsabs.harvard.edu/abs/2020MNRAS.498.5399H} {498, 5399}

\bibitem[\protect\citeauthoryear{{Hunter, J. D.}}{{Hunter, J.
  D.}}{2007}]{Hunter:2007aa}
{Hunter, J. D.} 2007, {Computing In Science \& Engineering}, 9, 90

\bibitem[\protect\citeauthoryear{{Jaiswal}, {Prince}, {Panda}  \&
  {Czerny}}{{Jaiswal} et~al.}{2023}]{2023A&A...670A.147J}
{Jaiswal} V.~K.,  {Prince} R.,  {Panda} S.,   {Czerny} B.,  2023, \mn@doi
  [\aap] {10.1051/0004-6361/202244352}, \href
  {https://ui.adsabs.harvard.edu/abs/2023A&A...670A.147J} {670, A147}

\bibitem[\protect\citeauthoryear{{Kammoun}, {Papadakis}  \&
  {Dov{\v{c}}iak}}{{Kammoun} et~al.}{2019}]{2019ApJ...879L..24K}
{Kammoun} E.~S.,  {Papadakis} I.~E.,   {Dov{\v{c}}iak} M.,  2019, \mn@doi
  [\apjl] {10.3847/2041-8213/ab2a72}, \href
  {https://ui.adsabs.harvard.edu/abs/2019ApJ...879L..24K} {879, L24}

\bibitem[\protect\citeauthoryear{Kammoun, Papadakis  \& Dovčiak}{Kammoun
  et~al.}{2021}]{10.1093/mnras/stab725}
Kammoun E.~S.,  Papadakis I.~E.,   Dovčiak M.,  2021, \mn@doi [MNRAS]
  {10.1093/mnras/stab725}, 503, 4163

\bibitem[\protect\citeauthoryear{Kammoun, Robin, Papadakis, Dovčiak  \&
  Panagiotou}{Kammoun et~al.}{2023}]{10.1093/mnras/stad2701}
Kammoun E.~S.,  Robin L.,  Papadakis I.~E.,  Dovčiak M.,   Panagiotou C.,
  2023, \mn@doi [MNRAS] {10.1093/mnras/stad2701}, 526, 138

\bibitem[\protect\citeauthoryear{{Korista} \& {Goad}}{{Korista} \&
  {Goad}}{2001}]{Korista2001}
{Korista} K.~T.,  {Goad} M.~R.,  2001, \mn@doi [\apj] {10.1086/320964}, \href
  {https://ui.adsabs.harvard.edu/abs/2001ApJ...553..695K} {553, 695}

\bibitem[\protect\citeauthoryear{{Korista} \& {Goad}}{{Korista} \&
  {Goad}}{2019}]{Korista2019}
{Korista} K.~T.,  {Goad} M.~R.,  2019, \mn@doi [\mnras]
  {10.1093/mnras/stz2330}, \href
  {https://ui.adsabs.harvard.edu/abs/2019MNRAS.489.5284K} {489, 5284}

\bibitem[\protect\citeauthoryear{{Kormendy} \& {Ho}}{{Kormendy} \&
  {Ho}}{2013}]{Kormendy2013}
{Kormendy} J.,  {Ho} L.~C.,  2013, \mn@doi [\araa]
  {10.1146/annurev-astro-082708-101811}, \href
  {https://ui.adsabs.harvard.edu/abs/2013ARA&A..51..511K} {51, 511}

\bibitem[\protect\citeauthoryear{{Kubota} \& {Done}}{{Kubota} \&
  {Done}}{2018}]{2018MNRAS.480.1247K}
{Kubota} A.,  {Done} C.,  2018, \mn@doi [\mnras] {10.1093/mnras/sty1890}, \href
  {https://ui.adsabs.harvard.edu/abs/2018MNRAS.480.1247K} {480, 1247}

\bibitem[\protect\citeauthoryear{{Landt}}{{Landt}}{2023}]{2023FrASS..1056088L}
{Landt} H.,  2023, \mn@doi [Frontiers in Astronomy and Space Sciences]
  {10.3389/fspas.2023.1256088}, \href
  {https://ui.adsabs.harvard.edu/abs/2023FrASS..1056088L} {10, 1256088}

\bibitem[\protect\citeauthoryear{Lawther, Goad, Korista, Ulrich  \&
  Vestergaard}{Lawther et~al.}{2018}]{10.1093/mnras/sty2242}
Lawther D.,  Goad M.~R.,  Korista K.~T.,  Ulrich O.,   Vestergaard M.,  2018,
  \mn@doi [MNRAS] {10.1093/mnras/sty2242}, 481, 533

\bibitem[\protect\citeauthoryear{{Mahmoud} \& {Done}}{{Mahmoud} \&
  {Done}}{2020}]{Mahmoud2020}
{Mahmoud} R.~D.,  {Done} C.,  2020, \mn@doi [\mnras] {10.1093/mnras/stz3196},
  \href {https://ui.adsabs.harvard.edu/abs/2020MNRAS.491.5126M} {491, 5126}

\bibitem[\protect\citeauthoryear{{McHardy} et~al.,}{{McHardy}
  et~al.}{2014}]{2014MNRAS.444.1469M}
{McHardy} I.~M.,  et~al., 2014, \mn@doi [\mnras] {10.1093/mnras/stu1636}, \href
  {https://ui.adsabs.harvard.edu/abs/2014MNRAS.444.1469M} {444, 1469}

\bibitem[\protect\citeauthoryear{{McHardy} et~al.,}{{McHardy}
  et~al.}{2018}]{2018MNRAS.480.2881M}
{McHardy} I.~M.,  et~al., 2018, \mn@doi [\mnras] {10.1093/mnras/sty1983}, \href
  {https://ui.adsabs.harvard.edu/abs/2018MNRAS.480.2881M} {480, 2881}

\bibitem[\protect\citeauthoryear{{Mehdipour} et~al.,}{{Mehdipour}
  et~al.}{2018}]{Mehdipour2018}
{Mehdipour} M.,  et~al., 2018, \mn@doi [\aap] {10.1051/0004-6361/201832604},
  \href {https://ui.adsabs.harvard.edu/abs/2018A&A...615A..72M} {615, A72}

\bibitem[\protect\citeauthoryear{Mirzaev, Bambi, Abdikamalov, Jiang, Liu, Riaz
  \& Shashank}{Mirzaev et~al.}{2024}]{Mirzaev_2024}
Mirzaev T.,  Bambi C.,  Abdikamalov A.~B.,  Jiang J.,  Liu H.,  Riaz S.,
  Shashank S.,  2024, \mn@doi [The Astrophysical Journal]
  {10.3847/1538-4357/ad8a63}, 976, 229

\bibitem[\protect\citeauthoryear{{Nandra}, {Clavel}, {Edelson}, {George},
  {Malkan}, {Mushotzky}, {Peterson}  \& {Turner}}{{Nandra}
  et~al.}{1998}]{Nandra1998ApJ...505..594N}
{Nandra} K.,  {Clavel} J.,  {Edelson} R.~A.,  {George} I.~M.,  {Malkan} M.~A.,
  {Mushotzky} R.~F.,  {Peterson} B.~M.,   {Turner} T.~J.,  1998, \mn@doi [\apj]
  {10.1086/306181}, \href
  {https://ui.adsabs.harvard.edu/abs/1998ApJ...505..594N} {505, 594}

\bibitem[\protect\citeauthoryear{Netzer}{Netzer}{2021}]{10.1093/mnras/stab3133}
Netzer H.,  2021, \mn@doi [MNRAS] {10.1093/mnras/stab3133}, 509, 2637

\bibitem[\protect\citeauthoryear{{Netzer}}{{Netzer}}{2022}]{Netzer2022}
{Netzer} H.,  2022, \mn@doi [\mnras] {10.1093/mnras/stab3133}, \href
  {https://ui.adsabs.harvard.edu/abs/2022MNRAS.509.2637N} {509, 2637}

\bibitem[\protect\citeauthoryear{{Pahari}, {McHardy}, {Vincentelli}, {Cackett},
  {Peterson}, {Goad}, {G{\"u}ltekin}  \& {Horne}}{{Pahari}
  et~al.}{2020}]{2020MNRAS.494.4057P}
{Pahari} M.,  {McHardy} I.~M.,  {Vincentelli} F.,  {Cackett} E.,  {Peterson}
  B.~M.,  {Goad} M.,  {G{\"u}ltekin} K.,   {Horne} K.,  2020, \mn@doi [\mnras]
  {10.1093/mnras/staa1055}, \href
  {https://ui.adsabs.harvard.edu/abs/2020MNRAS.494.4057P} {494, 4057}

\bibitem[\protect\citeauthoryear{{Pal} \& {Naik}}{{Pal} \&
  {Naik}}{2018}]{2018MNRAS.474.5351P}
{Pal} M.,  {Naik} S.,  2018, \mn@doi [\mnras] {10.1093/mnras/stx3103}, \href
  {https://ui.adsabs.harvard.edu/abs/2018MNRAS.474.5351P} {474, 5351}

\bibitem[\protect\citeauthoryear{{Peterson}}{{Peterson}}{2014}]{2014SSRv..183..253P}
{Peterson} B.~M.,  2014, \mn@doi [\ssr] {10.1007/s11214-013-9987-4}, \href
  {https://ui.adsabs.harvard.edu/abs/2014SSRv..183..253P} {183, 253}

\bibitem[\protect\citeauthoryear{{Peterson} et~al.,}{{Peterson}
  et~al.}{2004}]{2004ApJ...613..682P}
{Peterson} B.~M.,  et~al., 2004, \mn@doi [\apj] {10.1086/423269}, \href
  {https://ui.adsabs.harvard.edu/abs/2004ApJ...613..682P} {613, 682}

\bibitem[\protect\citeauthoryear{{Roming} et~al.,}{{Roming}
  et~al.}{2005}]{2005SSRv..120...95R}
{Roming} P. W.~A.,  et~al., 2005, \mn@doi [\ssr] {10.1007/s11214-005-5095-4},
  \href {https://ui.adsabs.harvard.edu/abs/2005SSRv..120...95R} {120, 95}

\bibitem[\protect\citeauthoryear{{Schlegel}, {Finkbeiner}  \&
  {Davis}}{{Schlegel} et~al.}{1998}]{1998ApJ...500..525S}
{Schlegel} D.~J.,  {Finkbeiner} D.~P.,   {Davis} M.,  1998, \mn@doi [\apj]
  {10.1086/305772}, \href
  {https://ui.adsabs.harvard.edu/abs/1998ApJ...500..525S} {500, 525}

\bibitem[\protect\citeauthoryear{{Seyfert}}{{Seyfert}}{1943}]{Seyfert1943}
{Seyfert} C.~K.,  1943, \mn@doi [\apj] {10.1086/144488}, \href
  {https://ui.adsabs.harvard.edu/abs/1943ApJ....97...28S} {97, 28}

\bibitem[\protect\citeauthoryear{{Shakura} \& {Sunyaev}}{{Shakura} \&
  {Sunyaev}}{1973}]{1973SS}
{Shakura} N.~I.,  {Sunyaev} R.~A.,  1973, \aap, \href
  {https://ui.adsabs.harvard.edu/abs/1973A&A....24..337S} {24, 337}

\bibitem[\protect\citeauthoryear{{Shappee} et~al.,}{{Shappee}
  et~al.}{2014}]{2014ApJ...788...48S}
{Shappee} B.~J.,  et~al., 2014, \mn@doi [\apj] {10.1088/0004-637X/788/1/48},
  \href {https://ui.adsabs.harvard.edu/abs/2014ApJ...788...48S} {788, 48}

\bibitem[\protect\citeauthoryear{{Shen} et~al.,}{{Shen}
  et~al.}{2023}]{Shen2023}
{Shen} Y.,  et~al., 2023, \mn@doi [arXiv e-prints] {10.48550/arXiv.2305.01014},
  \href {https://ui.adsabs.harvard.edu/abs/2023arXiv230501014S} {p.
  arXiv:2305.01014}

\bibitem[\protect\citeauthoryear{{Shimura} \& {Takahara}}{{Shimura} \&
  {Takahara}}{1995}]{1995ApJ...445..780S}
{Shimura} T.,  {Takahara} F.,  1995, \mn@doi [\apj] {10.1086/175740}, \href
  {https://ui.adsabs.harvard.edu/abs/1995ApJ...445..780S} {445, 780}

\bibitem[\protect\citeauthoryear{{Springob}, {Haynes}, {Giovanelli}  \&
  {Kent}}{{Springob} et~al.}{2005}]{Springob2005}
{Springob} C.~M.,  {Haynes} M.~P.,  {Giovanelli} R.,   {Kent} B.~R.,  2005,
  \mn@doi [\apjs] {10.1086/431550}, \href
  {https://ui.adsabs.harvard.edu/abs/2005ApJS..160..149S} {160, 149}

\bibitem[\protect\citeauthoryear{{Starkey}, {Huang}, {Horne}  \&
  {Lin}}{{Starkey} et~al.}{2023}]{Starkey2023}
{Starkey} D.~A.,  {Huang} J.,  {Horne} K.,   {Lin} D. N.~C.,  2023, \mn@doi
  [\mnras] {10.1093/mnras/stac3579}, \href
  {https://ui.adsabs.harvard.edu/abs/2023MNRAS.519.2754S} {519, 2754}

\bibitem[\protect\citeauthoryear{{Sun}, {Grier}  \& {Peterson}}{{Sun}
  et~al.}{2018}]{Sun2018}
{Sun} M.,  {Grier} C.~J.,   {Peterson} B.~M.,  2018, {PyCCF: Python Cross
  Correlation Function for reverberation mapping studies} (\mn@eprint {ascl}
  {1805.032})

\bibitem[\protect\citeauthoryear{{Sun} et~al.,}{{Sun} et~al.}{2020}]{Sun2020}
{Sun} M.,  et~al., 2020, \mn@doi [\apj] {10.3847/1538-4357/abb1c4}, \href
  {https://ui.adsabs.harvard.edu/abs/2020ApJ...902....7S} {902, 7}

\bibitem[\protect\citeauthoryear{Tie \& Kochanek}{Tie \&
  Kochanek}{2017}]{10.1093/mnras/stx2348}
Tie S.~S.,  Kochanek C.~S.,  2017, \mn@doi [MNRAS] {10.1093/mnras/stx2348},
  473, 80

\bibitem[\protect\citeauthoryear{{Troyer}, {Starkey}, {Cackett}, {Bentz},
  {Goad}, {Horne}  \& {Seals}}{{Troyer} et~al.}{2016}]{2016MNRAS.456.4040T}
{Troyer} J.,  {Starkey} D.,  {Cackett} E.~M.,  {Bentz} M.~C.,  {Goad} M.~R.,
  {Horne} K.,   {Seals} J.~E.,  2016, \mn@doi [\mnras] {10.1093/mnras/stv2862},
  \href {https://ui.adsabs.harvard.edu/abs/2016MNRAS.456.4040T} {456, 4040}

\bibitem[\protect\citeauthoryear{{Wanders} et~al.,}{{Wanders}
  et~al.}{1997}]{Wanders1997}
{Wanders} I.,  et~al., 1997, \mn@doi [\apjs] {10.1086/313054}, \href
  {https://ui.adsabs.harvard.edu/abs/1997ApJS..113...69W} {113, 69}

\bibitem[\protect\citeauthoryear{{White} \& {Peterson}}{{White} \&
  {Peterson}}{1994}]{White1994}
{White} R.~J.,  {Peterson} B.~M.,  1994, \mn@doi [\pasp] {10.1086/133456},
  \href {https://ui.adsabs.harvard.edu/abs/1994PASP..106..879W} {106, 879}

\makeatother
\end{thebibliography}




\appendix

\section{Comparison of PyROA with PyCCF}

\begin{figure*}
    \includegraphics[scale=0.7]{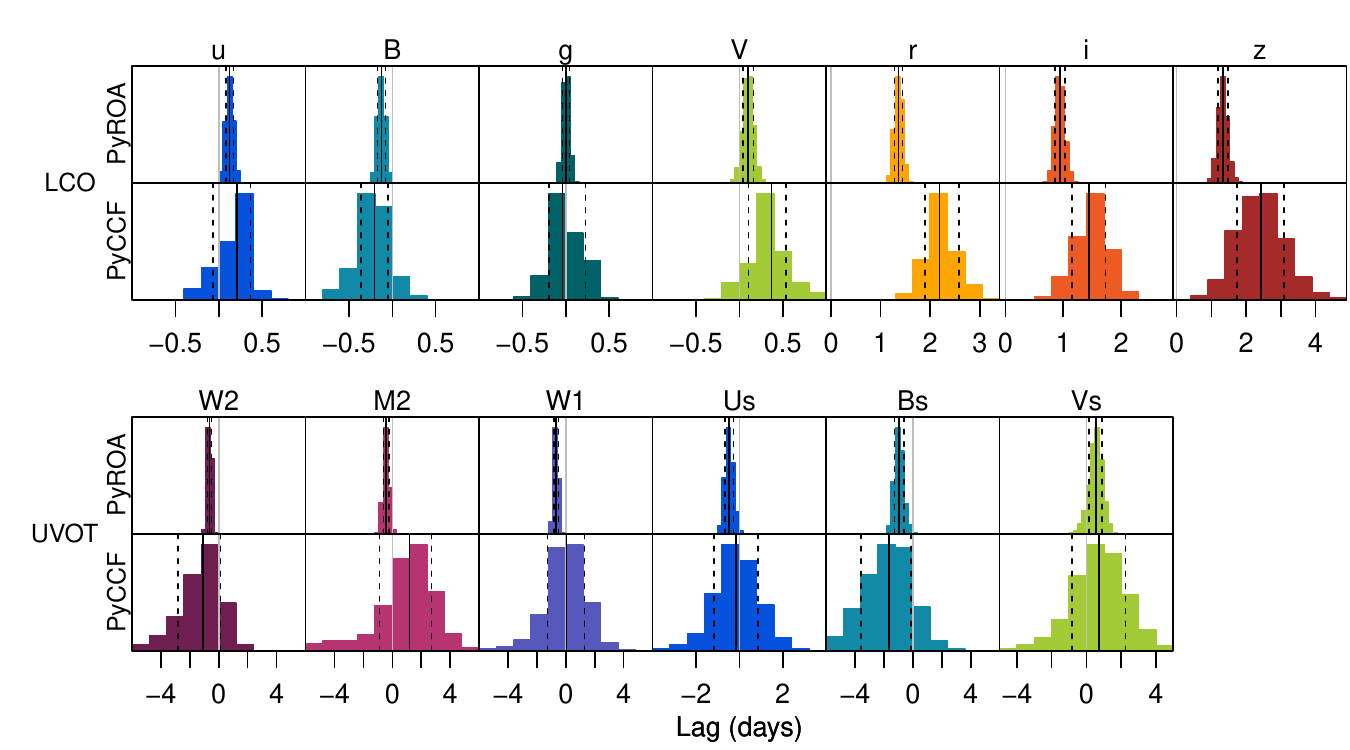}
    \caption{Comparison of derived lags from PYROA and ICCF method for LCO data ({\it top panel}) and for UVOT/{\it Swift} ({\it bottom panel}). The lags, $1\sigma$ uncertainties and the ratio between the uncertainties for each method are tabulated in Table \ref{tab:iccf}. 
    \label{fig:iccf} }
\end{figure*}

{In this appendix we compare the PyROA lag measurements derived by fitting the multi-band lightcurve data (Section\, \ref{sec:timeseries}) 
with lag measurements from the more traditional Interpolated Cross-Correlation Function (ICCF) methodology \citep{White1994}, as implemented in
the PyCCF code \citep{Sun2018}.
} The reported lags were determined by measuring the centroid of the CCF for values where the correlation coefficient $R> 0.8R_{\rm max}$, where $R_{\rm max}$ is the maximum correlation coefficient. The lag uncertainties were determined by $10^4$ Monte Carlo simulations using the flux randomization/random subset selection (FR/RSS). 
{The PyCCF and PyROA lags are listed in Table \ref{tab:iccf}.
Their probability distributions can be compared in Fig\,\ref{fig:iccf}.}

{
The PyCCF and PyROA lags are consistent within their uncertainty estimates for all bands.
However, the PyROA uncertainties are significantly smaller than those derived from PyCCF.
For this reason we opted to use PyROA lags rather than PyCCF lags in our analysis.
There are several reasons to consider for this difference:
\\
1) The FR/RSS method used to estimate ICCF lag uncertainties creates mock datasets using both bootstrap resampling (RSS=Random Subset Selection) and Monte-Carlo resampling of the measurement errors (FR=Flux Resampling). 
This standard procedure double-counts the statistical uncertainties.
Tests with simulated light curve data (random walk plus Gaussian noise) indicate that
this double-counting can increase the ICCF lag uncertainty estimates by a factor $\sqrt{2}$ for well-sampled datasets similar to these.
\\
2) ICCF lag estimates give equal weight to each datum, rather than optimal inverse-variance weights used with the ROA and other lightcurve fitting methods. 
We therefore expect smaller uncertainties for ROA lags.
This makes little difference if the uncertainties are close to the same value, but could degrade ICCF lags if data with larger uncertainties are given undue weight.
\\
3) The ROA light curve shape $X(t)$ is defined by fitting all of the data. 
With many light curves defining $X(t)$, it is effectively noise free (except in data gaps).
ICCF linearly interpolates noisy data to define one or both light curves.
Consequently, ICCF lags should be noisier by up to an additional factor of $\sqrt{2}$ (less if one light curve is much noisier than the other).
\\
4) Noise in the reference light curve introduces a correlated error in the ICCF lags across all the echo light curves.
Such correlations should be weaker for ROA lags where $X(t)$ is effectively noise free.
This may be unimportant if the reference light curve is much higher S/N than the echo light curves.
\\
5) The ROA model assumes that all light curves have the same shape, $X(t)$, and thus a delta distribution for each of the inter-band lags.
ICCF interpolates noisy data to define light curves with different shapes.
As we expect a finite width and asymmetric delay distribution, with positive skew,
the ROA lags should be shorter than the ICCF lags.  
Fig.\,\ref{fig:iccf} shows some evidence of this, particularly for the redder bands. However, large ICCF uncertainties tend to obscure this trend.
\\
6)
ROA lags from a fit that omits parameters for the width and shape of the delay distribution can be biased and their uncertainties can be under-estimated, due to neglect of parameter covariances.
This is important to consider when very good quality data provide information on higher moments of the delay distribution, such as significant smoothing or asymmetry of sharp light curve features at longer wavelengths.
The PyROA fit residuals suggest that the delta distribution is not a bad approximation.
\\
7) The ROA smoothing parameter $\Delta$ is a fit parameter optimised by minimising the BIC. 
A more flexible $X(t)$ introduces more parameters $p(\Delta)$.
Rapid peaks and dips in the highest quality light curves may require a small $\Delta$.
When the delay distribution is asymmetric, with positive skew, the ROA lag has a positive correlation with $\Delta$.
The $\tau(\Delta)$ trend can be investigated to quantify and interpret this effect
, but we have not pursued this here, since the simpler delta distribution model works well enough for our purposes.
}

\begin{table}
    \centering
    \begin{tabular}{cccc}
    \hline
      Bands   & PyCCF & PyROA &  $\sigma_{\rm PyCCF} / \sigma_{\rm PyROA}$ \\
      \hline
       $u$  & $0.115\pm0.148$   & $0.082\pm-0.041$ & 3.6     \\
       $B$  & $-0.204\pm0.158$   & $-0.156\pm-0.043$ & 3.7   \\
       $g$  & $-0.039\pm0.154$   &  $-0.039\pm-0.038$ & 4.0   \\
       $V$   & $0.369\pm0.265$    & $0.044\pm-0.061$  & 4.3  \\
       $r$   &  $2.185\pm0.293$   & $1.266\pm-0.076$  & 3.8  \\
       $i$  &  $1.442\pm0.292$   & $0.844\pm-0.086$   & 3.4  \\
       $z$  &  $2.423\pm0.679$   &  $1.223\pm-0.134$   & 5.1 \\
       \hline
    \end{tabular}
    \caption{Comparison of lags derived from PyCCF and PyROA. Lag uncertainties from PyCCF are larger by factors of 3 to 5 than those from PyROA, consistent with what is derived in \citet{Donnan2021}}
    \label{tab:iccf}
\end{table}
\bsp	
\label{lastpage}
\end{document}